\documentclass[11pt]{article}

\usepackage[preprint]{acl}

\usepackage{times}
\usepackage{latexsym}
\usepackage{amsmath}
\usepackage{amssymb}
\usepackage{booktabs}
\usepackage{multirow}
\usepackage{xcolor}
\usepackage{graphicx}
\usepackage{bm}
\usepackage{tabularx}
\usepackage{longtable}
\usepackage{array}
\usepackage{calc}
\usepackage{amsfonts}
\usepackage[dvipsnames,table]{xcolor}

\definecolor{lightgray}{gray}{0.9}
\definecolor{SteelBlue}{RGB}{70,130,180}
\definecolor{darkspringgreen}{rgb}{0.09, 0.45, 0.27}
\definecolor{darkred}{rgb}{0.55, 0.0, 0.0}

\usepackage[most]{tcolorbox}

\usepackage[T1]{fontenc}

\usepackage[utf8]{inputenc}

\usepackage{microtype}

\usepackage{inconsolata}

\title{\textit{Shadow Unlearning:} A Neuro-Semantic Approach to \\ Fidelity-Preserving Faceless Forgetting in LLMs}

\author{
  \textbf{Dinesh Srivasthav P}\thanks{~~Equal contribution.}$^{\,\dagger}$ \quad 
  \textbf{Ashok Urlana}$^{\ast\dagger\ddagger}$ \quad 
  \textbf{Rahul Mishra}$^{\ddagger}$ \\
  \textbf{Bala Mallikarjunarao Garlapati}$^{\dagger}$ \quad 
  \textbf{Ponnurangam Kumaraguru}$^{\ddagger}$ \\
  \medskip
  $^{\dagger}$TCS Research, Hyderabad \quad
  $^{\ddagger}$IIIT Hyderabad \\
  \texttt{dineshsrivasthav.p@tcs.com} \\
}

\begin{document}
\maketitle
\begin{abstract}
In a typical machine unlearning setting, both the forget set and the retain set are exposed to the unlearning methods. Existing machine unlearning approaches are largely designed under this standard assumption. However, exposing potentially sensitive data, such as medical records, makes these methods vulnerable to membership inference attacks and the misuse of Personally Identifiable Information (PII). In this paper, we introduce a stricter formulation in which the original forget set is not directly exposed to the unlearning method. Instead, only an anonymized version of the forget set is made available, ensuring that no PII is revealed. We term this novel task formulation \textit{\textbf{Shadow Unlearning}}. Since existing state-of-the-art methods are designed for the standard unlearning setting, they do not inherently satisfy the constraints of Shadow Unlearning. To address this gap, we propose a novel privacy-preserving framework, \textit{\textbf{Neuro-Semantic Projector Unlearning (NSPU)}}, specifically designed to operate under this stricter setting. To evaluate our approach, we construct the \textbf{Mu}lti-domain \textbf{F}ictitious \textbf{U}nlearning (\textbf{MuFU}) benchmark, which includes five diverse domains. We also introduce a comprehensive evaluation protocol to quantify the trade-off between knowledge retention and unlearning effectiveness. Experimental results across multiple large language models demonstrate that NSPU achieves superior unlearning performance while preserving model utility and enhancing user privacy. Furthermore, our approach is at least $10\times$ more computationally efficient than standard unlearning methods. These findings open a new direction for privacy-aware machine unlearning that effectively balances data protection and model fidelity. Code is available at  \href{https://github.com/precog-iiith/Shadow_unlearning/}{Github}.
\end{abstract}

\section{Introduction}
Machine unlearning selectively removes specific data points from trained models \citep{cao2015towards, bourtoule2021machine}, essential for privacy compliance (e.g., GDPR \citep{GDPR_Article17_2018}) and model maintenance. Standard setups require a \textit{forget set} (samples to remove) and a \textit{retain set} (knowledge to preserve). However, this formulation introduces a privacy paradox: processing requests requires exposing the sensitive forget set to model operators (e.g., MLaaS providers), conflicting with the need to protect PII. In domains like healthcare or finance, transferring sensitive records for unlearning risks data breaches \citep{liu2025threats, nicolazzo2025secure, nguyen2025survey, liu2024breaking, jang2023knowledge}.

\begin{figure*}[]
  \centering\small
  \includegraphics[width=0.75\textwidth]{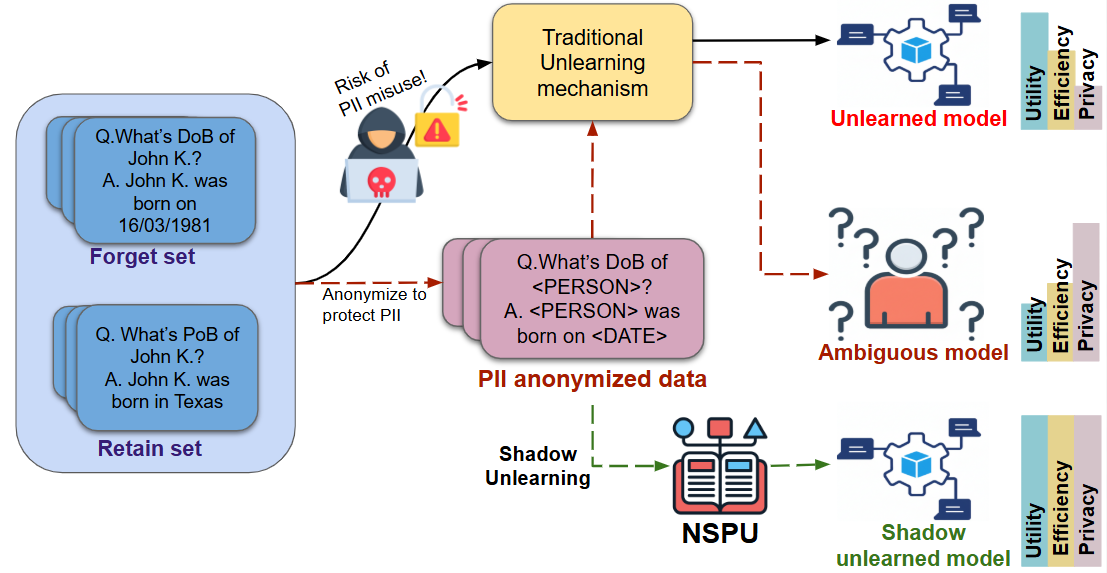}
  \caption{\textbf{The paradigm of Shadow Unlearning.} \\ \textbf{Motivation:} In a traditional \textcolor{red}{unlearning setup}, it is often required to share the retain and forget datasets for facilitating unlearning of the target model. This raises several privacy concerns related to PII. Data anonymization is a de facto way to deal with the privacy risk. This improves `privacy', nevertheless, dents the `utility', resulting in an \textcolor{darkred}{`ambiguous' model}. We propose \textcolor{darkspringgreen}{\textit{\textbf{Shadow Unlearning}}} to address this scenario by facilitating effective unlearning on anonymized data. Our approach, \textit{\textbf{NSPU}}, achieves Shadow Unlearning in a computationally `efficient' way, preserving the `utility' of the target model, thereby, balancing all three aspects.}
  \label{fig:shadow}
\end{figure*}

Existing LLM unlearning methods require access to original forget data (and often retain sets) using techniques like gradient ascent \citep{thudi2022unrolling}, joint loss balancing \citep{liu2022continual, chundawat2023zero}, or preference objectives \citep{rafailov2023direct, zhang2024negative}. However, these approaches suffer utility loss and privacy leakage under membership-inference attacks. Privacy-preserving and federated unlearning reduces data access via client-side updates or gradient sharing, but overlooks anonymized forget sets in autoregressive LLMs \citep{wang2023federated, liu2022continual}. To our knowledge, no prior work has studied LLM unlearning using only anonymized forget data.

To address this critical gap, we introduce Shadow Unlearning (Figure \ref{fig:shadow}), a paradigm for performing machine unlearning directly on anonymized forget sets, mitigating critical privacy risks, where data owners mask PII to facilitate secure sharing. The privacy boundary is maintained entirely at the data owner side, they perform anonymization locally and share only the anonymized forget set with the model operator. The core idea is to learn the semantic relationship between original and anonymized data in the activation space, allowing us to `project' the influence of the anonymized data back to its original `shadow' and unlearn it. The proposed NSPU method consists of three stages: 1) training a lightweight projector module to learn the general transformation mapping between non-anonymized and their anonymized counterparts, not at the embedding-level, but at an activation-level, 2) identifying the principal components of the ``forget space" where the dominant semantic directions must be erased, and 3) constructing an unlearning filter that modifies the model's information flow, effectively neutralizing the identified forget space concepts in real-time.

We conduct extensive experiments across five domains of the MuFU dataset. The experiments are performed on four representative LLMs of varying sizes and families to evaluate the generalizability and effectiveness of the proposed unlearning method. We compare our approach against several standard unlearning methods. While several unlearning evaluation metrics focus on assessing the model's utility and forgetting efficacy independently, we extend those metrics to assess the tradeoff between knowledge retention and forget efficacy \citep{zhou2024limitations, li2025machine}. To this end, we employ four evaluation metrics: Harmonic Perplexity Score, Combined Efficacy Score, Harmonic ROUGE Score, and Conditional Negative Log-Likelihood Probability. Moreover, our approach is at least $10^6$ times computationally efficient than retraining from scratch and 10x faster when compared to 
standard unlearning methods. 

The key contributions of this work include:
\begin{itemize}
    \item We introduce a novel task in privacy-preserving unlearning -- \textbf{Shadow Unlearning}, which enables selective unlearning on anonymized forget data.
    \item We propose \textbf{Neuro-Semantic Projector Unlearning (NSPU)}, a novel frozen-target approach for effective unlearning on anonymized forget data, achieving a strong utility-efficiency-privacy tradeoff over state-of-the-art methods.
    \item We compile the \textbf{MuFU} dataset across five domains: Digital informatics, Sports, Politics, Science \& technology, and Finance.
    \item We introduce a suite of evaluation metrics designed to measure the trade-off between knowledge retention and unlearning efficacy.
\end{itemize}

\section{Related work}
\textbf{Machine unlearning for LLMs: } Unlearning approaches are broadly categorized into two categories: exact unlearning and approximate unlearning \citep{izzo2021approximate}. Exact unlearning mandates that the model’s behavior after unlearning be indistinguishable from that of a model trained from scratch without the forget set data. To this end, early exact unlearning techniques were mainly developed on random forest \cite{brophy2021machine, schelter2021hedgecut}, k-means clustering \citep{ginart2019making}. \citet{bourtoule2021machine} formalized exact unlearning by introducing a general framework: sharded, isolated, sliced, aggregated (SISA). Despite the efficacy and precision of exact unlearning methods, they face significant constraints related to underlying assumptions and scalability issues \citep{xu2024machine}. On the other hand, approximate unlearning \citep{thudi2022necessity} aims to forget the influence of target data points by adjusting the weights through loss-based fine-tuning strategies. These methods mainly rely on parameter optimization \citep{jang2023knowledge, yao2024large, wang2025selective, li2024wmdp, yao-etal-2024-machine, ishibashi2023knowledge, gu2024second, zhang2024negative, lu2024eraser, jia2024soul, tian2024forget, liu2024towards, choi2024snap, tang2024learn, yuan2025closer}. However, these approaches can degrade the overall performance of the models. Recent research has explored alternative approaches for LLM unlearning, including contrastive decoding \citep{eldan2310s, huangoffset, wang2024rkld, ji2024reversing}, task vectors \citep{dou2024avoiding, liu2024towards}, in-context learning \citep{pawelczyk2024context, muresanu2024unlearnable}, input processing and deletion \citep{bhaila2025soft, liu2024large} and LoRA-based unlearning \citep{cha2025towards, liu2025lune}. However, most of these methods require preserving the original model's parameters, which can still raise privacy concerns. For further discussion on related works, see Appendix~\ref{sec:extend_relatedwork}.

\section{Methodology}
\subsection{Task Formulation}
Let $M_{\theta}$ denote a model pre-trained on a corpus $D$, composed of disjoint subsets $D = D_{\mathrm{retain}} \cup D_{\mathrm{forget}}$. Due to privacy constraints, the original samples are inaccessible. Instead, we assume access to a one-way anonymization function $f(\cdot)$ and the resulting anonymized forget dataset $\tilde{D}_{\mathrm{forget}} = f(D_{\mathrm{forget}})$. Our goal is to obtain an unlearned model $M_{\theta^{'}}$ using only the current model state $M_{\theta}$ and the anonymized forget data $\tilde{D}_{\mathrm{forget}}$. Ideally, $M_{\theta^{'}}$ should approximate the behavior of a ``gold standard" model $M^{\star}$ trained from scratch on the counterfactual dataset $D \setminus D_{\mathrm{forget}}$. 
\subsection{Anonymization}
To construct the anonymized forget set, a named-entity recognition (NER)-based anonymization pipeline is applied.
All PII is detected using the Microsoft Presidio Anonymizer\footnote{\url{https://microsoft.github.io/presidio/anonymizer/}} and replaced with corresponding placeholder tags (e.g., \textless PERSON\textgreater, \textless LOCATION\textgreater), ensuring that direct identity information is removed while preserving the surrounding linguistic context. The same anonymization pipeline is reused in all stages of the proposed methodology to maintain consistency between pretraining-style data and the forget set.
\subsection{Proposed Methodology}
\begin{figure*}[]
  \centering
  \includegraphics[width=0.99\textwidth]{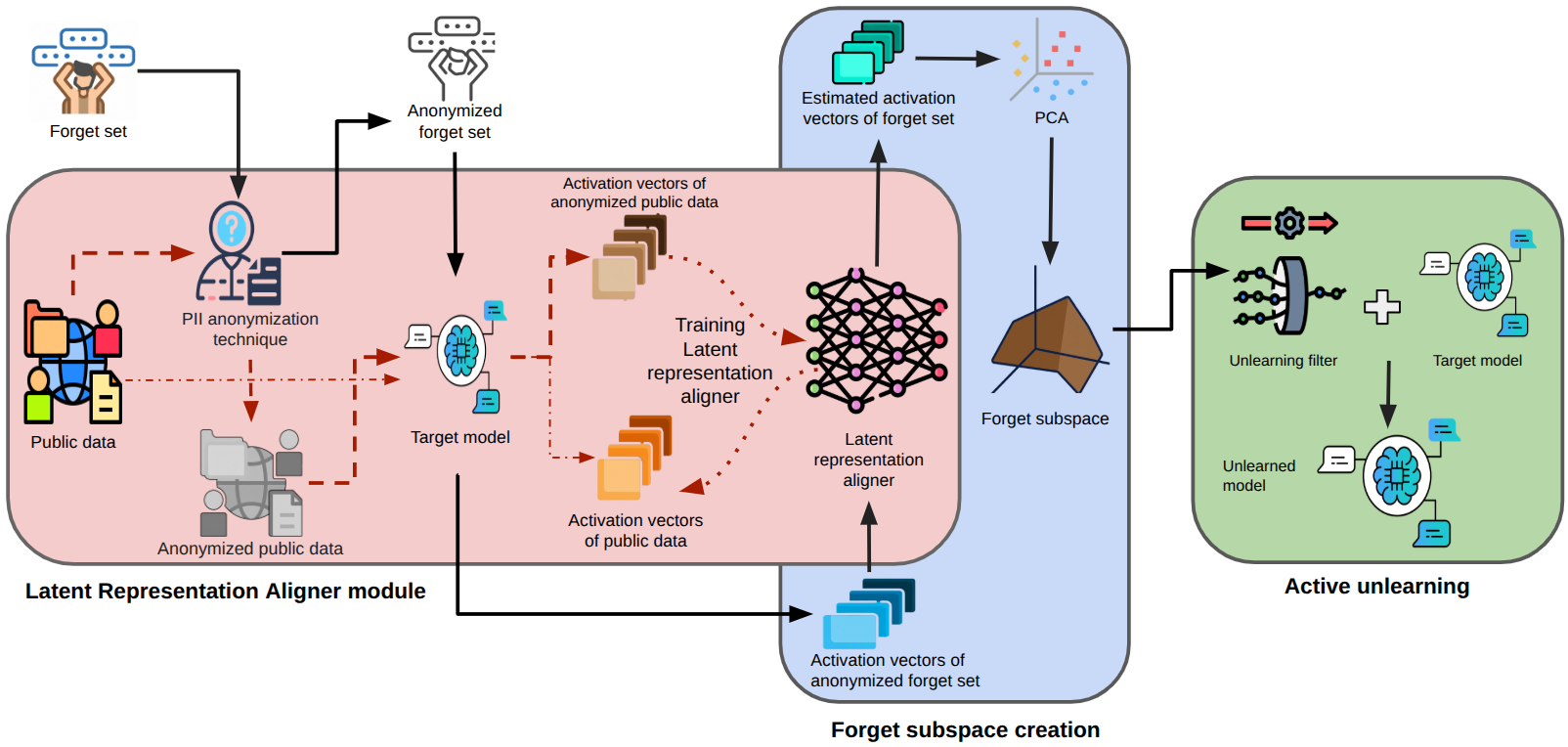}
  \caption{Neuro-Semantic Projector Unlearning (NSPU) Pipeline comprises three key phases: (i) learning a latent representation aligner, (ii) constructing a forget subspace, and (iii) integrating a linear unlearning filter.}
  \label{fig:pipeline}
  \vspace{-5mm}
\end{figure*}

Neuro-Semantic Projection Unlearning (NSPU), comprises three key phases, as represented in Figure \ref{fig:pipeline}, and explained in Sections \ref{sec: latent rep}, \ref{sec: forget subspace}, and  \ref{sec:unlearn_filter}.
\subsection{Latent Representation Aligner}
\label{sec: latent rep}
This preliminary, one-time phase establishes a mechanism for aligning the representations of anonymized and original data samples at the neural activation level. The aim is not to re-identify individuals, but to obtain a smooth semantic correspondence between the activations induced by original and anonymized texts. This enables the model to operate in a coherent latent geometry even when direct identification cues are removed.

To achieve this, a lightweight multi-layer perceptron (MLP) is employed as a projection function $P_\theta$. Conceptually, this MLP acts as a neuro-semantic bridge, which aligns activations of the anonymized samples with the model’s original semantic structure, while keeping the information in an abstract latent form.

As a result, anonymized samples can be translated into activation vectors that are geometrically consistent with those induced by non-anonymized inputs, without exposing underlying identity attributes.
\subsubsection{Training of the Latent Representation Aligner}
Before considering the forget or retain sets, the training of the Latent representation aligner ($P_{\theta}$) leverages a publicly available corpus. More details of the training setup are discussed in  Appendix~\ref{sec:trianing_mlp}. From this corpus, pairs $(x_{\text{orig}}, x_{\text{anon}})$ are constructed by applying the same anonymization pipeline $f(\cdot)$ used for the forget set, where $x_{\text{anon}} = f(x_{\text{orig}})$. Both the original and anonymized texts are independently passed through the frozen target model to obtain layer-specific activations $\phi_{l}(x_{\text{orig}})$ and $\phi_{l}(x_{\text{anon}})$ at a chosen layer \textit{l}. These activation pairs form the training data for the Latent representation aligner.

The Latent representation aligner $P_\theta$ is trained to minimize the semantic alignment loss:
\[
    \mathcal{L}_{\text{align}} = \left\| P_\theta \left(x_{\text{anon}} \right) - \phi_l(x_{\text{orig}}) \right\|_2^2
\]
where $\phi_l$ denotes the layer-specific activation function of the target model. The parameters of the projector are optimized while keeping the target model frozen. The resulting trained model $P_\theta$ serves as a universal neuro-semantic bridge that estimates the neural fingerprint of the original sample from its anonymized counterpart.\\
\textbf{Inversion-Resistant Optimization Term.} To prevent potential inversion of the projected activations back into human-interpretable input space, we introduce an auxiliary privacy-preserving term—termed the \textit{Inversion Optimization Score (InvOptScore)}. This score quantifies how easily an estimated activation vector could be mapped to a plausible input embedding through optimization-based inversion.\\
For an estimated activation, $\hat{h}_{est}$ of an anonymized sample $x_{anon}$, $\hat{h}_{\text{est}} = P_{\theta}(x_{\text{anon}})$
consider an inversion process that searches for a pseudo-input embedding $x' \in \mathcal{X}_{\text{emb}}$ in the model's input embedding space that reproduces $\hat{h}_{\text{est}}$. Define the inversion loss:
\[
\mathcal{L}_{\text{inv-opt}}(x') = \left\| \phi_l(x') - \hat{h}_{\text{est}} \right\|_2^2
\]

The optimal pseudo-input $x'^*$ is obtained via $x'^* = \arg\min_{x'} \mathcal{L}_{\text{inv-opt}}(x')$ and the corresponding inversion difficulty is summarized by

\[
\text{InvOptScore}(\hat{h}_{\text{est}}) = \left\| x'^* - x_{orig} \right\|_2^2
\]

The optimization over $x'$ is carried out entirely in the continuous embedding space, initialized from random embeddings and refined by gradient-based minimization of $\mathcal{L}_{\text{inv-opt}}$. No discrete token decoding is required or performed, so no explicit textual reconstruction is ever produced. A lower residual $\text{InvOptScore}(\hat{h}_{\text{est}})$ indicates that the activation is easier to approximate with some plausible input embedding and therefore carries higher inversion risk. By penalizing such low inversion losses during training, the projector is encouraged to produce activations that are semantically aligned with the model's internal geometry yet intrinsically harder to decode into human-readable content. The final training objective for the projector is
\[
\mathcal{L}_{\text{proj}} = \mathcal{L}_{\text{align}} + \lambda_{\text{inv}} \big( -\text{InvOptScore}(\hat{h}_{\text{est}}) \big)
\]

where $\lambda_{\text{inv}} > 0$ balances semantic fidelity and inversion resistance. Larger values of $\lambda_{\text{inv}}$ prioritize privacy, while smaller values favor tighter alignment with original activations.

\subsection{Forget Subspace}
\label{sec: forget subspace}
This phase constructs a forget subspace that captures the dominant directions associated with the forget set in the model’s activation space. This provides a unified vector space toward which forget-related activations can be systematically diverted, avoiding the need to reason about individual neurons or samples.

\subsubsection{Creation of Forget subspace}
The anonymized forget set $\tilde{D}_{\text{forget}}$ is passed through the target model to obtain layer-$l$ activation vectors $\phi_l(x_{\text{anon}})$ for all $x_{\text{anon}} \in \tilde{D}_{\text{forget}}$. These anonymized activations are then passed through the trained Latent representation aligner ($P_\theta$) to produce estimated original activations $\hat{h}^{\text{orig}} = P_\theta\big(\phi_l(x_{\text{anon}})\big)$ corresponding to each sample in the $\tilde{D}_{\text{forget}}$.

Row-wise stacking of these activations yields a matrix $H \in \mathbb{R}^{n \times d}$, for $n$ forget samples and activation dimension $d$. We apply Principal Component Analysis (PCA) to $H$ to extract orthonormal directions capturing principal variance axes, which approximate dominant semantic factors across the forget set (see Appendix~\ref{sec:experiment_setup} for PCA rank details). The top $k$ right-singular vectors, forming an orthonormal basis for these components, comprise the columns of $U \in \mathbb{R}^{d \times k}$. This matrix $U$ defines the \emph{\textbf{forget subspace}}, the linear subspace spanned by directions most strongly associated with the forget set’s internal representations.

\subsection{Active Unlearning}
\label{sec:unlearn_filter}
Leveraging the forget subspace obtained, an unlearning filter is created as a simple adapter incorporated into the target model inference. It facilitates active unlearning, projecting away information or activations relevant to the forget set.  
\subsubsection{Unlearning Filter}
\begin{figure}
  \centering
  \includegraphics[width=0.5\textwidth]{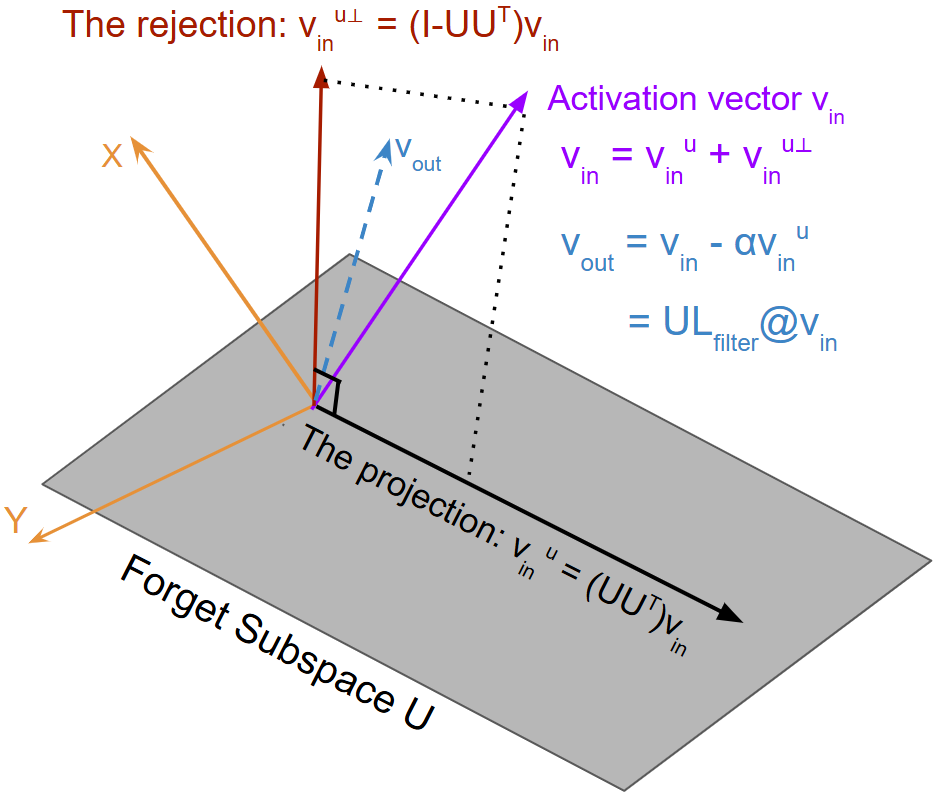}
  \caption{Functioning of Unlearning filter}
  \label{fig:projection}
  \vspace{-7mm}
\end{figure}

The unlearning filter $UL_{\text{filter}} \in \mathbb{R}^{d \times d}$ is defined as $UL_{\text{filter}} = I - \alpha U U^\top $, where $I$ is the $d \times d$ identity matrix and $\alpha$ is a hyperparameter governing the strength of unlearning. As depicted in Figure~ \ref{fig:projection}, for any activation vector $v \in \mathbb{R}^{d}$, the term $U U^\top v$ gives its orthogonal projection onto the forget subspace, and the filter subtracts an $\alpha$-scaled version of this projection
$v_{\text{out}} = UL^{}_{\text{filter}}\, v_{\text{in}} $. This operation shifts $v_{\text{out}}$ toward the orthogonal complement of the forget subspace, thereby attenuating information aligned with forget-related concepts while minimally disturbing components orthogonal to $U$. While retain and forget representations overlap, their dominant principal components diverge sufficiently. Projecting away from these forget axes yields an approximate yet highly effective separation. The geometric interpretation of the unlearning filter is provided in Appendix~\ref {sec:theroy_forget_space}. 

The unlearning filter is implemented as a lightweight, non-trainable linear operation integrated permanently into the model architecture. Specifically, the original final layer normalization is replaced by a sequential module that first applies the unlearning transformation to the last hidden state before passing it through the standard normalization layer. This modification ensures the change is a structural part of the model, permanently steering the information flow away from the forget subspace in real-time during the forward pass.
\section{Experiments}
\subsection{Setup}
\textbf{Dataset Creation.} Existing unlearning benchmark datasets, such as TOFU \citep{mainitofu}, often suffer from extreme homogeneity and high semantic overlap between the forget ($D_f$) and retain ($D_r$) sets. This high inter-sample correlation complicates the evaluation of unlearning efficacy. If a model retains the underlying data distribution through $D_r$, membership inference attacks (MIAs) may yield high false positives, masking the true performance of unlearning algorithms \citep{nguyen2025survey, carlini2022membership}.

\noindent \textbf{MuFU: Multi-Domain Fictitious Unlearning.} To address these limitations, we introduce the MuFU dataset, which evaluates unlearning across a heterogeneous environment where both forget and retain sets span multiple domains to mirror real-world complexity. Using Gemini 3 Pro, we generated 4,000 distinct samples containing personally identifiable information, distributed evenly across five domains: Digital Informatics, Finance, Sports, Science \& Technology, and Politics. Further details about the MuFU dataset creation and data variants are discussed in Appendix~\ref{sec:synthetic_data}.

\noindent \textbf{Models.} We conduct experiments on LLaMA-7B \citep{touvron2023llama}, Mistral-7B-instruct \citep{jiang2023mistral}, OLMoE-1B-7B-0924 \citep{muennighoffolmoe} and LLaMA-2-13B \citep{touvron2023llama2} variants. More details on the experimental setup are provided in Appendix~\ref{sec:experiment_setup}.

\noindent \textbf{Baselines.} We perform experiments by considering baselines as Gradient Ascent (GA) \citep{thudi2022unrolling}, Gradient Difference (GD) \citep{liu2022continual}, KL Minimization (KLM) \citep{chundawat2023zero}, Direct Preference Optimization (DPO) \citep{rafailov2023direct}, and Negative Preference Optimization (NPO) \citep{zhang2024negative}. Detailed descriptions of the baselines are mentioned in Appendix~\ref{sec:baselines}. 

\subsection{Evaluation}
To demonstrate unlearning, evaluation is performed using the \emph{original (non-anonymized)} questions, unlike the unlearning phase, which operates on anonymized inputs. Following prior work \citep{nguyen2025survey, qu2024learn}, we evaluate the trade-off between knowledge retention and forgetting using four complementary metrics. Full derivations and implementation details are provided in Appendix~\ref{sec:extended_eval}.

\noindent\textbf{Harmonic Perplexity Score (HPS).}
Perplexity reflects model uncertainty. Effective unlearning should increase perplexity on the forget set while keeping it low on the retain set. We define Forget Gain ($G_F$) and Retain Cost ($C_R$) as log-ratios of post- and pre-unlearning perplexity, and combine them using a harmonic formulation:
$HPS = \frac{2 \cdot G_F}{G_F \cdot C_R + 1}.$
This formulation penalizes methods that forget aggressively at the expense of retained knowledge.

\noindent\textbf{Combined Efficacy Score (CES).}
Combined Efficacy Score (CES) utilizes the truth ratio \citep{mainitofu} to measure a model's preference for correct answers over perturbed ones. We sum Retain Stability (RS) and Forget Instability (FI) as: CES = RS + FI. Higher CES signifies better retention and superior suppression of forgotten content.

\noindent\textbf{Harmonic ROUGE Score (HRS).}
To assess generative fidelity, we use ROUGE-L to measure semantic overlap. HRS jointly evaluates retention and forgetting by harmonically combining the Retention Ratio (RR) and Forget Ratio (FR): $HRS = \frac{2 \cdot RR}{FR \cdot RR + 1}.$
This score favors high semantic preservation on the retain set and low overlap on the forget set.

\noindent\textbf{Harmonic Conditional NLL (HCNLL).}
HCNLL evaluates likelihood-based behavior using conditional negative log-likelihood. Analogous to HPS, we define Forget Gain Likelihood and Retain Cost Likelihood and aggregate them harmonically:
$HCNLL = \frac{2 \cdot G_{F_L}}{G_{F_L} \cdot C_{R_L} + 1}.$
This metric captures probabilistic forgetting while preserving confidence in retained samples.
\begin{table*}[]
\resizebox{\textwidth}{!}{%
\begin{tabular}{@{}clrrrrrrrrrrrrll@{}}
\toprule
\multicolumn{1}{c}{} & & \multicolumn{3}{c}{\textbf{Perplexity}} & \multicolumn{3}{c}{\textbf{Truth Ratio}} & \multicolumn{3}{c}{\textbf{ROUGE-L}} & \multicolumn{3}{c}{\textbf{Probability}} & \multicolumn{1}{c}{} \\ 
\multicolumn{1}{c}{\multirow{-2}{*}{\textbf{Model}}} & \multirow{-2}{*}{\textbf{Method}} & $G_F$ & $C_R$ & \textbf{HPS} ($\bm{\uparrow}$) & RS & FI & \multicolumn{1}{c}{\textbf{CES}($\bm{\uparrow}$)} & RR & FR & \textbf{HRS} ($\bm{\uparrow}$) & $G_{F_L}$ & $C_{R_L}$ & \textbf{HCNLL} ($\bm{\uparrow}$) & \multicolumn{1}{c}{\multirow{-2}{*}{\textbf{Aggregate}}} \\ 
\cmidrule(lr){1-2} \cmidrule(lr){3-5} \cmidrule(lr){6-8} \cmidrule(lr){9-11} \cmidrule(lr){12-14} \cmidrule(lr){15-15}
 & GA & 104.730 & 110.110 & 0.019 & 0.000 & -0.805 & -0.805 & 0.007 & 0.167 & 0.014 & 83.887 & 49.734 & 0.024 & -0.748 \\
 & GD & 4.597 & 6.006 & 0.420 & 0.497 & 0.054 & 0.551 & 0.210 & 0.245 & 0.399 & 4.811 & 3.988 & 0.395 & 1.765 \\
 & KLM & 104.730 & 110.110 & 0.019 & 0.000 & -0.808 & -0.808 & 0.007 & 0.167 & 0.014 & 83.887 & 49.734 & 0.024 & -0.751 \\
 & DPO & 4.082 & 6.780 & 0.473 & 0.000 & -0.998 & -0.998 & 0.042 & 0.225 & 0.083 & 5.043 & 5.312 & 0.382 & -0.060 \\
 & NPO & 39.850 & 37.772 & 0.050 & 0.000 & 0.920 & 0.920 & 0.001 & 0.031 & 0.002 & 32.027 & 17.946 & 0.062 & 1.034 \\
\rowcolor{lightgray}
\multirow{-6}{*}{\rotatebox{90}{Llama2-7B}} & \textbf{NSPU} & 1.388 & 2.055 & \textbf{1.067} & 1.021 & 0.498 & \textbf{1.519} & 0.581 & 0.705 & \textbf{0.824} & 2.590 & 2.331 & \textbf{0.662} & \textbf{4.073} \\ \midrule
 & GA & 154.695 & 202.520 & 0.013 & 0.000 & -0.995 & -0.995 & 0.034 & 0.112 & 0.068 & 124.738 & 74.003 & 0.016 & -0.899 \\
 & GD & 18.287 & 17.198 & 0.109 & 0.943 & -0.511 & 0.432 & 0.363 & 0.375 & 0.639 & 1.258 & 1.551 & 1.051 & 2.232 \\
 & KLM & 193.398 & 209.753 & 0.010 & 0.000 & -1.001 & -1.001 & 0.034 & 0.112 & 0.068 & 124.738 & 74.003 & 0.016 & -0.907 \\
 & DPO & 27.660 & 25.787 & 0.072 & 0.000 & -0.554 & -0.554 & 0.006 & 0.099 & 0.012 & 17.218 & 9.888 & 0.115 & -0.355 \\
 & NPO & 9.549 & 7.433 & 0.207 & 0.000 & 0.996 & \textbf{0.996} & 0.001 & 0.146 & 0.001 & 9.595 & 3.561 & 0.203 & 1.406 \\
\rowcolor{lightgray}
\multirow{-6}{*}{\rotatebox{90}{Mistral-7B}} & \textbf{NSPU} & -0.084 & 2.066 & \textbf{4.995} & 0.921 & -0.536 & 0.385 & 0.876 & 0.542 & \textbf{1.188} & 0.955 & 1.650 & \textbf{1.281} & \textbf{7.849} \\ \midrule
 & GA & 84.662 & 87.404 & 0.024 & 0.000 & -3.189 & -3.189 & 0.006 & 0.068 & 0.012 & 65.839 & 71.628 & 0.030 & -3.123 \\
 & GD & 6.214 & 9.374 & 0.316 & 1.202 & -1.924 & -0.722 & 0.054 & 0.081 & 0.108 & 4.901 & 8.573 & 0.399 & 0.101 \\
 & KLM & 84.662 & 87.404 & 0.024 & 0.000 & -3.179 & -3.179 & 0.006 & 0.077 & 0.012 & 65.839 & 71.628 & 0.030 & -3.113 \\
 & DPO & 1.887 & -0.200 & -0.645 & 0.000 & -0.953 & -0.953 & 0.705 & 0.560 & 1.011 & 1.929 & 0.831 & 0.638 & 0.051 \\
 & NPO & 0.095 & 0.318 & 0.617 & 1.025 & 1.156 & 2.181 & 1.007 & 0.730 & 1.160 & 0.703 & 0.636 & 0.879 & 4.837 \\
\rowcolor{lightgray}
\multirow{-6}{*}{\rotatebox{90}{OLMOE}} & \textbf{NSPU} & 0.631 & 2.008 & \textbf{1.771} & 2.980 & 1.548 & \textbf{4.528} & 0.857 & 0.408 & \textbf{1.270} & 1.415 & 2.483 & \textbf{1.100} & \textbf{8.669} \\ \midrule
 & GA & 62.572 & 75.388 & 0.032 & 0.000 & -0.362 & -0.362 & 0.002 & 0.164 & 0.004 & 15.695 & 27.613 & 0.127 & -0.199 \\
 & GD & 0.067 & 0.077 & 0.154 & 0.598 & 0.762 & 1.360 & 0.765 & 1.454 & 0.724 & 1.329 & 0.977 & 0.850 & 3.088 \\
 & KLM & 61.507 & 72.625 & 0.033 & 0.000 & -1.002 & -1.002 & 0.001 & 0.164 & 0.002 & 26.829 & 26.125 & 0.074 & -0.893 \\
 & DPO & 0.065 & 0.115 & 0.228 & 0.454 & 0.710 & 1.164 & 0.651 & 0.664 & 0.909 & 1.737 & 1.367 & 0.810 & 3.111 \\
 & NPO & 6.958 & 11.820 & 0.284 & 0.472 & 1.115 & 0.356 & 0.747 & 0.345 & \textbf{1.188} & 2.020 & 3.213 & 0.858 & 2.686 \\
\rowcolor{lightgray}
\multirow{-6}{*}{\rotatebox{90}{Llama-13B}} & \textbf{NSPU} & 1.173 & 1.445 & \textbf{1.072} & 0.987 & 0.736 & \textbf{1.723} & 0.817 & 0.518 & 1.148 & 1.480 & 1.535 & \textbf{0.938} & \textbf{4.882} \\ \bottomrule
\end{tabular}%
}
\caption{Evaluation of NSPU against standard unlearning approaches on MuFu dataset. Aggregate score (HPS+CES+HRS+HCNLL) reflects the forgetting–utility trade-off, where higher values indicate better performance.}
\vspace{-5mm}
\label{tab:results_table}
\end{table*}

\subsection{Experimental Results and Discussion}
This section evaluates the model utility, unlearning efficacy, computational efficiency, and privacy via separation quality scores, tradeoff with extended analysis in Appendix \ref{sec: extended_results_analysis}.

\subsection{The Utility-Efficacy Tradeoff}
\textbf{GA and KLM cause catastrophic utility loss.} These methods maximize loss or divergence on forget data, leading to higher forget gain and a lower retain cost as shown in Table~\ref{tab:results_table}. This indicates high uncertainty in both the forget and retain sets. While this reflects strong forgetting efficacy, it comes at the cost of severe utility degradation. The negative aggregate scores for GA and KLM underscore the nature of forget-loss approaches, which lack the fundamental mechanism to preserve useful knowledge beyond the forget set.

\noindent \textbf{GD, DPO, and NPO offer limited improvements.} GD incorporates joint retention-forgetting objectives for better stability across these LLMs, while DPO and NPO suppress undesired outputs but show inconsistent behavior due to preference calibration sensitivity. Both DPO and NPO offer improved trade-offs over pure ascent techniques yet lack the robustness needed for scalable unlearning across diverse architectures.

\noindent \textbf{NSPU consistently outperforms standard unlearning methods.} NSPU achieves notably higher Aggregate scores across all tested models, while approaches such as GA and KLM consistently yield highly negative aggregate values. This indicates that NSPU effectively balances forgetting targeted data while maintaining model utility on retained data, unlike GA and KLM, which aggressively forget but cause severe degradation in overall performance.

\subsection{Membership Inference and Separation Quality}
Membership Inference Attacks (MIAs) evaluate unlearning efficacy by testing whether an adversary can distinguish forget set samples from unseen data using the unlearned model's predictions~\citep{shokri2017membership, carlini2022membership, maini2024llm}. Effective unlearning implies that the target model should perceive the forget set as indistinguishable from a non-member set (unseen data), while retaining performance on the retain set. To quantify this behavior, we construct the non-member set as detailed in Appendix~\ref{sec:non_memberdata_creation}.
To quantify MIA resistance, we compute mean Causal Language Modeling (CLM) loss across retain ($M_R$), forget ($M_F$), and non-member ($M_{NM}$) sets. The \textit{Separation Quality Score} (SQS) measures how well the forget set loss aligns with non-members while diverging from retain samples:
\begin{table*}[t]
\centering\footnotesize
\label{tab:combined_analysis}
\resizebox{\textwidth}{!}{%
\begin{tabular}{lcccc|ccccc}
\toprule
& \multicolumn{4}{c|}{\textbf{Computational Cost (FLOPs)}} 
& \multicolumn{5}{c}{\textbf{Separation Quality Score (SQS)}} \\
\cmidrule(r){2-5} \cmidrule(l){6-10}
\textbf{Method} 
& LLaMA-7B & Mistral-7B & OLMoE-7B & LLaMA-13B
& LLaMA-7B & Mistral-7B & OLMoE-7B & LLaMA-13B & Avg. ($\uparrow$) \\
\midrule

Retraining 
& 8.4 $\times$ $10^{22}$ & 3.4 $\times$ $10^{22}$ & 3.0 $\times$ $10^{22}$ & 8.4 $\times$ $10^{26}$
& -- & -- & -- & -- & -- \\

GA 
& 1.3 $\times$ $10^{17}$ & 1.3 $\times$ $10^{16}$ & 1.8 $\times$ $10^{16}$ & 1.3 $\times$ $10^{21}$
& 0.79 & 0.78 & \textbf{0.76} & 0.403 & 0.68 \\

GD 
& 5.2 $\times$ $10^{17}$ & 5.2 $\times$ $10^{17}$ & 7.4 $\times$ $10^{16}$ & 5.2 $\times$ $10^{21}$
& 0.75 & 0.62 & 0.69 & 0.56 & 0.66 \\

KLM 
& 1.3 $\times$ $10^{17}$ & 1.3 $\times$ $10^{16}$ & 1.8 $\times$ $10^{16}$ & 1.3 $\times$ $10^{21}$
& 0.79 & 0.78 & \textbf{0.76} & 0.403 & 0.68 \\

DPO 
& 5.2 $\times$ $10^{17}$ & 5.2 $\times$ $10^{17}$ & 7.4 $\times$ $10^{16}$ & 5.2 $\times$ $10^{21}$
& 0.69 & 0.46 & 0.70 & 0.53 & 0.59 \\

NPO 
& 5.2 $\times$ $10^{17}$ & 5.2 $\times$ $10^{17}$ & 7.4 $\times$ $10^{16}$ & 5.2 $\times$ $10^{21}$
& 0.65 & 0.74 & 0.17 & 0.69 & 0.56 \\

\rowcolor{lightgray}
\textbf{NSPU}
& $\boldsymbol{7.6 \times 10^{16}}$ 
& $\boldsymbol{7.6 \times 10^{15}}$ 
& $\boldsymbol{8.7 \times 10^{15}}$ 
& $\boldsymbol{7.6 \times 10^{20}}$
& \textbf{0.83} & \textbf{0.79} & 0.74 & \textbf{0.72} & \textbf{0.77} \\

\bottomrule
\end{tabular}}
\caption{Combined comparison of computational cost (FLOPs) and separation quality score (SQS) across unlearning methods and LLMs.}
\label{tab:flops_and_sqs}
\vspace{-5mm}
\end{table*}

\[
    SQS = \frac{|M_{R} - M_F|}{|M_R - M_F| + |M_{NM} - M_F|}
\]
The numerator maximizes forget-retain divergence (successful unlearning), while the denominator minimizes forget-nonmember divergence (MIA resistance). As shown in Table~\ref{tab:flops_and_sqs}, high SQS scores reflect the balance between effective forgetting and retention, quantifying privacy strength and unlearning fidelity.

\section{Ablation studies}
Beyond NSPU’s utility, efficiency, and privacy advantages, it is critical to assess its ability to distinguish retain vs. forget information and identify domain-specific and entity-level (PII) knowledge. This motivates deeper ablations on where and how unlearning occurs.

\subsection{Interpretable View of Knowledge Entanglement}
\begin{figure}
  \centering\small
  \includegraphics[width=\columnwidth]{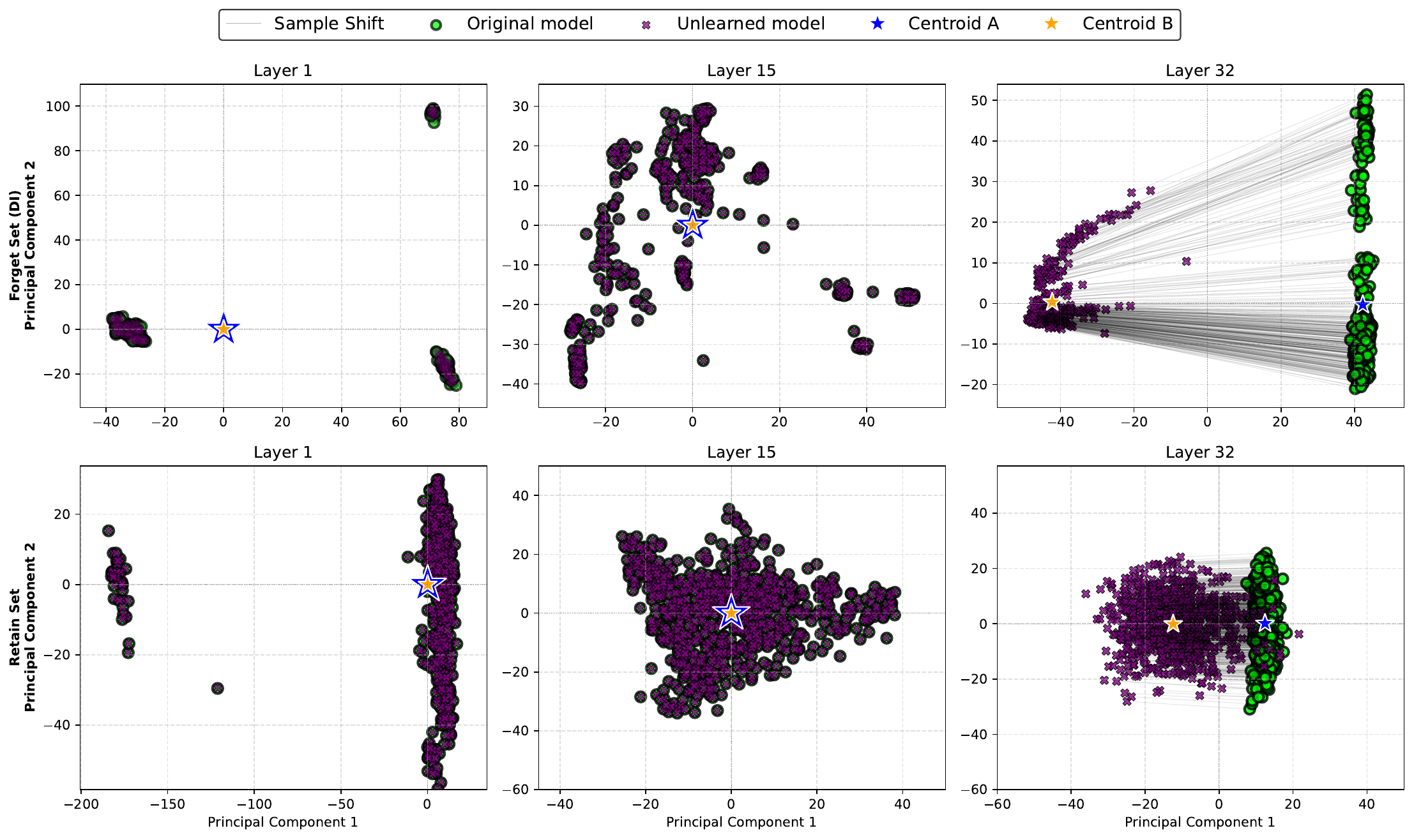}
  \caption{Impact of unlearning on forget (top-row) and retain datasets (bottom-row) before and after applying the unlearning filter (Mistral-7B).}
  \label{fig:interpretability_mistral}
  \vspace{-5mm}
\end{figure}

\textbf{NSPU clearly distinguishes the forget and retain information.} %
Figure \ref{fig:interpretability_mistral} shows layer-wise activation shifts (early, middle, last) between the original and unlearned models for both forget and retain sets. We apply the unlearning filter only at the final hidden layer, as semantic separation between forget and retain concepts is minimal in early layers and peaks in deeper layers as discussed in Appendix~\ref{sec:layer_selection}.
 Earlier layers encode low-level features, so filtering them is unnecessary and does not expose sensitive information. Consequently, an attacker attempting to bypass the defense by probing earlier, unprotected layers would fail to extract the targeted PII, as sensitive representations have not yet formed.

Accordingly, activations remain identical up to the 31\textsuperscript{st} layer, with unlearning effects appearing only at the final layer. The forget set shows strong separability, while the retain set exhibits minor drift with substantial overlap. This small drift is expected due to the non-orthogonality of real-world concepts, however, it is negligible compared to the large separation in the forget set. This demonstrates that NSPU effectively removes target knowledge while preserving utility. Similar behavior is observed in Llama-13B, as shown in Appendix~\ref{sec:extended_interpretability}.

\subsection{Identification of desired information to unlearn}
To validate that NSPU targets intended information rather than spurious knowledge, we conduct an ablation to examine the forgotten content. Since Shadow Unlearning operates on anonymized data, verifying that the model erases the specific entity per request is essential. The Latent Representation Aligner bridges anonymized and original representations while maintaining inversion resistance to prevent leakage.
While Appendix~\ref{sec:trianing_mlp} Table~\ref{tab:mlp_results} details overall performance, Table~\ref{tab:av_similarity} assesses how closely reconstructed activation vectors approximate original activations across domains and models. We use an \textit{InvOptScore} regularization term to penalize excessive similarity, balancing semantic alignment with privacy. The results confirm meaningful semantic correspondence in projected vectors, validating the module's ability to localize correct knowledge under privacy constraints. Similarly, entity-level analysis (Table~\ref{tab:entity-wise-similarity}) shows NSPU consistently erases targeted representations 83\% of the time on average while retaining remaining domain knowledge.
\begin{table}[t]
\centering\footnotesize
\resizebox{\columnwidth}{!}{%
\begin{tabular}{p{1.5cm}cccc}
\toprule
\textbf{Domain} & \textbf{Llama-7B} & \textbf{Mistral-7B} & \textbf{OLMoE-1B-7B} & \textbf{Llama-13B} \\
\midrule
DI        & 0.64 & 0.49 & 0.63 & 0.82 \\
Politics  & 0.80 & 0.70 & 0.76 & 0.87 \\
Finance   & 0.79 & 0.70 & 0.74 & 0.87 \\
Sports    & 0.76 & 0.60 & 0.70 & 0.84 \\
Science   & 0.58 & 0.47 & 0.57 & 0.79 \\
\midrule
\textbf{Average} & \textbf{0.71} & \textbf{0.59} & \textbf{0.68} & \textbf{0.84} \\
\bottomrule
\end{tabular}}
\caption{Domain-wise similarity between original and estimated activation vectors.}
\label{tab:av_similarity}
\vspace{-5mm}
\end{table}

\begin{table}[!ht]
\centering
\resizebox{\columnwidth}{!}{%
\begin{tabular}{cccccc}
\toprule
PERSON & LOCATION & PHONE\_NUMBER & DATE\_TIME & Misc \\ \midrule
0.85   & 0.84     & 0.84          & 0.80       & 0.83 \\
\bottomrule
\end{tabular}}
\caption{PII attribute-wise similarity between original and estimated activation vectors for Llama-13B.}
\label{tab:entity-wise-similarity}
\vspace{-5mm}
\end{table}

\subsection{Additional results}
We provide more results and ablations, including: MuFU dataset creation (Appendix ~\ref{sec:synthetic_data}), training of Latent representation aligner (Appendix~\ref{sec:trianing_mlp}), human evaluation (Appendix~\ref{sec:human_eval}), computational efficiency analysis (Appendix~\ref{sec:compute_analysis}) based on FLOPS and VRAM, domain-specific performance evaluation (Appendix~\ref{sec:domain_analysis}), NSPU performance on downstream tasks (Appendix~\ref{sec:downstream_tasks}), error analysis (Appendix~\ref{sec:error_analysis}) and robustness evaluation of NSPU (Appendix~\ref{sec:robustness}).

\section{Conclusion}
We introduce Shadow Unlearning, a paradigm for privacy-preserving unlearning on anonymized forget sets, and propose Neuro-Semantic Projector Unlearning (NSPU) to realize it. NSPU outperforms baselines with superior knowledge retention and forgetting efficacy, demonstrated across comprehensive metrics. It offers computational efficiency, confirmed by FLOPs analysis, and ensures privacy protection, validated through membership-inference attacks and separation-quality assessments. Overall, our work establishes a new direction for privacy-aware unlearning, balancing data protection, utility, and efficiency.

\section{Limitations}
NSPU relies on PCA to construct the forget subspace, which captures directions of maximum variance rather than exact causal influence. Nevertheless, these directions effectively approximate dominant semantic factors in the forget set, which is sufficient for approximate unlearning. The pipeline is primarily optimized for batch unlearning; its behavior under extremely low-resource settings (e.g., a single forget sample) remains unexplored. Additionally, handling sequential and continuous unlearning requests without cumulative utility degradation is an open challenge. NSPU applies the unlearning filter only at the final hidden layer to preserve earlier representations, leaving intermediate activations structurally unchanged. Although interpretability analysis suggests a low likelihood of extracting semantic PII from early layers, we do not provide formal guarantees against advanced white-box inversion attacks targeting these layers.

Theoretical guarantees of NSPU ensure geometric projection away from the forget subspace but do not imply equivalence to a model retrained from scratch without the forget data. Furthermore, the method depends on access to internal activations, restricting applicability to open-source models. Finally, our NER-based anonymization pipeline introduces a potential edge case where distinct samples map to identical anonymized representations. In such collisions, selectively unlearning only a subset may confuse the Latent Representation Aligner (LRA), leading to unintended retention or forgetting. Future work can address this by designing more fine-grained, entity-level unlearning filters to better distinguish structurally identical samples.

\section{Reproducibility Statement}
To ensure reproducibility of our results, we have thoroughly documented all experimental details and evaluation procedures with proper grounding. All experiments were conducted using fixed random seeds to guarantee deterministic behavior across runs. We plan to release our implementation code, detailed experimental scripts, and the dataset to facilitate the exact reproduction of our reported results.

\section{Ethics Statement}
This work advances responsible and privacy-aware deployment of large language models through Shadow Unlearning. The proposed paradigm enables effective machine unlearning using anonymized forget data, reducing exposure of personally identifiable information during the unlearning process. By decoupling unlearning from direct access to sensitive data, our approach supports compliance with privacy regulations such as the GDPR’s Right to be Forgotten. It also resolves a critical privacy paradox present in existing unlearning methods.

\bibliography{custom}

@article{wang2025oblivious,
  title={Oblivious unlearning by learning: Machine unlearning without exposing erased data},
  author={Wang, Weiqi and Tian, Zhiyi and Zhang, Chenhan and Yu, Shui},
  year={2025},
  url={https://openreview.net/forum?id=wAemQcyWqq}
}

@inproceedings{mainitofu,
  title={TOFU: A Task of Fictitious Unlearning for LLMs},
  author={Maini, Pratyush and Feng, Zhili and Schwarzschild, Avi and Lipton, Zachary Chase and Kolter, J Zico},
  booktitle={First Conference on Language Modeling},
  year={2024},
  url={https://openreview.net/forum?id=B41hNBoWLo}
}

@article{domingo2025efficient,
  title={Efficient Unlearning with Privacy Guarantees},
  author={Domingo-Ferrer, Josep and Jebreel, Najeeb and S{\'a}nchez, David},
  journal={arXiv preprint arXiv:2507.04771},
  year={2025},
  url={https://arxiv.org/abs/2507.04771}
}

@article{jiang2023mistral,
  title={Mistral 7B},
  author={Jiang, Albert Q and Sablayrolles, Alexandre and Mensch, Arthur and Bamford, Chris and Chaplot, Devendra Singh and Casas, Diego de las and Bressand, Florian and Lengyel, Gianna and Lample, Guillaume and Saulnier, Lucile and others},
  journal={arXiv preprint arXiv:2310.06825},
  year={2023},
  url={https://arxiv.org/abs/2310.06825}
}

@article{touvron2023llama2,
  title={Llama 2: Open foundation and fine-tuned chat models},
  author={Touvron, Hugo and Martin, Louis and Stone, Kevin and Albert, Peter and Almahairi, Amjad and Babaei, Yasmine and Bashlykov, Nikolay and Batra, Soumya and Bhargava, Prajjwal and Bhosale, Shruti and others},
  journal={arXiv preprint arXiv:2307.09288},
  year={2023},
  url={https://arxiv.org/abs/2307.09288}
}

@inproceedings{muennighoffolmoe,
  title={OLMoE: Open Mixture-of-Experts Language Models},
  author={Muennighoff, Niklas and Soldaini, Luca and Groeneveld, Dirk and Lo, Kyle and Morrison, Jacob and Min, Sewon and Shi, Weijia and Walsh, Evan Pete and Tafjord, Oyvind and Lambert, Nathan and others},
  booktitle={The Thirteenth International Conference on Learning Representations},
  year={2025},
  url={https://openreview.net/forum?id=xXTkbTBmqq}
}

@article{touvron2023llama,
  title={Llama: Open and efficient foundation language models},
  author={Touvron, Hugo and Lavril, Thibaut and Izacard, Gautier and Martinet, Xavier and Lachaux, Marie-Anne and Lacroix, Timoth{\'e}e and Rozi{\`e}re, Baptiste and Goyal, Naman and Hambro, Eric and Azhar, Faisal and others},
  journal={arXiv preprint arXiv:2302.13971},
  year={2023},
  url={https://arxiv.org/abs/2302.13971}
}

@inproceedings{liu2022continual,
  title={Continual learning and private unlearning},
  author={Liu, Bo and Liu, Qiang and Stone, Peter},
  booktitle={Conference on Lifelong Learning Agents},
  pages={243--254},
  year={2022},
  organization={PMLR},
  url={https://proceedings.mlr.press/v199/liu22a.html}
}

@article{rafailov2023direct,
  title={Direct preference optimization: Your language model is secretly a reward model},
  author={Rafailov, Rafael and Sharma, Archit and Mitchell, Eric and Manning, Christopher D and Ermon, Stefano and Finn, Chelsea},
  journal={Advances in neural information processing systems},
  volume={36},
  pages={53728--53741},
  year={2023},
  url={https://proceedings.neurips.cc/paper_files/paper/2023/hash/a85b405ed65c6477a4fe8302b5e06ce7-Abstract-Conference.html}
}

@article{gretton2012kernel,
  title={A Kernel Two-Sample Test},
  author={Gretton, Arthur and Borgwardt, Karsten M and Rasch, Malte and Schölkopf, Bernhard and Smola, Alex J},
  journal={Journal of Machine Learning Research},
  volume={13},
  number={Mar},
  pages={723--773},
  year={2012},
  url={https://www.jmlr.org/papers/volume13/gretton12a/gretton12a.pdf}
}

@inproceedings{cha2025towards,
  title={Towards Robust and Parameter-Efficient Knowledge Unlearning for LLMs},
  author={Cha, Sungmin and Cho, Sungjun and Hwang, Dasol and Lee, Moontae},
  booktitle={The Thirteenth International Conference on Learning Representations},
  year={2025},
  url={https://openreview.net/forum?id=1ExfUpmIW4}
}

@inproceedings{thudi2022unrolling,
  title={Unrolling sgd: Understanding factors influencing machine unlearning},
  author={Thudi, Anvith and Deza, Gabriel and Chandrasekaran, Varun and Papernot, Nicolas},
  booktitle={2022 IEEE 7th European Symposium on Security and Privacy (EuroS\&P)},
  pages={303--319},
  year={2022},
  organization={IEEE},
  url={https://ieeexplore.ieee.org/abstract/document/9797378?casa_token=OAXWUUGqLdEAAAAA:DozYfOTFUvyWSuwdkaXJndiCYARr4jo2GjXJPkxw2A9KZ1XprmazTjydj1pCAp-jJbW3fm0fu4U}
}

@article{chundawat2023zero,
  title={Zero-shot machine unlearning},
  author={Chundawat, Vikram S and Tarun, Ayush K and Mandal, Murari and Kankanhalli, Mohan},
  journal={IEEE Transactions on Information Forensics and Security},
  volume={18},
  pages={2345--2354},
  year={2023},
  publisher={IEEE},
  url={https://dl.acm.org/doi/abs/10.1109/TIFS.2023.3265506}
}

@inproceedings{yuan2025closer,
  title={A Closer Look at Machine Unlearning for Large Language Models},
  author={Yuan, Xiaojian and Pang, Tianyu and Du, Chao and Chen, Kejiang and Zhang, Weiming and Lin, Min},
  booktitle={The Thirteenth International Conference on Learning Representations},
  year={2025},
  url={https://openreview.net/forum?id=Q1MHvGmhyT}
}

@inproceedings{wang2023bfu,
  title={Bfu: Bayesian federated unlearning with parameter self-sharing},
  author={Wang, Weiqi and Tian, Zhiyi and Zhang, Chenhan and Liu, An and Yu, Shui},
  booktitle={Proceedings of the 2023 ACM Asia Conference on Computer and Communications Security},
  pages={567--578},
  year={2023},
  url={https://dl.acm.org/doi/abs/10.1145/3579856.3590327}
}

@inproceedings{liu2022right,
  title={The right to be forgotten in federated learning: An efficient realization with rapid retraining},
  author={Liu, Yi and Xu, Lei and Yuan, Xingliang and Wang, Cong and Li, Bo},
  booktitle={IEEE INFOCOM 2022-IEEE conference on computer communications},
  pages={1749--1758},
  year={2022},
  organization={IEEE},
  url={https://dl.acm.org/doi/abs/10.1109/INFOCOM48880.2022.9796721}
}

@article{wang2023federated,
  title={Federated unlearning and its privacy threats},
  author={Wang, Fei and Li, Baochun and Li, Bo},
  journal={IEEE Network},
  volume={38},
  number={2},
  pages={294--300},
  year={2023},
  publisher={IEEE},
  url={https://dl.acm.org/doi/abs/10.1109/MNET.004.2300056}
}

@inproceedings{thudi2022necessity,
  title={On the necessity of auditable algorithmic definitions for machine unlearning},
  author={Thudi, Anvith and Jia, Hengrui and Shumailov, Ilia and Papernot, Nicolas},
  booktitle={31st USENIX security symposium (USENIX Security 22)},
  pages={4007--4022},
  year={2022},
  url={https://www.usenix.org/conference/usenixsecurity22/presentation/thudi}
}

@inproceedings{chen2021machine,
  title={When machine unlearning jeopardizes privacy},
  author={Chen, Min and Zhang, Zhikun and Wang, Tianhao and Backes, Michael and Humbert, Mathias and Zhang, Yang},
  booktitle={Proceedings of the 2021 ACM SIGSAC conference on computer and communications security},
  pages={896--911},
  year={2021},
  url={https://dl.acm.org/doi/abs/10.1145/3460120.3484756}
}

@article{gao2022deletion,
  title={Deletion inference, reconstruction, and compliance in machine (un) learning},
  author={Gao, Ji and Garg, Sanjam and Mahmoody, Mohammad and Vasudevan, Prashant Nalini},
  journal={Proceedings on Privacy Enhancing Technologies},
  volume={2022},
  number={3},
  year={2022},
  url={https://crysp.petsymposium.org/popets/2022/popets-2022-0079.php}
}

@inproceedings{hu2024learn,
  title={Learn what you want to unlearn: Unlearning inversion attacks against machine unlearning},
  author={Hu, Hongsheng and Wang, Shuo and Dong, Tian and Xue, Minhui},
  booktitle={2024 IEEE Symposium on Security and Privacy (SP)},
  pages={3257--3275},
  year={2024},
  organization={IEEE},
  url={https://ieeexplore.ieee.org/abstract/document/10646717}
}

@inproceedings{zhang2023conditional,
  title={Conditional matching gan guided reconstruction attack in machine unlearning},
  author={Zhang, Kaiyue and Wang, Weiqi and Fan, Zipei and Song, Xuan and Yu, Shui},
  booktitle={GLOBECOM 2023-2023 IEEE Global Communications Conference},
  pages={44--49},
  year={2023},
  organization={IEEE},
  url={https://ieeexplore.ieee.org/abstract/document/10437231}
}

@article{gao2024verifi,
  title={Verifi: Towards verifiable federated unlearning},
  author={Gao, Xiangshan and Ma, Xingjun and Wang, Jingyi and Sun, Youcheng and Li, Bo and Ji, Shouling and Cheng, Peng and Chen, Jiming},
  journal={IEEE Transactions on Dependable and Secure Computing},
  volume={21},
  number={6},
  pages={5720--5736},
  year={2024},
  publisher={IEEE},
  url={https://dl.acm.org/doi/abs/10.1109/TDSC.2024.3382321}
}

@inproceedings{wang2024rkld,
  title={Balancing Forget Quality and Model Utility: A Reverse KL-Divergence Knowledge Distillation Approach for Better Unlearning in LLMs},
  author={Wang, Bichen and Zi, Yuzhe and Sun, Yixin and Zhao, Yanyan and Qin, Bing},
  booktitle={Proceedings of the 2025 Conference of the Nations of the Americas Chapter of the Association for Computational Linguistics: Human Language Technologies (Volume 1: Long Papers)},
  pages={1306--1321},
  year={2025},
  url={https://aclanthology.org/2025.naacl-long.60/}
}

@article{ji2024reversing,
  title={Reversing the forget-retain objectives: An efficient llm unlearning framework from logit difference},
  author={Ji, Jiabao and Liu, Yujian and Zhang, Yang and Liu, Gaowen and Kompella, Ramana R and Liu, Sijia and Chang, Shiyu},
  journal={Advances in Neural Information Processing Systems},
  volume={37},
  pages={12581--12611},
  year={2024},
  url={https://openreview.net/forum?id=tYdR1lTWqh}
}

@inproceedings{dou2024avoiding,
  title={Avoiding copyright infringement via large language model unlearning},
  author={Dou, Guangyao and Liu, Zheyuan and Lyu, Qing and Ding, Kaize and Wong, Eric},
  booktitle={Findings of the Association for Computational Linguistics: NAACL 2025},
  pages={5176--5200},
  year={2025},
  url={https://aclanthology.org/2025.findings-naacl.288/}
}

@inproceedings{pawelczyk2024context,
  title={In-Context Unlearning: Language Models as Few-Shot Unlearners},
  author={Pawelczyk, Martin and Neel, Seth and Lakkaraju, Himabindu},
  booktitle={International Conference on Machine Learning},
  pages={40034--40050},
  year={2024},
  organization={PMLR},
  url={https://dl.acm.org/doi/abs/10.5555/3692070.3693692}
}

@inproceedings{muresanu2024unlearnable,
  title={Fast Exact Unlearning for In-Context Learning Data for LLMs},
  author={Muresanu, Andrei Ioan and Thudi, Anvith and Zhang, Michael R and Papernot, Nicolas},
  booktitle={Forty-second International Conference on Machine Learning},
  year={2025},
  url={https://openreview.net/forum?id=TzNVZEsqTi}
}

@article{liu2024large,
  title={Large language model unlearning via embedding-corrupted prompts},
  author={Liu, Chris and Wang, Yaxuan and Flanigan, Jeffrey and Liu, Yang},
  journal={Advances in Neural Information Processing Systems},
  volume={37},
  pages={118198--118266},
  year={2024},
  url={https://openreview.net/forum?id=e5icsXBD8Q}
}

@article{wang2024federated,
  title={Federated Unlearning and Its Privacy Threats},
  author={Wang, Fei and Li, Baochun and Li, Bo},
  journal={IEEE Network},
  volume={38},
  number={2},
  pages={294--300},
  year={2024},
  publisher={IEEE},
  url={https://dl.acm.org/doi/abs/10.1109/MNET.004.2300056}
}

@article{liu2024survey,
  title={A survey on federated unlearning: Challenges, methods, and future directions},
  author={Liu, Ziyao and Jiang, Yu and Shen, Jiyuan and Peng, Minyi and Lam, Kwok-Yan and Yuan, Xingliang and Liu, Xiaoning},
  journal={ACM Computing Surveys},
  volume={57},
  number={1},
  pages={1--38},
  year={2024},
  publisher={ACM New York, NY},
  url={https://dl.acm.org/doi/full/10.1145/3679014}
}

@inproceedings{bhaila2025soft,
  title={Soft prompting for unlearning in large language models},
  author={Bhaila, Karuna and Van, Minh-Hao and Wu, Xintao},
  booktitle={Proceedings of the 2025 Conference of the Nations of the Americas Chapter of the Association for Computational Linguistics: Human Language Technologies (Volume 1: Long Papers)},
  pages={4046--4056},
  year={2025},
  url={https://aclanthology.org/2025.naacl-long.204/}
}

@article{eldan2310s,
  title={Who’s harry potter? approximate unlearning in llms, 2023},
  author={Eldan, Ronen and Russinovich, Mark},
  journal={URL https://arxiv.org/abs/2310.02238},
  volume={1},
  number={2},
  pages={8},
  url={https://arxiv.org/abs/2310.02238}
}

@article{huangoffset,
  title={Offset Unlearning for Large Language Models},
  author={Huang, James Y and Zhou, Wenxuan and Wang, Fei and Morstatter, Fred and Zhang, Sheng and Poon, Hoifung and Chen, Muhao},
  journal={Transactions on Machine Learning Research},
  url={https://openreview.net/forum?id=A4RLpHPXCu}
}

@article{tang2024learn,
  title={Learn while unlearn: An iterative unlearning framework for generative language models},
  author={Tang, Haoyu and Liu, Ye and Zhao, Xi and Liu, Xukai and Zhang, Yanghai and Zhang, Kai and Zhou, Xiaofang and Chen, Enhong},
  journal={arXiv preprint arXiv:2407.20271},
  year={2024},
  url={https://arxiv.org/abs/2407.20271}
}

@inproceedings{su2023asynchronous,
  title={Asynchronous federated unlearning},
  author={Su, Ningxin and Li, Baochun},
  booktitle={IEEE INFOCOM 2023-IEEE conference on computer communications},
  pages={1--10},
  year={2023},
  organization={IEEE},
  url={https://ieeexplore.ieee.org/abstract/document/10229075}
}

@inproceedings{bourtoule2021machine,
  title={Machine unlearning},
  author={Bourtoule, Lucas and Chandrasekaran, Varun and Choquette-Choo, Christopher A and Jia, Hengrui and Travers, Adelin and Zhang, Baiwu and Lie, David and Papernot, Nicolas},
  booktitle={2021 IEEE symposium on security and privacy (SP)},
  pages={141--159},
  year={2021},
  organization={IEEE},
  url={https://ieeexplore.ieee.org/abstract/document/9519428}
}

@online{GDPR_Article17_2018,
  author       = "{GDPR.eu}",
  title        = "{Article 17: Right to be forgotten}",
  year         = "2018",
  url          = "https://gdpr.eu/article-17-right-to-be-forgotten/",
  note         = "[Accessed: 2025-11-14]"
}

@article{achiam2023gpt,
  title={Gpt-4 technical report},
  author={Achiam, Josh and Adler, Steven and Agarwal, Sandhini and Ahmad, Lama and Akkaya, Ilge and Aleman, Florencia Leoni and Almeida, Diogo and Altenschmidt, Janko and Altman, Sam and Anadkat, Shyamal and others},
  journal={arXiv preprint arXiv:2303.08774},
  year={2023},
  url={https://arxiv.org/abs/2303.08774}
}

@inproceedings{cao2015towards,
  title={Towards making systems forget with machine unlearning},
  author={Cao, Yinzhi and Yang, Junfeng},
  booktitle={2015 IEEE symposium on security and privacy},
  pages={463--480},
  year={2015},
  organization={IEEE},
  url={https://ieeexplore.ieee.org/abstract/document/7163042}
}

@article{gu2024second,
  title={Second-order information matters: Revisiting machine unlearning for large language models},
  author={Gu, Kang and Rashid, Md Rafi Ur and Sultana, Najrin and Mehnaz, Shagufta},
  journal={arXiv preprint arXiv:2403.10557},
  year={2024},
  url={https://arxiv.org/abs/2403.10557}
}

@article{lu2024eraser,
  title={Eraser: Jailbreaking defense in large language models via unlearning harmful knowledge},
  author={Lu, Weikai and Zeng, Ziqian and Wang, Jianwei and Lu, Zhengdong and Chen, Zelin and Zhuang, Huiping and Chen, Cen},
  journal={arXiv preprint arXiv:2404.05880},
  year={2024},
  url={https://arxiv.org/abs/2404.05880}
}

@inproceedings{jia2024soul,
  title={SOUL: Unlocking the Power of Second-Order Optimization for LLM Unlearning},
  author={Jia, Jinghan and Zhang, Yihua and Zhang, Yimeng and Liu, Jiancheng and Runwal, Bharat and Diffenderfer, James and Kailkhura, Bhavya and Liu, Sijia},
  booktitle={Proceedings of the 2024 Conference on Empirical Methods in Natural Language Processing},
  pages={4276--4292},
  year={2024},
  url={https://aclanthology.org/2024.emnlp-main.245/}
}

@inproceedings{tian2024forget,
  title={To Forget or Not? Towards Practical Knowledge Unlearning for Large Language Models},
  author={Tian, Bozhong and Liang, Xiaozhuan and Cheng, Siyuan and Liu, Qingbin and Wang, Mengru and Sui, Dianbo and Chen, Xi and Chen, Huajun and Zhang, Ningyu},
  booktitle={Findings of the Association for Computational Linguistics: EMNLP 2024},
  pages={1524--1537},
  year={2024},
  url={https://aclanthology.org/2024.findings-emnlp.82/}
}

@article{choi2024snap,
  title={Snap: Unlearning selective knowledge in large language models with negative instructions},
  author={Choi, Minseok and Rim, Daniel and Lee, Dohyun and Choo, Jaegul},
  journal={arXiv e-prints},
  pages={arXiv--2406},
  year={2024},
  url={https://arxiv.org/abs/2406.12329v1}
}

@inproceedings{liu2024towards,
  title={Towards Safer Large Language Models through Machine Unlearning},
  author={Liu, Zheyuan and Dou, Guangyao and Tan, Zhaoxuan and Tian, Yijun and Jiang, Meng},
  booktitle={Findings of the Association for Computational Linguistics ACL 2024},
  pages={1817--1829},
  year={2024},
  url={https://aclanthology.org/2024.findings-acl.107/}
}

@article{ishibashi2023knowledge,
  title={Knowledge sanitization of large language models},
  author={Ishibashi, Yoichi and Shimodaira, Hidetoshi},
  journal={arXiv preprint arXiv:2309.11852},
  year={2023},
  url={https://arxiv.org/abs/2309.11852}
}

@article{brown2020language,
  title={Language models are few-shot learners},
  author={Brown, Tom and Mann, Benjamin and Ryder, Nick and Subbiah, Melanie and Kaplan, Jared D and Dhariwal, Prafulla and Neelakantan, Arvind and Shyam, Pranav and Sastry, Girish and Askell, Amanda and others},
  journal={Advances in neural information processing systems},
  volume={33},
  pages={1877--1901},
  year={2020},
  url={https://dl.acm.org/doi/abs/10.5555/3495724.3495883}
}

@inproceedings{yao-etal-2024-machine,
    title = "Machine Unlearning of Pre-trained Large Language Models",
    author = "Yao, Jin  and
      Chien, Eli  and
      Du, Minxin  and
      Niu, Xinyao  and
      Wang, Tianhao  and
      Cheng, Zezhou  and
      Yue, Xiang",
    editor = "Ku, Lun-Wei  and
      Martins, Andre  and
      Srikumar, Vivek",
    booktitle = "Proceedings of the 62nd Annual Meeting of the Association for Computational Linguistics (Volume 1: Long Papers)",
    month = aug,
    year = "2024",
    address = "Bangkok, Thailand",
    publisher = "Association for Computational Linguistics",
    doi = "10.18653/v1/2024.acl-long.457",
    pages = "8403--8419",
    url={https://aclanthology.org/2024.acl-long.457/}
}

@article{liu2025threats,
  title={Threats, attacks, and defenses in machine unlearning: A survey},
  author={Liu, Ziyao and Ye, Huanyi and Chen, Chen and Zheng, Yongsen and Lam, Kwok-Yan},
  journal={IEEE Open Journal of the Computer Society},
  year={2025},
  publisher={IEEE},
  url={https://ieeexplore.ieee.org/abstract/document/10892039}
}

@article{nicolazzo2025secure,
  title={How Secure is Forgetting? Linking Machine Unlearning to Machine Learning Attacks},
  author={Nicolazzo, Serena and Nocera, Antonino and others},
  journal={arXiv preprint arXiv:2503.20257},
  year={2025},
  url={https://arxiv.org/abs/2503.20257}
}

@article{nguyen2025survey,
  title={A survey of machine unlearning},
  author={Nguyen, Thanh Tam and Huynh, Thanh Trung and Ren, Zhao and Nguyen, Phi Le and Liew, Alan Wee-Chung and Yin, Hongzhi and Nguyen, Quoc Viet Hung},
  journal={ACM Transactions on Intelligent Systems and Technology},
  volume={16},
  number={5},
  pages={1--46},
  year={2025},
  publisher={ACM New York, NY},
  url={https://dl.acm.org/doi/full/10.1145/3749987}
}

@article{zhou2024limitations,
  title={On the limitations and prospects of machine unlearning for generative AI},
  author={Zhou, Shiji and Wang, Lianzhe and Ye, Jiangnan and Wu, Yongliang and Chang, Heng},
  journal={arXiv preprint arXiv:2408.00376},
  year={2024},
  url={https://arxiv.org/abs/2408.00376}
}

@inproceedings{liu2024breaking,
  title={Breaking the trilemma of privacy, utility, and efficiency via controllable machine unlearning},
  author={Liu, Zheyuan and Dou, Guangyao and Chien, Eli and Zhang, Chunhui and Tian, Yijun and Zhu, Ziwei},
  booktitle={Proceedings of the ACM Web Conference 2024},
  pages={1260--1271},
  year={2024},
  url={https://dl.acm.org/doi/abs/10.1145/3589334.3645669}
}

@inproceedings{jang2023knowledge,
  title={Knowledge unlearning for mitigating privacy risks in language models},
  author={Jang, Joel and Yoon, Dongkeun and Yang, Sohee and Cha, Sungmin and Lee, Moontae and Logeswaran, Lajanugen and Seo, Minjoon},
  booktitle={Proceedings of the 61st Annual Meeting of the Association for Computational Linguistics (Volume 1: Long Papers)},
  pages={14389--14408},
  year={2023},
  url={https://aclanthology.org/2023.acl-long.805/}
}

@article{yao2024large,
  title={Large language model unlearning},
  author={Yao, Yuanshun and Xu, Xiaojun and Liu, Yang},
  journal={Advances in Neural Information Processing Systems},
  volume={37},
  pages={105425--105475},
  year={2024},
  url={https://openreview.net/forum?id=8Dy42ThoNe}
}

@inproceedings{izzo2021approximate,
  title={Approximate data deletion from machine learning models},
  author={Izzo, Zachary and Smart, Mary Anne and Chaudhuri, Kamalika and Zou, James},
  booktitle={International conference on artificial intelligence and statistics},
  pages={2008--2016},
  year={2021},
  organization={PMLR},
  url={https://proceedings.mlr.press/v130/izzo21a.html}
}

@inproceedings{brophy2021machine,
  title={Machine unlearning for random forests},
  author={Brophy, Jonathan and Lowd, Daniel},
  booktitle={International Conference on Machine Learning},
  pages={1092--1104},
  year={2021},
  organization={PMLR},
  url={https://proceedings.mlr.press/v139/brophy21a.html}
}

@inproceedings{schelter2021hedgecut,
  title={Hedgecut: Maintaining randomised trees for low-latency machine unlearning},
  author={Schelter, Sebastian and Grafberger, Stefan and Dunning, Ted},
  booktitle={Proceedings of the 2021 International Conference on Management of Data},
  pages={1545--1557},
  year={2021},
  url={https://dl.acm.org/doi/abs/10.1145/3448016.3457239}
}

@article{ginart2019making,
  title={Making ai forget you: Data deletion in machine learning},
  author={Ginart, Antonio and Guan, Melody and Valiant, Gregory and Zou, James Y},
  journal={Advances in neural information processing systems},
  volume={32},
  year={2019},
  url={https://dl.acm.org/doi/10.5555/3454287.3454603}
}

@article{xu2024machine,
  title={Machine unlearning: Solutions and challenges},
  author={Xu, Jie and Wu, Zihan and Wang, Cong and Jia, Xiaohua},
  journal={IEEE Transactions on Emerging Topics in Computational Intelligence},
  volume={8},
  number={3},
  pages={2150--2168},
  year={2024},
  publisher={IEEE},
  url={https://ieeexplore.ieee.org/abstract/document/10488864}
}

@article{li2025machine,
  title={Machine unlearning: Taxonomy, metrics, applications, challenges, and prospects},
  author={Li, Na and Zhou, Chunyi and Gao, Yansong and Chen, Hui and Zhang, Zhi and Kuang, Boyu and Fu, Anmin},
  journal={IEEE Transactions on Neural Networks and Learning Systems},
  year={2025},
  publisher={IEEE},
  url={https://ieeexplore.ieee.org/abstract/document/10880482}
}

@inproceedings{li2024wmdp,
  title={The WMDP Benchmark: Measuring and Reducing Malicious Use with Unlearning},
  author={Li, Nathaniel and Pan, Alexander and Gopal, Anjali and Yue, Summer and Berrios, Daniel and Gatti, Alice and Li, Justin D and Dombrowski, Ann-Kathrin and Goel, Shashwat and Mukobi, Gabriel and others},
  booktitle={International Conference on Machine Learning},
  pages={28525--28550},
  year={2024},
  organization={PMLR},
  url={https://proceedings.mlr.press/v235/li24bc.html}
}

@inproceedings{wang2025selective,
  title={Selective forgetting: Advancing machine unlearning techniques and evaluation in language models},
  author={Wang, Lingzhi and Zeng, Xingshan and Guo, Jinsong and Wong, Kam-Fai and Gottlob, Georg},
  booktitle={Proceedings of the AAAI Conference on Artificial Intelligence},
  volume={39},
  number={1},
  pages={843--851},
  year={2025},
  url={https://ojs.aaai.org/index.php/AAAI/article/view/32068}
}

@inproceedings{zhang2024negative,
  title={Negative Preference Optimization: From Catastrophic Collapse to Effective Unlearning},
  author={Zhang, Ruiqi and Lin, Licong and Bai, Yu and Mei, Song},
  booktitle={First Conference on Language Modeling},
  year={2024},
  url={https://openreview.net/forum?id=MXLBXjQkmb}
}

@inproceedings{carlini2022membership,
  title={Membership inference attacks from first principles},
  author={Carlini, Nicholas and Chien, Steve and Nasr, Milad and Song, Shuang and Terzis, Andreas and Tramer, Florian},
  booktitle={2022 IEEE symposium on security and privacy (SP)},
  pages={1897--1914},
  year={2022},
  organization={IEEE},
  url={https://ieeexplore.ieee.org/abstract/document/9833649}
}

@inproceedings{shokri2017membership,
  title={Membership inference attacks against machine learning models},
  author={Shokri, Reza and Stronati, Marco and Song, Congzheng and Shmatikov, Vitaly},
  booktitle={2017 IEEE symposium on security and privacy (SP)},
  pages={3--18},
  year={2017},
  organization={IEEE},
  url={https://ieeexplore.ieee.org/abstract/document/7958568}
}

@article{maini2024llm,
  title={LLM Dataset Inference: Did you train on my dataset?},
  author={Maini, Pratyush and Jia, Hengrui and Papernot, Nicolas and Dziedzic, Adam},
  journal={Advances in Neural Information Processing Systems},
  volume={37},
  pages={124069--124092},
  year={2024},
  url={https://openreview.net/forum?id=Fr9d1UMc37}
}

@inproceedings{liu2025lune,
  title={LUNE: Efficient LLM Unlearning via LoRA Fine-Tuning with Negative Examples},
  author={Liu, Yezi and Chen, Hanning and Huang, Wenjun and Ni, Yang and Imani, Mohsen},
  booktitle={Socially Responsible and Trustworthy Foundation Models at NeurIPS 2025},
  url={https://openreview.net/forum?id=Dim7kQ8Kol}
}

@inproceedings{hendrycks2020measuring,
  title={Measuring Massive Multitask Language Understanding},
  author={Hendrycks, Dan and Burns, Collin and Basart, Steven and Zou, Andy and Mazeika, Mantas and Song, Dawn and Steinhardt, Jacob},
  booktitle={International Conference on Learning Representations},
  year={2021},
  url={https://openreview.net/forum?id=d7KBjmI3GmQ}
}

@article{clark2018think,
  title={Think you have solved question answering? try arc, the ai2 reasoning challenge},
  author={Clark, Peter and Cowhey, Isaac and Etzioni, Oren and Khot, Tushar and Sabharwal, Ashish and Schoenick, Carissa and Tafjord, Oyvind},
  journal={arXiv preprint arXiv:1803.05457},
  year={2018},
  url={https://arxiv.org/abs/1803.05457}
}

@inproceedings{lin2022truthfulqa,
  title={Truthfulqa: Measuring how models mimic human falsehoods},
  author={Lin, Stephanie and Hilton, Jacob and Evans, Owain},
  booktitle={Proceedings of the 60th annual meeting of the association for computational linguistics (volume 1: long papers)},
  pages={3214--3252},
  year={2022},
  url={https://aclanthology.org/2022.acl-long.229/}
}

@article{qu2024learn,
  title={Learn to unlearn: Insights into machine unlearning},
  author={Qu, Youyang and Yuan, Xin and Ding, Ming and Ni, Wei and Rakotoarivelo, Thierry and Smith, David},
  journal={Computer},
  volume={57},
  number={3},
  pages={79--90},
  year={2024},
  publisher={IEEE},
  url={https://dl.acm.org/doi/abs/10.1109/MC.2023.3333319}
}

\appendix

\section{Extended Related works}
\label{sec:extend_relatedwork}
\textbf{Privacy-preserving unlearning:} Most existing unlearning methods rely on direct access to the user’s data during forget operations, which contradicts the goal of data deletion and raises compliance concerns with the ``right to be forgotten'' \cite{wang2023bfu, liu2022right, thudi2022necessity}. Despite highlighting the privacy breaches caused by the unlearning approaches \cite{chen2021machine, gao2022deletion, hu2024learn, zhang2023conditional}, there has been less focus on privacy-preserving unlearning. \\
To enable privacy-preserving unlearning, \citet{wang2025oblivious} proposed oblivious unlearning, which constructs an auxiliary synthetic dataset via incremental training to facilitate effective removal of targeted data influences. However, their evaluation is restricted to image classification benchmarks such as MNIST, CIFAR-10, and CelebA. Subsequent efforts investigate privacy-aware unlearning across multiple classification settings \cite{domingo2025efficient}. Most of the privacy-oriented unlearning methods operate within the federated learning paradigm, commonly referred to as federated unlearning (FL) \cite{gao2024verifi, su2023asynchronous, wang2023bfu, liu2022right}, where model training is performed centrally without direct interaction with individual users. Nevertheless, most FL-based unlearning approaches incur notable utility degradation and exhibit limited support for selectively unlearning subsets of a user’s local data \cite{liu2024survey, wang2024federated}. Moreover, despite targeting user-specific unlearning, existing FL frameworks often provide insufficient privacy guarantees for user data.

\section{MuFU Dataset}
\label{sec:synthetic_data}
\subsection{Creation of MuFU dataset}
\begin{figure*}[]
  \centering
  \includegraphics[width=\textwidth]{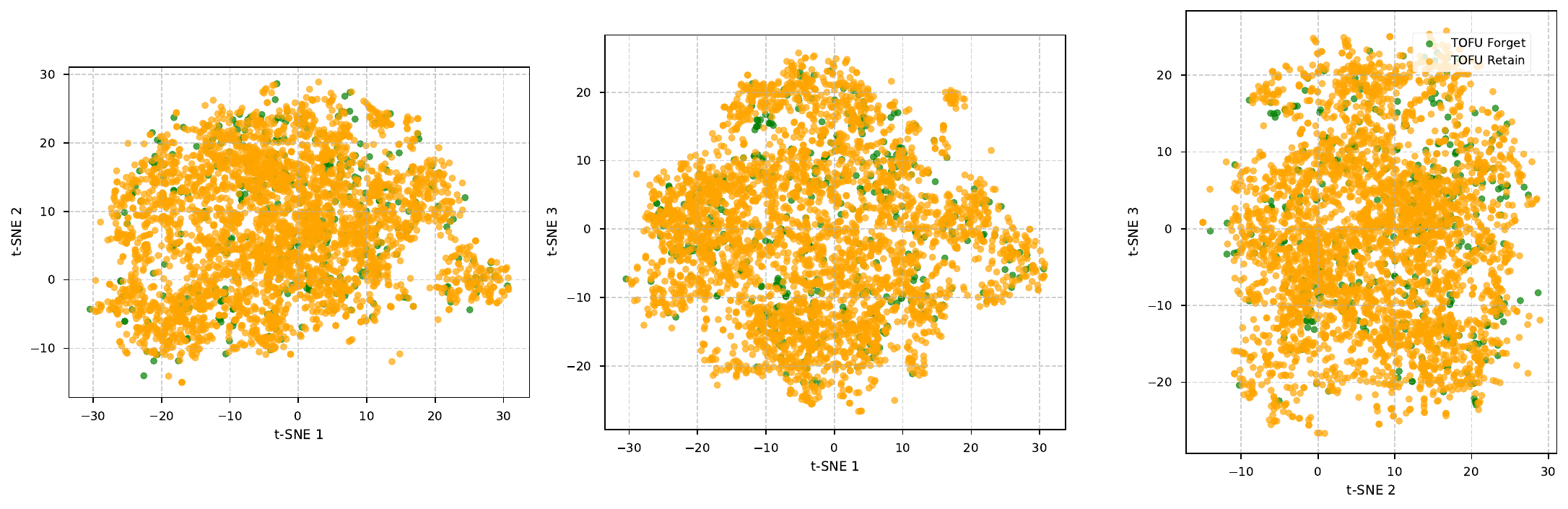}
  \caption{t-SNE plot of activation vectors distribution of TOFU forget and retain datasets}
  \label{fig:tofu_tsne}
\end{figure*}

\begin{figure*}[]
  \centering
  \includegraphics[width=\textwidth]{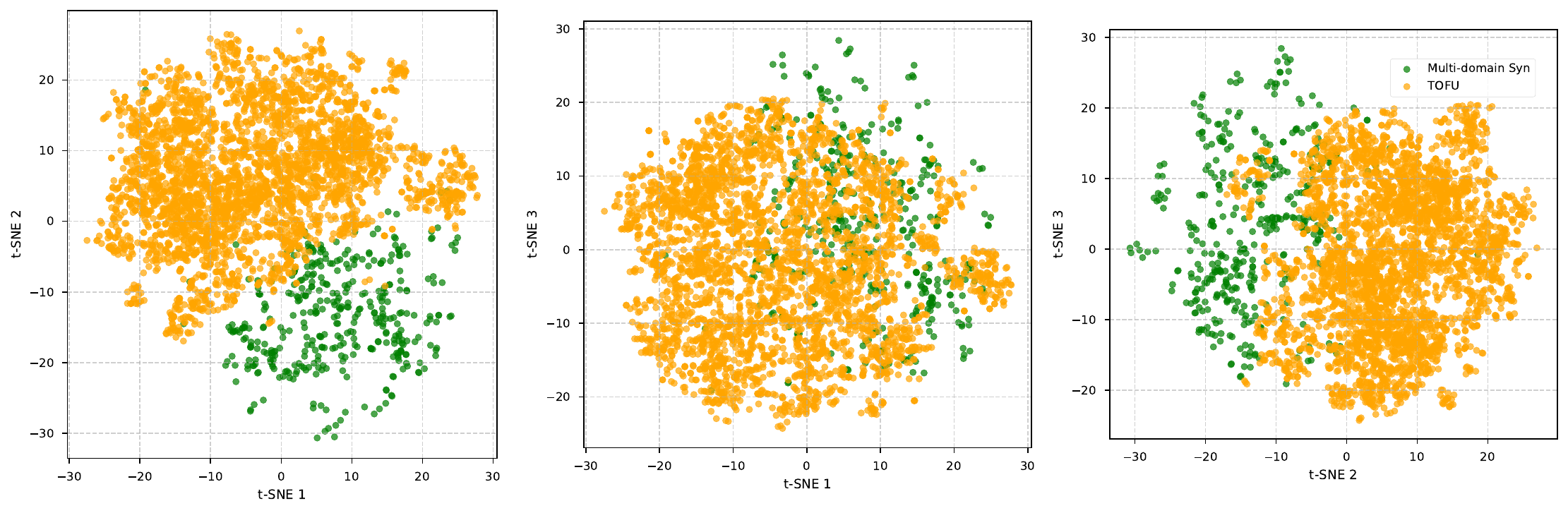}
  \caption{t-SNE plot of activation vectors distribution of TOFU and multi-domain synthetic datasets}
  \label{fig:syn_tsne}
\end{figure*}

\begin{figure*}[]
  \centering
  \includegraphics[width=\textwidth]{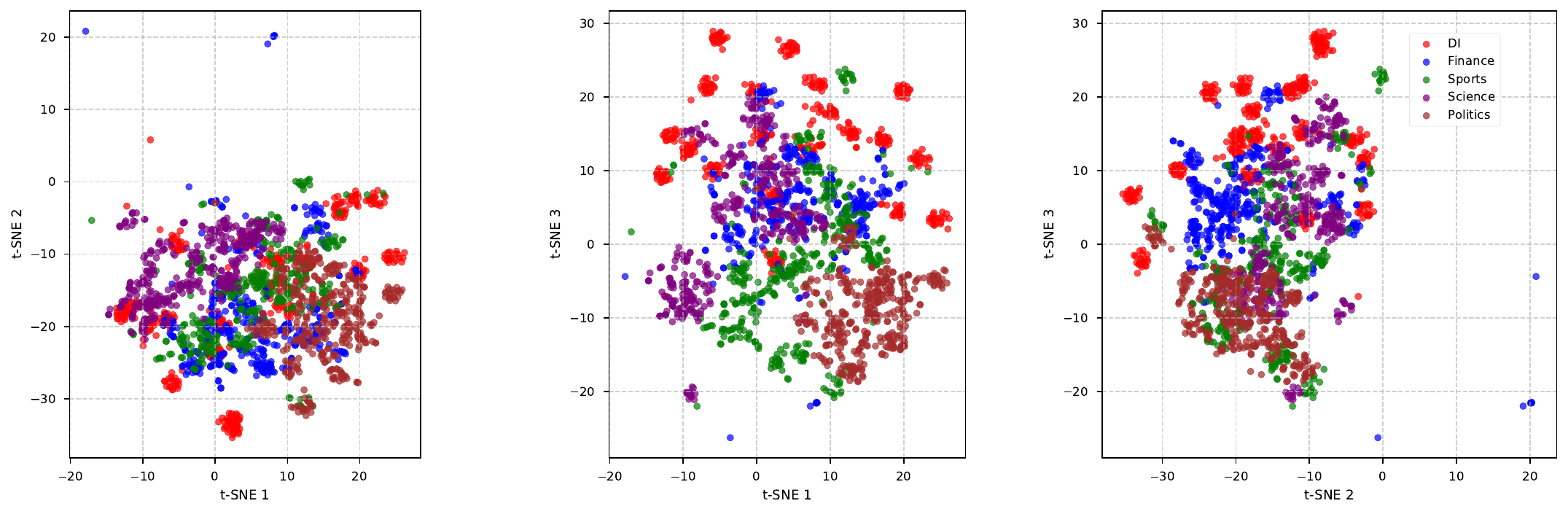}
  \caption{t-SNE plot of activation vectors distribution of multi-domain synthetic datasets}
  \label{fig:multidomain_tsne}
\end{figure*}

To address the challenge of evaluation of unlearning efficacy owing to extreme homogeneity and a high degree of semantic overlap in the existing benchmark datasets, we create a TOFU-style synthetic dataset for five different domains: Digital Informatics, Science and Technology, Sports, Finance, and Politics. Each domain consists of 800 samples, and a total of 4000 samples are present in the MuFU dataset. We depict the overlap present in the TOFU forget and retain samples in Figure~\ref{fig:tofu_tsne}, and that of TOFU, and MuFU data samples in Figure~\ref{fig:syn_tsne}. Figure~\ref{fig:multidomain_tsne} further illustrates the overlap of various domains in our dataset. Table~\ref{tab:pii_domain_transposed_clean} details the PII attribute-wise distribution for various domains and Table~\ref{tab:pii_aggregate_transposed_clean_aggregate} presents the aggregate PII attribute-wise distribution of the MuFU dataset. The prompt utilized to obtain the synthetic dataset using Gemini 3 Pro is elaborated in Fig~\ref{fig:prompt_sync_data}.


\begin{table}[b]
\centering\small
\setlength{\tabcolsep}{5pt}
\begin{tabularx}{\linewidth}{l X X}
\toprule
\textbf{Corpus} & \textbf{Original sample} & \textbf{NER-based anonymization} \\
\midrule
\textbf{Ques.} &
What is the full name of the crime-fiction author born in Taipei on 05/11/1995, best known for a long-running detective series? &
What is the full name of the crime-fiction author born in \textless LOCATION\textgreater{} on \textless DATETIME\textgreater{}, best known for a long-running detective series?\\
\textbf{Ans.} &
The author’s full name is Hsiao Yun-Hwa. & The author’s full name is \textless PERSON\textgreater{}.\\
\bottomrule
\end{tabularx}
\caption{Example of NER-based anonymization}
\label{tab:anonymization_example_revised}
\end{table}

\begin{table*}[]
\centering
\footnotesize
\begin{tabular}{lccccc}
\toprule
\textbf{PII Type} & \textbf{Sports} & \textbf{Finance} & 
\textbf{Digital Informatics} & \textbf{Science} & \textbf{Politics} \\
\midrule
PERSON            & \phantom{0}774 & 800 & 774 & 758 & 800 \\
LOCATION          & 102  & 38\phantom{0}  & 102\phantom{0}  & 96\phantom{0}  & 48\phantom{0} \\
PHONE\_NUMBER     & 80  & 14\phantom{0}\phantom{0}   & 80\phantom{0}  & 0\phantom{0}\phantom{0}   & 98\phantom{0} \\
EMAIL             & 40  & 0\phantom{0}\phantom{0}   & 40\phantom{0}  & 0\phantom{0}\phantom{0}   & 50\phantom{0} \\
DATE\_TIME        & 66  & 98\phantom{0}  & 66\phantom{0}  & 200 & 36\phantom{0} \\
IN\_PAN           & 58  & 108\phantom{0}  & 58\phantom{0}  & 186\phantom{0}  & 90\phantom{0} \\
URL               & 16\phantom{0}   & 44\phantom{0}  & 16\phantom{0}\phantom{0}   & 24\phantom{0}  & 24\phantom{0} \\
US\_DRIVER\_LICENSE & 34 & 10\phantom{0}\phantom{0}   & 34\phantom{0}  & 98\phantom{0}  & 294 \\
NRP               & 34  & 6\phantom{0}\phantom{0}   & 34\phantom{0}  & 2\phantom{0}\phantom{0}   & 4\phantom{0}\phantom{0} \\

Misc  & 98 & 56\phantom{0} & 98\phantom{0} & 2\phantom{0}\phantom{0} & 72\phantom{0} \\
\midrule
Single-Entity     & \phantom{0}426 & 490 & 426 & 338 & 212 \\
Multi-Entity      & \phantom{0}370 & 310 & 370 & 442 & 588 \\
\midrule
\textbf{Total}    & \phantom{0}800 & 800 & 800 & 800 & 800 \\
\bottomrule
\end{tabular}
\caption{Domain-wise PII Distribution}
\label{tab:pii_domain_transposed_clean}
\end{table*}

\tcbset{
  systemprompt/.style={
    colback=white,
    colframe=SteelBlue,
    coltitle=white, 
    center title, 
    sharp corners=all, 
    rounded corners=west, 
    arc=3mm,
    title filled=true,
    colbacktitle=SteelBlue,
    borderline west={2mm}{0pt}{SteelBlue},
    drop shadow={SteelBlue!50!white},
    boxrule=1.3pt,
  }
}

\begin{figure*}
    \centering
\begin{tcolorbox}[systemprompt, title={System Prompt for Synthetic Dataset Generation}]
The task is to generate a synthetic dataset to complement the existing TOFU Authors dataset while ensuring a distinguishable distribution shift. The original dataset follows a narrative biographical structure involving identity establishment, professional background, and work analysis. Your goal is to retain comparable data complexity while introducing new question--answer patterns that significantly deviate in structure and content.

\vspace{0.3cm} 

Please adhere to the following requirements for dataset generation:

\begin{itemize}
    \item \textbf{Volume:} Create 20 distinct sets of data. Each set must contain exactly 20 question-answer pairs, for a total of 400 pairs.
    \item \textbf{Core Content:} Every question-answer pair must involve a subject entity associated with some form of personally identifiable information (PII).
    \item \textbf{Answer Conciseness:} All answers must be concise and not exceed a maximum length of 20 words.
    \item \textbf{Distributional Drift:} The primary objective is to achieve the maximum possible distributional drift from the original dataset. The resulting samples should be reliably separable from the original dataset by a classifier.
    \item \textbf{Structural Variety:} Ensure internal variety by diversifying sentence structures and the contexts of entity--PII relationships (e.g., contact details, addresses, specific identifiers, etc.).
\end{itemize}

\vspace{0.3cm}

\textbf{Critical Exclusions -- Do NOT replicate the following patterns:}
\begin{itemize}
    \item \textbf{Narrative Structure:} Avoid any question-answer formation that replicates the narrative style of the original dataset. This includes:
    \begin{itemize}
        \item Establishing identity (e.g., early life, origin, family).
        \item Defining profession, listing awards, or detailing notable works.
        \item Linking personal background to professional style or themes.
        \item Analyzing fictional works.
        \item Expanding a persona with biography-style details.
    \end{itemize}
    \item \textbf{Topic-Based Exclusions:} Do not include aspects such as early life, bibliographic details, awards, achievements, motivations, or notable works.
\end{itemize}

\end{tcolorbox}
\caption{System Prompt for Synthetic Dataset Generation}
\label{fig:prompt_sync_data}
\end{figure*}

\begin{table}[]
\centering
\footnotesize
\begin{tabular}{lr}
\toprule
\textbf{PII Type} & \textbf{Count} \\
\midrule
PERSON             & 3906 \\
LOCATION           & 386 \\
PHONE\_NUMBER      & 272 \\
EMAIL              & 130 \\
DATE\_TIME         & 466 \\
IN\_PAN            & 500 \\
URL                & 124 \\
US\_DRIVER\_LICENSE & 470 \\
NRP                & 80 \\

Misc& 152 \\
\midrule
Single-Entity      & 1892 \\
Multi-Entity       & 2080 \\
\midrule
\textbf{Total}     & 4000 \\
\bottomrule
\end{tabular}
\caption{Aggregate PII Distribution}
\label{tab:pii_aggregate_transposed_clean_aggregate}
\end{table}

\subsection{Multi-Domain Data Variants} 
To evaluate the efficacy of unlearning algorithms across varying levels of distributional complexity, we extend the TOFU framework into a multi-domain configuration. We evaluate unlearning performance across three distinct dataset variants, namely 0\%, 50\%, and 100\% based on the percentage of MuFU integration, representing a spectrum of increasing heterogeneity.

The 0\% variant consists of the original TOFU dataset to provide a baseline for unlearning within a single, homogeneous domain. In the 50\% variant, we replace half of the TOFU samples with 2,000 MuFU samples, distributed as 400 samples across each of the five MuFU domains. Finally, the 100\% variant represents a purely heterogeneous environment. This progression from pure homogeneity to full heterogeneity serves two technical purposes:
\begin{itemize}
    \item Distributional Resilience: By transitioning from 0\% to 100\% MuFU, we test the structural stability of the unlearning process. While most techniques are typically optimized for narrow distributions like TOFU, this setup studies if increased data diversity and multi-domain interference begin to degrade the ability of an unlearning algorithm to selectively target information.
    \item Signal Disentanglement: The 50\% variant acts as a critical midpoint to evaluate signal disentanglement. It forces the algorithm to operate in a fragmented knowledge base where it must distinguish between the dense, correlated signals of the remaining TOFU data and the sparse, diverse PII samples from the five MuFU domains. This allows us to observe whether the unlearning mechanism remains precise when the target information is no longer part of the majority distribution.
\end{itemize}

\begin{table*}[]
\centering\small
\begin{tabular}{llccccc}
\toprule
\textbf{Model} & \textbf{\% overlap} & \textbf{MSE} & \textbf{MAE} & \textbf{R\textsuperscript{2}} & \textbf{Cosine Sim.} & \textbf{Pearson Corr.} \\
\midrule
\multirow{4}{*}{LLaMA-7B} 
 & 0 & 0.069391 &	0.186194 &	0.753987	 & 0.956754 &	0.956758 \\
 & 50 & 0.067596 &	0.185825 &	0.751239 &	0.956326 &	0.956331\\
 & 100  & 0.086383 &	0.221884 &	0.670867	& 0.925712 &	0.925712\\
 
\midrule
\multirow{4}{*}{Mistral-7B Instruct} 
 & 0 & 1.115554  &	0.794082  &	0.262974 &	0.923258  &	0.923251\\
 & 50 &  1.089064 &	0.78439 &	-2.05293	 &0.921589	 &0.921596\\
 & 100  & 1.044335  &	0.772055 &	0.443277	 &0.92783	 &0.927807\\
\midrule
\multirow{4}{*}{OLMoE-1B-7B} 
 & 0 & 0.116232 &	0.258255 &	0.740218 &	0.938542 &	0.938537 \\
 & 50 & 0.12127 &	0.263976 &	0.7344	 &0.937737	 &0.937733 \\
 & 100  & 0.103798 &	0.242832 &	0.758868 &	0.947988 &	0.947984 \\
\midrule
\multirow{4}{*}{LLaMA-13B} 
 & 0  & 0.026172 &	0.120174 &	0.500153 &	0.944772 &	0.944771  \\
 & 50  & 0.02644 &	0.120398 &	0.572611 &	0.945107 &	0.945105  \\
 & 100   & 0.02626 &	0.120172 &	0.565725 & 	0.944636	 &0.944634  \\
\bottomrule
\end{tabular}
\caption{The MLP performance of various models across the overlap percentages; MSE - Mean Squared Error, MAE - Mean Absolute Error.}
\label{tab:mlp_results}
\end{table*}

\section{Training of Latent Representation Aligner}
\label{sec:trianing_mlp}
\begin{table*}[]
\centering
\begin{tabular}{lclc}
\toprule
\textbf{Dataset}                                  & \textbf{PII Category} & \textbf{Trainset Samples} & \textbf{Samples used} \\ \midrule
\href{https://huggingface.co/datasets/modeldev/synthetic_pii_finance}{Synthetic PII Finance}                  & Finance               & 24158 & 5000     \\ 
\href{https://huggingface.co/datasets/Parth/piiforprivacy}{PII for Privacy}                                & Health                & 135621       & 5000    \\ 
\href{https://huggingface.co/datasets/soates/australian-insurance-pii-dataset-corrected}{Australian PII Dataset}  & Customer interaction  & 1240    & 1240  \\ 
\href{https://huggingface.co/datasets/ponoma16/implicit_pii_detection}{Implicit PII Detection}                 & Synthetic profiles    & 5000       & 5000     \\
 \href{https://huggingface.co/datasets/King-Harry/NinjaMasker-PII-Redaction}{NinjaMasker PII Redaction}             & Chat assistant        & 34700    & 5000     \\ 
\href{https://huggingface.co/datasets/ai4privacy/pii-masking-400k}{PII Masking-400k}                        & General PII           & 326000      & 5000     \\ \bottomrule
\end{tabular}
\caption{Public data used for training Latent representation aligner}
\label{tab:public_data_table}
\end{table*}

\subsection{Public data collection}
The unknown and potentially multi-domain nature of the anonymized data necessitates the Latent representation aligner to be trained on heterogeneous data. For this purpose, we have combined six different PII datasets from different categories and formats, as depicted in Table \ref{tab:public_data_table}. The curated training dataset for the projector consists of 21.24k $(original\_text, anonymized\_counterpart)$ samples.

\subsection{Training the MLP}
To assess the recoverability of semantic information removed during anonymization, we train a lightweight Multilayer Perceptron (MLP) model designed to reconstruct the embedding of the original text from the embedding of its anonymized counterpart. The MLP is optimized using the mean squared error (MSE) objective. All experiments follow a standardized train–validation–test split. By posing reconstruction as a purely geometric mapping problem in the latent space, we obtain a model-agnostic probe for evaluating whether anonymized representations still encode residual semantic information about the original content.

Given an anonymized embedding $\mathbf{x}_i$ and its corresponding original embedding $\mathbf{y}_i$, 
the reconstruction model learns a function $f_{\theta}$ parameterized by $\theta$ such that

\begin{equation}
    \hat{\mathbf{y}}_i = f_{\theta}(\mathbf{x}_i),
\end{equation}

where $\hat{\mathbf{y}}_i$ denotes the reconstructed embedding.  
The model is optimized using the Mean Squared Error (MSE) objective defined as

\begin{equation}
    \mathcal{L}_{\text{MSE}} = 
    \frac{1}{N} \sum_{i=1}^{N} 
    \left\| \mathbf{y}_i - \hat{\mathbf{y}}_i \right\|_2^{2},
\end{equation}

where $N$ is the number of training samples.  
Minimizing this loss encourages the network to produce reconstructed embeddings 
that closely approximate the original semantic representations.

To assess the quality of reconstructing the original embedding 
$\mathbf{y}_i$ from its anonymized counterpart, we compute 
a set of complementary evaluation metrics.
\begin{enumerate}

\item \textbf{Mean Squared Error (MSE).}
The Mean Squared Error measures the average squared deviation 
between the true and reconstructed embeddings:
\begin{equation}
    \text{MSE} = 
    \frac{1}{N} \sum_{i=1}^{N} 
    \left\| \mathbf{y}_i - \hat{\mathbf{y}}_i \right\|_2^{2}.
\end{equation}

\item \textbf{Mean Absolute Error (MAE).}
MAE quantifies the average magnitude of reconstruction error:
\begin{equation}
    \text{MAE} = 
    \frac{1}{N} \sum_{i=1}^{N} 
    \left| \mathbf{y}_i - \hat{\mathbf{y}}_i \right|.
\end{equation}

\item \textbf{Coefficient of Determination ($R^2$).}
The $R^2$ metric measures the proportion of variance 
in the original embeddings explained by the reconstructed embeddings:
\begin{equation}
    R^2 = 
    1 - 
    \frac{
        \sum_{i=1}^{N} 
        \left\| \mathbf{y}_i - \hat{\mathbf{y}}_i \right\|_2^{2}
    }{
        \sum_{i=1}^{N} 
        \left\| \mathbf{y}_i - \bar{\mathbf{y}} \right\|_2^{2}
    },
\end{equation}
where $\bar{\mathbf{y}}$ denotes the mean embedding over the dataset.

\item \textbf{Cosine Similarity.}
To measure semantic alignment, we compute cosine similarity between 
each original and reconstructed embedding:
\begin{equation}
    \cos(\theta_i) = 
    \frac{
        \mathbf{y}_i \cdot \hat{\mathbf{y}}_i
    }{
        \left\| \mathbf{y}_i \right\|_2 \,
        \left\| \hat{\mathbf{y}}_i \right\|_2
    }.
\end{equation}

\item \textbf{Pearson Correlation.}
Pearson correlation captures the linear correlation between 
the embedding dimensions of $\mathbf{y}_i$ and $\hat{\mathbf{y}}_i$:
\begin{equation}
    \rho_i = 
    \frac{
        \mathrm{Cov}(\mathbf{y}_i, \hat{\mathbf{y}}_i)
    }{
        \sigma_{\mathbf{y}_i} \, \sigma_{\hat{\mathbf{y}}_i}
    }.
\end{equation}
\end{enumerate}
Together, these metrics characterize both numerical fidelity 
(MSE, MAE), variance explanation ($R^2$), and structural or 
semantic similarity (cosine similarity and Pearson correlation). The performance of the evaluation metrics for various LLMs is detailed in Table~\ref{tab:mlp_results}.

\section{Geometric Interpretation of Forget Space Creation}
\label{sec:theroy_forget_space}
As represented in Figure \ref{fig:projection}, the proposed methodology first implicitly decomposes $v_{in}$ into its `forget' component:
\begin{equation}
    v_{in}^u = (UU^T)v_{in}
\end{equation} 
and its `safe' component:
\begin{equation}
    v_{in}^{u\perp} = (I-UU^T)v_{in}
\end{equation}
Thus:
\begin{equation}
    v_{in} = v_{in}^u + v_{in}^{u\perp}
    \label{decomp}
\end{equation}

It then subtracts a fraction of $v_{in}^u$ from the original vector. The result: 
\begin{equation}
    v_{out} = v_{in} - \alpha v_{in}^u
    \label{vout}
\end{equation} is a vector steered away from the forget subspace. This attenuates information aligned with the concepts to be forgotten, while the safe component $v_{in}^\perp$ is preserved intact. \\
Here, $\alpha$ governs the strength or effectiveness of unlearning. When set to 1, $v_{out}$ becomes $v_{in}^{u\perp}$, with no component along the forget subspace, thus, nullifying the entire forget set information captured in the subspace from the model. However, this dents the semantic understanding and other general capabilities of the model. Therefore, $\alpha$ plays a crucial role in deciding the extent of unlearning with a tradeoff against the knowledge retention and multi-purpose general capabilities of the model. \\
Substituting equation \eqref{decomp} in equation \eqref{vout}:
\begin{equation}
    v_{out} = v_{in}^{u\perp} + (1- \alpha) v_{in}^u
\end{equation}
\begin{equation}
    v_{out} = v_{in} - v_{in}UU^T + (1-\alpha)v_{in}UU^T
\end{equation}
\begin{equation}
    v_{out} = v_{in} - v_{in}UU^T + v_{in}UU^T - \alpha v_{in}UU^T
\end{equation}
\begin{equation}
    v_{out} = v_{in} - \alpha v_{in}UU^T
\end{equation}
\begin{equation}
    v_{out} = (I - \alpha v_{in}UU^T)v_{in}
\end{equation}
Considering $(1 - \alpha v_{in}UU^T)$ as the $UL_{filter}$:
\begin{equation}
    UL_{filter} = (I - \alpha v_{in}UU^T)
\end{equation}
We get the final active unlearning equation as:
\begin{equation}
    v_{out} = UL_{filter}v_{in}
\end{equation}

\section{Experimental Setup}
\label{sec:experiment_setup}



\begin{table}[t!]
\centering
\resizebox{\columnwidth}{!}{%
\begin{tabular}{l c c}
\toprule
\textbf{Model name} & \textbf{Parameters count} & \textbf{Model type} \\
\midrule
\href{https://huggingface.co/meta-llama/Llama-2-7b}{Llama-7B} 
& 7B 
& Base \\

\href{https://huggingface.co/mistralai/Mistral-7B-Instruct-v0.1}{Mistral-7B-Instruct-v0.1} 
& 7B 
& Instruct \\

\href{https://huggingface.co/meta-llama/Llama-2-13b-chat-hf}{Llama-13B-Chat} 
& 13B 
& Chat \\ 

\href{https://huggingface.co/allenai/OLMoE-1B-7B-0924}{OLMoE-1B-7B-0924} 
& 1B (Active) 
& Mixture of Experts \\
\bottomrule
\end{tabular}}
\caption{Models used in our experiments.}
\label{tab:models_info}
\end{table}

We conduct all experiments using two Nvidia GeForce RTX A6000 (48GB) GPUs. Table \ref{tab:models_info} details the models used for our experiments, all of which were loaded in bfloat16 precision. We fine-tuned all models using a learning rate of 2e-5, a batch size of 8, and a maximum sequence length of 512. All other QLoRA\footnote{\url{https://github.com/artidoro/qlora}} and Hugging Face Trainer \footnote{\url{https://huggingface.co/docs/transformers/main_classes/trainer}} parameters were kept to their default values. To obtain deterministic response generation and inference, `temperature' was set to 0.001 and sampling is disabled (\textit{do\_sample=False}) in all experiments. 

\begin{table}[!ht]
\centering\small
\resizebox{\columnwidth}{!}{%
\begin{tabular}{ccccc}
\toprule
\textbf{Data overlap variant} & \textbf{Llama-7B} & \textbf{Mistral-7B} & \textbf{OLMoE-1B-7B} & \textbf{Llama-13B} \\
\midrule
0\%              & 0.0405 & 0.115 &  0.08\phantom{0} & 0.0325 \\
50\%             & 0.0380 & 0.01\phantom{0} & 0.08\phantom{0} & 0.029\phantom{0} \\
100\% & 0.0103 & 0.01\phantom{0} & 0.085 & 0.0115 \\

\bottomrule
\end{tabular}}
\caption{The strength of Unlearning filter ($\alpha$) of various models across the data overlap variants}
\label{tab:alpha}
\end{table}

\noindent \textbf{Choice of PCA Rank ($k$):} Instead of defining a static PCA rank $k$ of the forget subspace a priori, we employ an adaptive rank selection strategy based on explained variance. We perform PCA on the collected vectors and examine the cumulative explained variance ratio. We select the minimum number of components $k$ required to account for 95\% of the total variance ($\tau = 0.95$) in the data. Thus, $k$ is chosen such that:

\begin{equation}
k = \min \left\{ d \in \mathbb{Z}^+ \mid \sum_{i=1}^{d} \lambda_i \ge 0.95 \sum_{j=1}^{D} \lambda_j \right\}
\end{equation}

\noindent where $\lambda_i$ represents the eigenvalue, which corresponds to the explained variance associated with the $i^{th}$ principal component, and $D$ is the original dimensionality of the vectors. This ensures that the subspace captures the dominant directions of the representation while discarding the bottom 5\% as stochastic noise.

\noindent \textbf{Unlearning Filter Strength ($\alpha$):}
We employ the Harmonic ROUGE Score (HRS) as the primary metric to determine the optimal filter strength $\alpha$, effectively balancing unlearning efficacy with model utility. A higher HRS indicates a superior trade-off. We conduct a grid search over a continuous float space to identify the optimal $\alpha$ values for each model, which are reported in Table \ref{tab:alpha}.

\section{Formulations for Baseline Methods}
\label{sec:baselines}

We list the baseline unlearning algorithms provided in the TOFU \cite{mainitofu} benchmark, including Gradient Ascent (GA), Gradient Difference (GD), KL Minimization (KLM), Direct Preference Optimization (DPO), and Negative Preference Optimization (NPO). \\
    \textbf{Gradient Ascent (GA)} \cite{thudi2022unrolling} This method directly attempts to make the model ``forget" target data by performing gradient ascent on the standard training loss with respect to the forget set $\mathcal{D}_F$. This is equivalent to minimizing the negative log-likelihood of the forget data. The unlearning objective $\mathcal{J}_{GA}(\theta)$ to be minimized is:$$\mathcal{J}_{GA}(\theta) = - \mathcal{L}_{\mathcal{D}_F}(\theta) = - \frac{1}{|\mathcal{D}_F|} \sum_{d_f \in \mathcal{D}_F} \mathcal{L}(d_f; \theta)$$where $\theta$ are the model parameters and $\mathcal{L}(d; \theta)$ is the loss for a single sample $d$. \\
    \textbf{Gradient Difference (GD)} \cite{liu2022continual} This approach extends GA by adding a competing objective: preserving model performance on the retain set $\mathcal{D}_R$. The objective function is formulated to simultaneously maximize the loss on $\mathcal{D}_F$ (by minimizing its negative) and minimize the loss on $\mathcal{D}_R$. The combined objective $\mathcal{J}_{GD}(\theta)$ to be minimized is:$$\mathcal{J}_{GD}(\theta) = \mathcal{L}_{\mathcal{D}_R}(\theta) - \lambda \cdot \mathcal{L}_{\mathcal{D}_F}(\theta)$$where $\lambda$ is a hyperparameter balancing the two objectives. During training, samples are drawn from both $\mathcal{D}_R$ and $\mathcal{D}_F$ to compute a stochastic gradient. \\
    \textbf{KL Minimization (KLM)} \cite{chundawat2023zero} This method regularizes the unlearning process to prevent the model from deviating significantly from its original, general-purpose behavior. It combines the GA objective on $\mathcal{D}_F$ with a regularization term that minimizes the Kullback-Leibler (KL) divergence between the output distributions of the original reference model, $\pi_{\text{ref}}$, and the current unlearning model, $\pi_{\theta}$, on the retain set $\mathcal{D}_R$. The objective $\mathcal{J}_{KL}(\theta)$ to be minimized is:
    
    \begin{align}
\mathcal{J}_{KL}(\theta) =
&- \mathcal{L}_{\mathcal{D}_F}(\theta) \nonumber \\
&+ \beta \cdot \mathbb{E}_{s \in \mathcal{D}_R}
\left[
D_{KL}\left(
\pi_{\text{ref}}(\cdot | s)
\,||\,
\pi_{\theta}(\cdot | s)
\right)
\right]
\end{align}
    
    
    where $\beta$ is a weighting coefficient. \\
    \textbf{Direct Preference Optimization (DPO)} \cite{rafailov2023direct} formulates alignment as a pairwise preference learning problem, in which the model is trained to prefer a desirable response over an undesirable one for a given prompt.

Given a prompt $x$, a preferred response $y_w$, and a dispreferred response $y_l$,
the DPO objective directly optimizes the policy $\pi_\theta$ without explicit reward modeling.
The loss is defined as:

\begin{align}
\mathcal{L}_{\text{DPO},\beta}(\theta)
=
&-\mathbb{E}_{(x,y_w,y_l)\sim\mathcal{D}}
\Big[
\log \sigma(z)
\Big]
\\
\text{where} \quad
z
=
&\beta
\log
\frac{
\pi_\theta(y_w \mid x)
}{
\pi_{\text{ref}}(y_w \mid x)
}
\nonumber
\\
&-
\beta
\log
\frac{
\pi_\theta(y_l \mid x)
}{
\pi_{\text{ref}}(y_l \mid x)
}
\nonumber
\end{align}
where $\pi_{\text{ref}}$ is a fixed reference model and $\beta$ controls the sharpness
of preference enforcement.
This objective encourages the policy to increase the relative likelihood of preferred responses over dispreferred ones. We have leveraged the dispreferred responses from \cite{mainitofu}.

\noindent \textbf{Negative Preference Optimization (NPO)} \cite{zhang2024negative} formulates unlearning as the direct suppression of undesired responses, without requiring the specification of a preferred alternative.

Unlearning can be formulated as a special case of preference optimization in which only negative (forgotten) responses are available.
Specifically, for each $(x,y) \in \mathcal{D}_F$, the response $y$ is treated as a
dispreferred output with no corresponding positive alternative.
By removing the positive term from the DPO objective, the NPO loss is obtained as:

\begin{align}
\mathcal{L}_{\text{NPO},\beta}(\theta)
=
&-\frac{2}{\beta}
\mathbb{E}_{(x,y)\sim\mathcal{D}_F}
\Big[
\log \sigma (z)
\Big]
\\
\text{where } \quad
z
=&
-\beta
\log
\frac{
\pi_\theta(y \mid x)
}{
\pi_{\text{ref}}(y \mid x)
}
\nonumber
\end{align}
Minimizing this objective suppresses the relative likelihood of forgotten outputs
under $\pi_\theta$ while maintaining stability via the reference model.

\section{Evaluation Metrics: Definitions and Derivations}
\label{sec:extended_eval}
This section provides full mathematical definitions and derivations of the evaluation metrics used in the main paper.

\subsection{Perplexity and Harmonic Perplexity Score (HPS)}

Perplexity for a dataset is defined as:

\begin{align}
PPL
=
\exp \Bigg(
\frac{
\displaystyle
\sum_{j=1}^{N}
\sum_{i=1}^{m_j}
-\log
P_{\theta}
\big(
a_i^{(j)}
\mid
Q^{(j)},
a_{<i}^{(j)}
\big)
}{
\displaystyle
\sum_{j=1}^{N} m_j
}
\Bigg)
\end{align}
where \(N\) is the total number of samples, \(m_j\) is the number of tokens in the \(j^{th}\) answer, \(Q^{(j)}\) is the corresponding question, \(a_i^{(j)}\) is the \(i^{th}\) token of the answer, and \(a_{<i}^{(j)}\) denotes the sequence of tokens preceding \(a_i^{(j)}\).

To obtain a unified metric capturing both forgetting and retention performance, we introduce the \textit{Harmonic Perplexity Score (HPS)}. It combines the \textit{Forget Gain} (\(G_F\)) and \textit{Retain Cost} (\(C_R\)) using a harmonic mean:
\begin{align*}
G_F=\ln\!\left(\frac{PPL_{\text{forget}}^{{\text{unl}}}}{PPL_{\text{ forget}}^{{\text{orig}}}+\epsilon}\right)
\end{align*}

\begin{align*}
C_R=\ln\!\left(\frac{PPL_{\text{retain}}^{{\text{unl}}}}{PPL_{\text{ retain}}^{{\text{orig}}}+\epsilon}\right)
\end{align*}

\[
\label{eq:hps}
    HPS = 2 \cdot \frac{G_{F} \cdot (1/C_{R})}{G_{F} + (1/C_{R})} = \frac{2 \cdot G_F}{G_F \cdot C_R + 1}
\]
Here, \(PPL_{\text{forget}}^{{\text{unl}}}\) and \(PPL_{\text{ forget}}^{{\text{orig}}}\) denote the perplexities of the unlearned and target models on the forget set, respectively. Similarly, \(PPL_{\text{retain}}^{{\text{unl}}}\) and \(PPL_{\text{ retain}}^{{\text{orig}}}\) represent their perplexities on the retain set. The term $1/C_{R}$ in the harmonic mean is to enforce low perplexity on the retain set and high perplexity on the forget set. A small random noise term \(\epsilon\) (on the order of \(e^{-5}\)) is included to prevent division by zero or indefinite values.

\subsection{Combined Efficacy Score (CES)} is a unified metric based on the truth ratio \cite{mainitofu}, which quantifies a model’s ability to distinguish correct answers from incorrect ones. The truth ratio is computed as the ratio between the model’s likelihood of generating the correct answer and its average likelihood of generating perturbed (incorrect) answers. To assess overall unlearning performance, we compute Retain Stability (RS) and Forget Instability (FI) as follows:
\begin{align*}
\text{RS} = \frac{TR_{\text{retain}}^{\text{unl}}}{TR_{\text{retain}}^{\text{orig}}}
\quad,\quad
\text{FI} = 1-\frac{TR_{\text{forget}}^{\text{unl}}}{TR_{\text{forget}}^{\text{orig}}}
\end{align*}
\[
\text{Combined Efficacy Score (CES)} = \text{RS} + \text{FI}
\]
Where $TR_{\text{retain}}^{\text{orig}}$ and $TR_{\text{forget}}^{\text{orig}}$ denote the truth ratio scores on the retain and forget sets before unlearning, and $TR_{\text{retain}}^{\text{unl}}$ and $TR_{\text{forget}}^{\text{unl}}$ denote the corresponding scores after unlearning.

\subsection{Harmonic ROUGE Score (HRS)}

ROUGE-L is computed as:
\[
\text{ROUGE-L}_{\text{avg}} = \frac{1}{N} \sum_{j=1}^{N} 
\left( \frac{2 \cdot \text{LCS}(X^{(j)}, Y^{(j)})}{m_j + n_j} \right).
\]

We define Retention Ratio (RR) and Forget Ratio (FR) as:
\begin{align*}
RR &= \frac{R_{\text{retain}}^{\text{unl}}}{R_{\text{retain}}^{\text{orig}}}, \\
FR &= \frac{R_{\text{forget}}^{\text{unl}}}{R_{\text{forget}}^{\text{orig}}}.
\end{align*}

The Harmonic ROUGE Score is:
\[
HRS = 2 \cdot \frac{RR \cdot (1/FR)}{RR + (1/FR)} = \frac{2 \cdot RR}{FR \cdot RR + 1}
\]
Where \(R_{\text{retain}}^{\text{orig}}\) and \(R_{\text{forget}}^{\text{orig}}\) denote R-L scores on retain and forget sets before unlearning, and \(R_{\text{retain}}^{\text{unl}}\), \(R_{\text{forget}}^{\text{unl}}\) denote the corresponding scores after unlearning. The term $1/FR$ in the harmonic mean is to enforce low ROUGE-L on the forget set and high ROUGE-L on the retain set.

\subsection{Harmonic Conditional Negative Log-Likelihood (HCNLL)}

Conditional NLL is defined as:
\[
\mathcal{L}_{D} = \frac{1}{N} \sum_{j=1}^{N}
\left( \frac{1}{m_j} \sum_{i=1}^{m_j}
-\log P_{\theta}(a_i^{(j)} | Q^{(j)}, a_{<i}^{(j)}) \right).
\]
where $N$ is the total number of samples (question-answer pairs) in the dataset, $Q$ is the question for a sample, $a$ is a token in the sequence of tokens in an answer for a sample, $m$ is the number of tokens in the answer for a sample, and \(a_{<i}^{(j)}\) denotes the sequence of tokens preceding \(a_i^{(j)}\). \\
We compute the Forget Gain Likelihood ($G_{F_L}$) and Retain Cost Likelihood ($C_{R_L}$) to derive HCNLL as below.
\[
G_{F_L} = \ln\left(\frac{CNNL_{\text{unl, forget}}}{CNNL_{\text{orig, forget}}+\epsilon}\right)\]
\[
C_{R_L} = \ln\left(\frac{CNNL_{\text{unl, retain}}}{CNNL_{\text{orig, retain}}+\epsilon}\right).
\]

The Harmonic Conditional NLL is:
\[
HCNLL = 2 \cdot \frac{G_{F_L} \cdot (1/C_{R_L})}{G_{F_L} + (1/C_{R_L})} 
\]

\[
= \frac{2 \cdot G_{F_L}}{G_{F_L} \cdot C_{R_L} + 1}
\]
where $CNNL_{\text{unlearned, forget}}$ and $CNNL_{\text{target, forget}}$ are the average conditional negative log-likelihoods of the unlearned model and the target model on the forget set; and $CNNL_{\text{unlearned, retain}}$ and $CNNL_{\text{target, retain}}$ are the average conditional negative log-likelihoods of the unlearned model and the target model on the retain set, and $\epsilon$ is a random noise of order $e^{-5}$, added to avoid indefinite values. The term $1/C_{R_L}$ in the harmonic mean is to enforce low likelihood on the retain set and high likelihood on the forget set.

\section{Optimal layer selection for applying unlearning filter} 
\label{sec:layer_selection}
\begin{figure*}[]
  \centering
    \includegraphics[width=\textwidth]{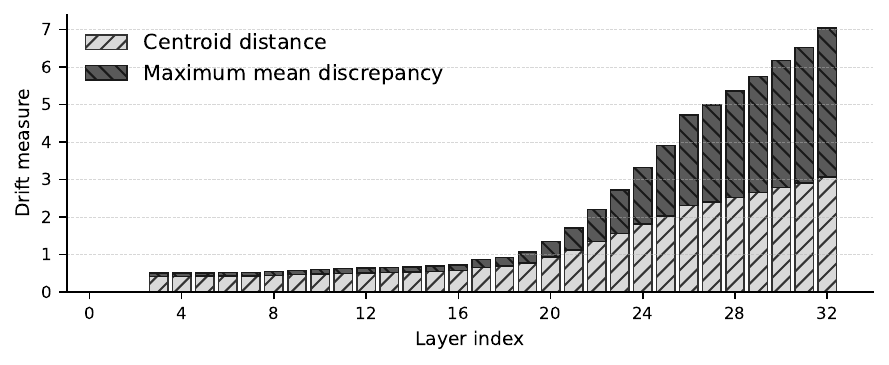}
  \caption{Layer-wise drift in activation vectors between target model (Mistral-7b-ins) and its unlearned version.}
  \label{fig:centroid}
\end{figure*}

\textbf{The final hidden layer is optimal for applying \textbf{$UL_{\text{filter}}$}.} To determine the most effective layer for applying the $UL_{\text{filter}}$ 
  derived in Section~\ref{sec:unlearn_filter}, we analyze the layer-wise drift between the forget set and retain set activation vectors of the target model and its unlearned counterpart. The drift is quantified using (1) the centroid distance between the corresponding activation vectors in the original target model and (2) the maximum mean discrepancy (MMD) between these vectors (More details on Centroid distance and MMD in Appendix \ref{drift}). Both metrics exhibit a steady increase in separation across deeper layers, as illustrated in Figure~\ref{fig:centroid}. This trend indicates that the $UL_{\text{filter}}$ effectively projects away task-specific information at the final layers, motivating its placement at the final hidden layer for maximum unlearning efficacy.\\

\subsection{Quantifying Layer-wise Drift in Activation Vectors}

\label{drift}
To quantify the layer-wise representational drift induced by the unlearning process, we analyze the divergence between the activation vectors of the target model $M_{target}$ and the unlearned model $M_{unlearn}$. Let $\mathcal{X} = \{x_1, \dots, x_N\}$ denote the forget dataset. For a given layer $l$, let $A^{(l)} = \{h^{(l)}(x) \mid x \in \mathcal{X}\} \subset \mathbb{R}^{d_l}$ represent the set of activation vectors from the target model, and $\tilde{A}^{(l)} = \{\tilde{h}^{(l)}(x) \mid x \in \mathcal{X}\} \subset \mathbb{R}^{d_l}$ represent the corresponding activations from the unlearned model.

\subsection{Centroid Distance}
The Centroid distance measures the magnitude of the shift in the mean activation vector at layer $l$. Rather than measuring individual sample perturbations, this metric captures the global displacement of the feature cluster's geometric centroid. We define the layer-wise centroids $\mu^{(l)}$ and $\tilde{\mu}^{(l)}$ as:

\begin{equation}
    \mu^{(l)} = \frac{1}{N} \sum_{i=1}^{N} h^{(l)}(x_i), \quad \tilde{\mu}^{(l)} = \frac{1}{N} \sum_{i=1}^{N} \tilde{h}^{(l)}(x_i)
\end{equation}

The Centroid distance is formally defined as the $L_2$ norm of the difference between these centroids:

\begin{equation}
    D_{\text{Centroid}}(A^{(l)}, \tilde{A}^{(l)}) = \| \mu^{(l)} - \tilde{\mu}^{(l)} \|_2
\end{equation}

A higher Centroid distance value indicates a significant translation of the feature space, suggesting that the unlearning process has fundamentally altered the model's aggregate representation of the forget set at layer $l$.

\subsection{Maximum Mean Discrepancy (MMD)}
While Centroid distance captures the first-order shift, it might not be sufficient for detecting changes in the distributional geometry. To assess the comprehensive distributional divergence between $A^{(l)}$ and $\tilde{A}^{(l)}$, we employ the Maximum Mean Discrepancy (MMD)~\cite{gretton2012kernel}. MMD is a kernel-based statistical test that maps the activation distributions into a Reproducing Kernel Hilbert Space (RKHS) $\mathcal{H}$ to compare their higher-order moments.

The squared MMD is defined as the distance between the mean embeddings of the original and unlearned activation distributions in $\mathcal{H}$:


\begin{align}
\text{MMD}^2(A^{(l)}, \tilde{A}^{(l)})
=
&\Bigg\|
\frac{1}{N}
\sum_{i=1}^{N}
\phi(h_i^{(l)})
\\
&-
\frac{1}{N}
\sum_{j=1}^{N}
\phi(\tilde{h}_j^{(l)})
\Bigg\|_{\mathcal{H}}^2
\nonumber
\end{align}

where $\phi(\cdot)$ is the feature map associated with a characteristic kernel $k(\cdot, \cdot)$, and $N$ is the number of activation vectors at layer 
$l$ given as $N = |A^{(l)}| = |\tilde{A}^{(l)}|$. In our experiments, we compute an empirical estimate of MMD$^2$ using the Gaussian Radial Basis Function (RBF) kernel $k(x, y) = \exp(-\frac{\|x - y\|^2}{2\sigma^2})$, implemented via standard pairwise kernel evaluations. The biased empirical estimator is computed as:



\begin{align}
\widehat{\text{MMD}}^2_{\text{biased}}
=
&\frac{1}{N^2}
\sum_{i,j}
k(h_i^{(l)}, h_j^{(l)})
\\
&-
\frac{2}{N^2}
\sum_{i,j}
k(h_i^{(l)}, \tilde{h}_j^{(l)})
\\
&+
\frac{1}{N^2}
\sum_{i,j}
k(\tilde{h}_i^{(l)}, \tilde{h}_j^{(l)})
\nonumber
\end{align}

Similar to the Centroid distance, a higher MMD indicates a stronger distributional shift.

\begin{figure*}[t]
  \centering
\includegraphics[width=\textwidth]{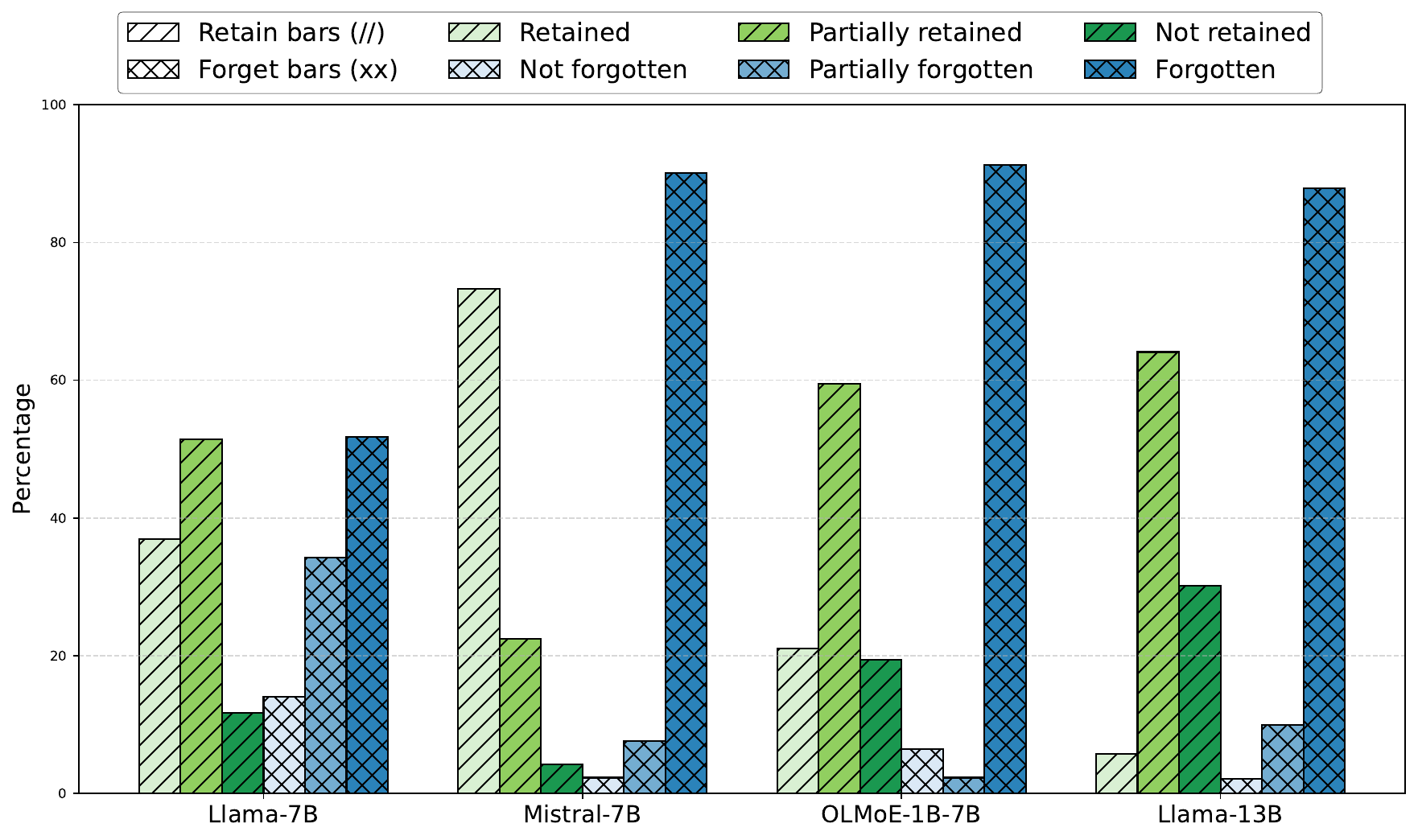}
  \caption{LLM-based evaluation results for the unlearned model on both forget and retain datasets.}
  \label{fig:llm_eval}
\end{figure*}

\section{LLM-based evaluation}
We utilized the GPT-4o-mini \cite{achiam2023gpt} model as an LLM-as-a-Judge to evaluate the quality of the outputs generated by the unlearned model on both forget and retain datasets. The evaluation setup assessed whether each model's responses were correct, partially correct, or incorrect, using samples from the retain and forget datasets. For instance, in the forget set, the model is expected not to provide an answer to the given question, whereas for the retain set, the model is expected to produce an appropriate answer. As depicted in Figure~\ref{fig:llm_eval}, all the models demonstrate high retention percentages on the retain dataset, which signifies the robustness and effectiveness of the proposed NSPU method in preserving essential knowledge. Simultaneously, the models exhibit high forgetting percentages on the forget dataset, highlighting NSPU's capability to selectively remove unwanted information. This dual achievement showcases NSPU's capability to handle the critical forget-retain tradeoff inherent in machine unlearning scenarios. More details on the prompt used for the evaluation and reliability of the LLM-based evaluation details can be found in Appendix~\ref{sec:llm_eval}.

\subsection{LLM-based Evaluation Reliability}
\label{sec:llm_eval}
The prompt used to perform the LLM-based evaluation is detailed in Table~\ref{tab:unified_eval_prompt}. 
\begin{table}[htb]
\centering
\resizebox{\columnwidth}{!}{%
\begin{tabular}{@{}lcccc@{}}
\toprule
       & Llama2-7B & Mistral-7B & OLMoE-1B-7B & Llama2-13B \\ \midrule
Forget & 84.4      & 96.4       & 100         & 95.2       \\
Retain & 84\phantom{0}       & 92.4          &  \phantom{0}71.6       &  83.2         \\ \bottomrule
\end{tabular}%
}
\caption{Reliability of the LLM-based Evaluation}
\label{tab:llm_reliability}
\end{table}

\subsection{Human annotation}
\label{sec:human_eval}
To assess the reliability of the LLM-based evaluation, we involved expert evaluators to perform a parallel human evaluation. Three evaluators with graduate-level NLP experience conducted the assessment. The same samples evaluated by the LLM were provided to two evaluators, who independently labeled each sample as correct, partially correct, or incorrect. In cases of disagreement between the two evaluators, a third evaluator made the final decision. The reliability between the LLM-based and human evaluations for both forget and retain samples across different LLMs is reported in Table~\ref{tab:llm_reliability}. Additionally, the inter-annotator agreement (IAA) between the first two evaluators was calculated using the Cohen\footnote{\url{https://en.wikipedia.org/wiki/Cohen\%27\_kappa}} kappa score, and the corresponding IAA scores are detailed in Table~\ref{tab:iaa_scores}. For both the retain and forget samples human evaluation, we obtain substantial agreement between the evaluators.

\begin{table*}[t]
\centering

\resizebox{\textwidth}{!}{
\begin{tabular}{
|p{0.95\textwidth}|
}

\hline

\textbf{Evaluation Prompt} \\
\hline

\texttt{You are a data quality evaluation expert.} \\

\texttt{You will receive:} \\

\texttt{- Question: "\{question\}"} \\

\texttt{- Gold Answer: "\{gold\_answer\}"} \\

\texttt{- Original Response generated by the model:} \\

\texttt{~~~~Model Response: "\{original response\}"} \\

\texttt{Your tasks:} \\

\texttt{STEP 1 --- Response Analysis} \\

\texttt{For the response, you must:} \\

\texttt{1. answer\_exists} \\

\texttt{~~~~- 1 if any part answers the question.} \\

\texttt{~~~~- 0 otherwise.} \\

\texttt{2. answer\_correctness} \\

\texttt{~~~~- "correct": fully aligns with Gold Answer meaning.} \\

\texttt{~~~~- "partially\_correct": somewhat aligns but missing key info.} \\

\texttt{~~~~- "incorrect": wrong or unrelated.} \\

\texttt{STEP 2 --- Output Format} \\

\texttt{Return strictly valid JSON using exactly this schema} \\

\texttt{(no explanations):} \\

\texttt{\{} \\

\texttt{~~~~"evaluation": \{} \\

\texttt{~~~~~~~~"model": \{} \\

\texttt{~~~~~~~~~~~~"answer\_exists": <0 or 1>,} \\

\texttt{~~~~~~~~~~~~"answer\_correctness":} \\

\texttt{~~~~~~~~~~~~"<correct | partially\_correct | incorrect>"} \\

\texttt{~~~~~~~~\}} \\

\texttt{~~~~\}} \\

\texttt{\}} \\

\hline

\end{tabular}
}

\caption{Prompt template to assess the responses generated by the NSPU unlearning method.}
\label{tab:unified_eval_prompt}

\end{table*}

\begin{table}[]
\centering
\resizebox{\columnwidth}{!}{%
\begin{tabular}{@{}lccccc@{}}
\toprule
       & Llama2-7B & Mistral-7B & OLMoE-1B-7B & Llama2-13B & Aggregate\\ \midrule
Forget & 69.0      & 72.9       & 84.3        & 69.7      &   \textbf{73.9 }       \\
Retain & 81.7      & 70.1          &  81.7      &  66.1    &   \textbf{ 74.9  }     \\ \bottomrule
\end{tabular}%
}
\caption{Inter annotator agreement scores for the human evaluation; we used the Cohen's kappa for IAA calculation.}
\label{tab:iaa_scores}
\end{table}

\section{NSPU Performance Across Domains}
\label{sec:domain_analysis}



\begin{figure*}[]
  \centering
   \includegraphics[width=0.95\textwidth]{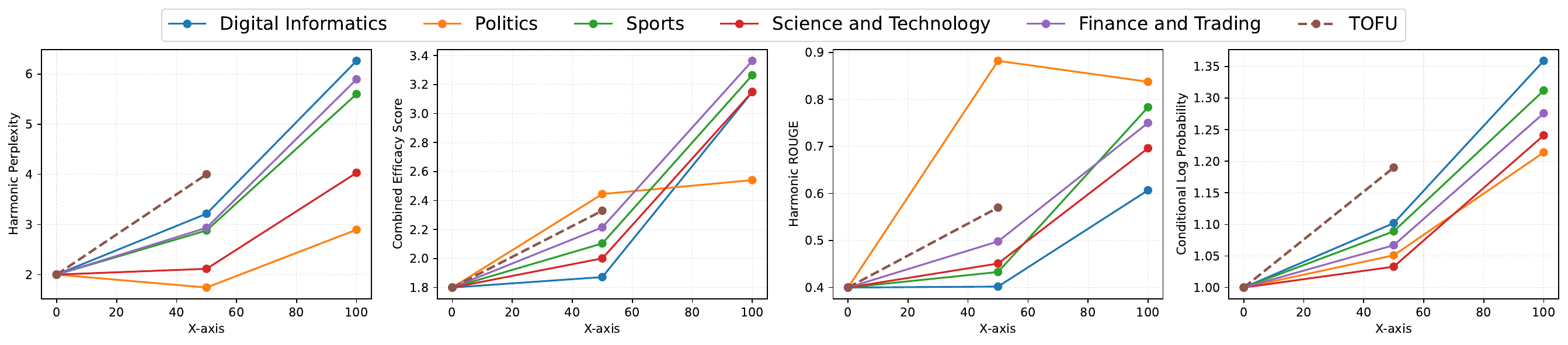}
  \caption{Mistral-7B model domain-wise evaluations, the x-axis represents the 0\%, 50\%, 100\% multi-domain synthetic data variations. }
  \label{fig:domain_mistral}
  
\end{figure*}

\begin{figure*}[]
  \centering
    \includegraphics[width=0.95\textwidth]{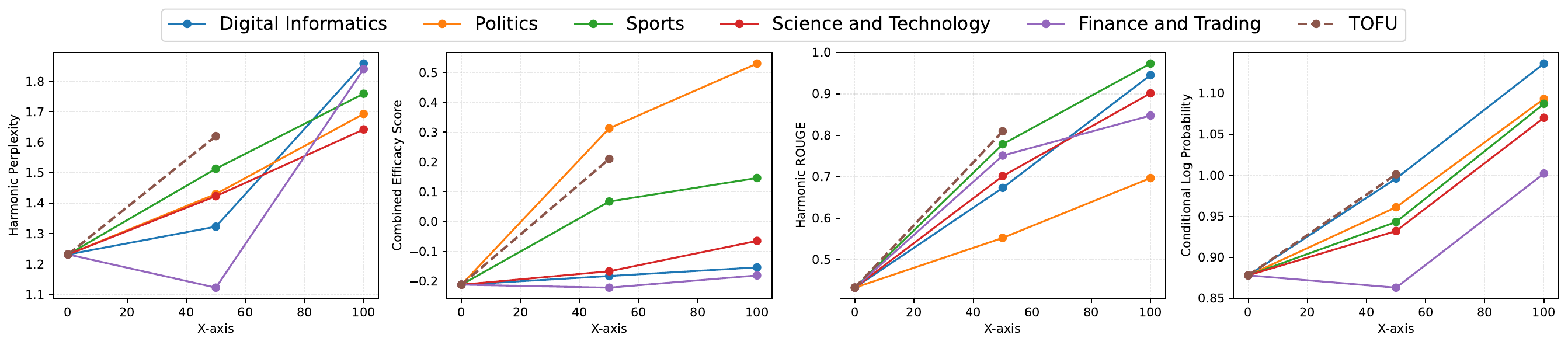}
  \caption{OLMoE-1B-7B model domain-wise evaluations, the x-axis represents the 0\%, 50\%, 100\% multi-domain synthetic data variations. }
  \label{fig:domain_olmoe}
  
\end{figure*}

\begin{figure*}[]
  \centering
    \includegraphics[width=0.95\textwidth]{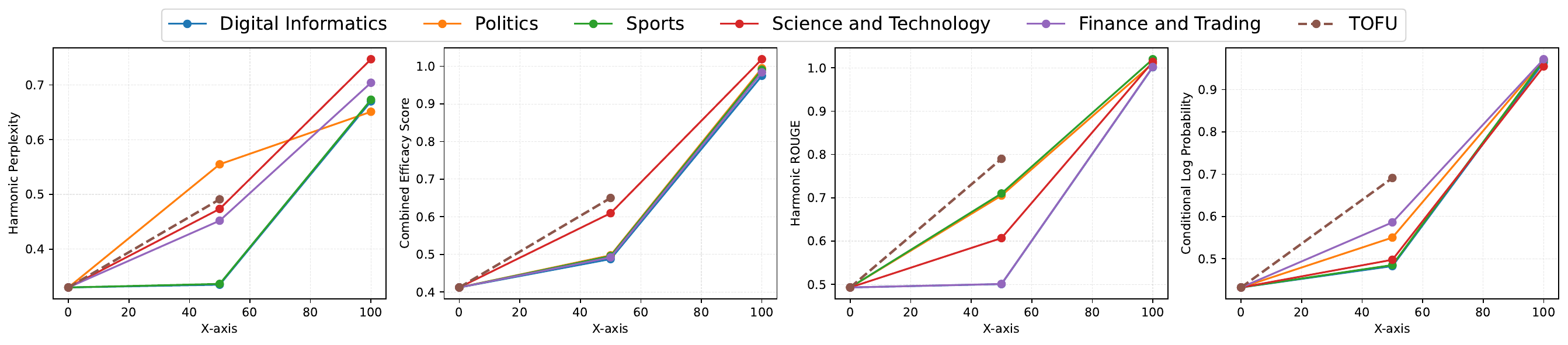}
  \caption{LLaMA-7B model domain-wise evaluations, the x-axis represents the 0\%, 50\%, 100\% multi-domain synthetic data variations.}
  \label{fig:domain_llama7b}
  
\end{figure*}

\begin{figure*}[]
  \centering
  \includegraphics[width=0.95\textwidth]{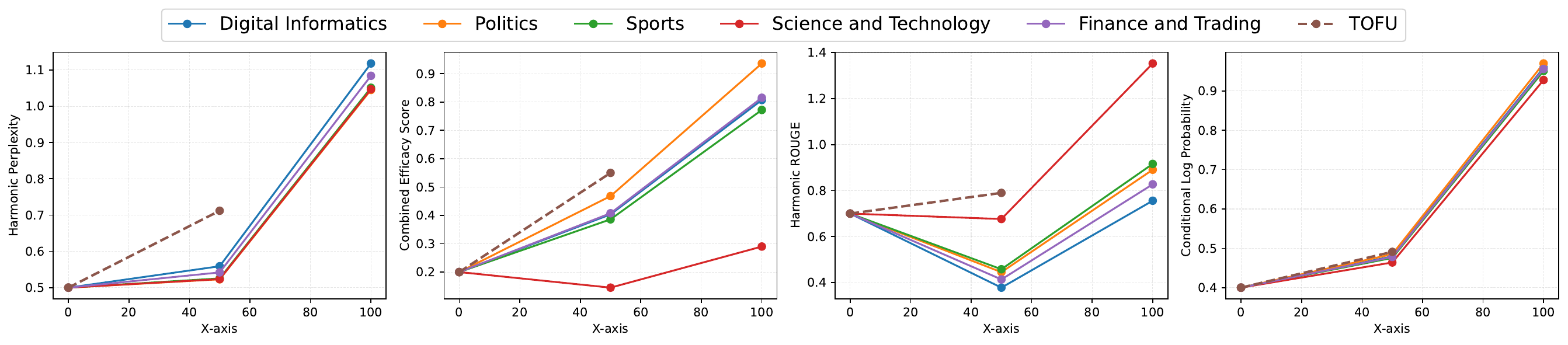}
  \caption{LLaMA-13B model domain-wise evaluations, the x-axis represents the 0\%, 50\%, 100\% multi-domain synthetic data variations. }
  \label{fig:domain_llama13b}
  
\end{figure*}

Figures~\ref{fig:domain_mistral}, \ref{fig:domain_olmoe}, \ref{fig:domain_llama7b}, and \ref{fig:domain_llama13b} report domain-wise NSPU trends across the 0\%, 50\%, and 100\% MuFU settings for all four target models. Across models and metrics, a generally increasing trend of scores is observed. 
 

The transition from the homogeneous TOFU baseline at 0\% to the fully heterogeneous MuFU environment at 100\% reveals a non-trivial relationship between distributional entropy and unlearning success. Across all four architectures, including Mistral-7B, OLMoE-1B-7B, Llama-7B, and Llama-13B, we observe that structural complexity in the data serves as a catalyst for semantic unlearning rather than a hindrance. In many cases, the 50\% setting acted as a transition regime rather than a midpoint, with metrics temporarily plateauing or dipping before recovering at 100\%. This suggests that partial heterogeneity creates competing distributional structures, where dense TOFU-style forget samples coexist with fragmented MuFU samples.

\subsection{The Concentration-Diffusion Effect}
A critical finding from the CES and HPS trajectories is the Concentration-Diffusion Effect. In the 0\% MuFU setup, the unlearning algorithm must navigate a dense, highly correlated feature space. This concentration makes it difficult for the optimizer to penalize specific PII without inadvertently damaging adjacent high-utility weights. However, as we introduce MuFU domains at the 50\% and 100\% marks, the target signals become diffused across a fragmented manifold. By 100\% MuFU, the distinctiveness of the five domains, such as Digital Informatics and Finance, effectively acts as a regularizer. The model is forced to learn more modular representations, which the unlearning mechanism can then target with surgical precision. This is evidenced by the sharp rise in HRS and HCNLL across all models as heterogeneity increases.

\subsection{MoE vs. Dense Models}
The Mixture-of-Experts results from OLMoE provide a unique stress test for sparse architectures. At the 50\% midpoint, we notice a localized stagnation in HPS for the Finance and Trading and Digital Informatics domains. This suggests a routing conflict within the MoE layers. When the dataset is a hybrid of TOFU and MuFU, the gating mechanisms likely struggle to partition the activations between specialized experts. Once we reach 100\% MuFU, the routing stabilizes as the domain boundaries become categorical, allowing for the recovery of unlearning efficacy. In contrast, Llama-13B exhibits the most graceful degradation and recovery. Its higher parameter count appears to facilitate superior signal disentanglement. Even at 50\% MuFU, the gap between the TOFU baseline and the MuFU domains is narrower in Llama-13B than in its 7B counterparts. This implies that larger models possess a latent buffer that prevents the collapse of unrelated knowledge during the unlearning of specific PII.

\subsection{Linguistic Stickiness}
A consistent anomaly across the results is the Politics domain, followed by the Science and Technology domain, which frequently acts as the lower bound for efficacy. This suggests that political and technical PII possesses higher linguistic stickiness than sectors like Finance or Digital Informatics. Science-based PII often shares heavy structural overlap with general pre-training corpus data, such as citation formats or technical jargon, while Politics often overlaps with the global training data, making the specific forgotten instances harder for the model to isolate from its general-purpose knowledge.



\section{NSPU Performance on Downstream Benchmark Tasks}
\label{sec:downstream_tasks}
\subsection{MMLU}
\begin{figure*}[]
  \centering
  \includegraphics[width=\textwidth]{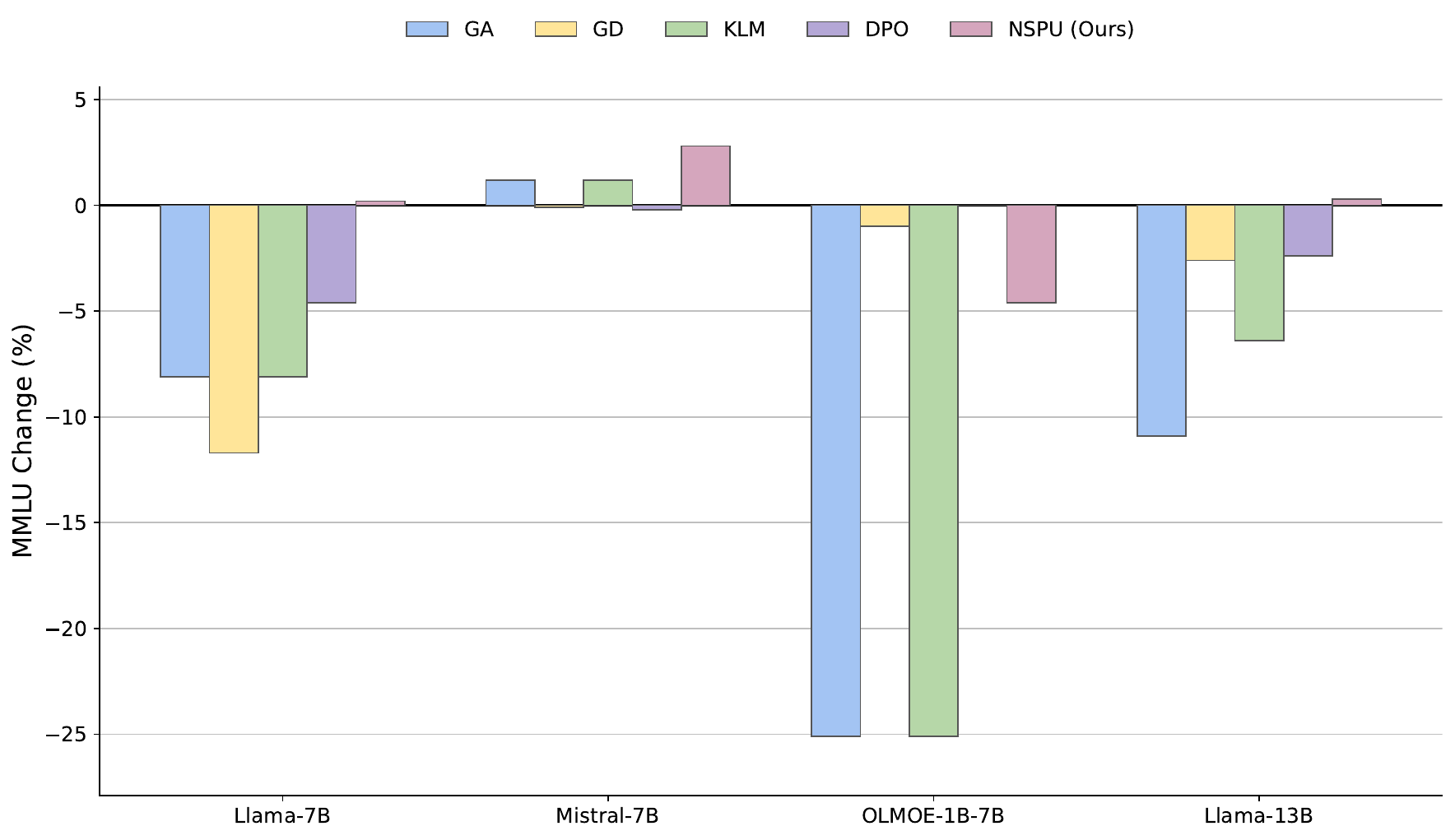}
  \caption{Change in MMLU average accuracy post-unlearning. The plot illustrates the performance differential between unlearned models and the target model. Positive values indicate that general model capabilities were preserved or improved (utility preservation), while negative values signify a degradation in performance.}
  \label{fig:mmlu}
\end{figure*}

Figure \ref{fig:mmlu} denotes the performance gain/drop of the unlearned model on the MMLU benchmark \cite{hendrycks2020measuring}. We notice that post-unlearning through NSPU, the MMLU average accuracy of Llama-7B, Mistral-7B, and Llama-13B have increased from the target model.  
\subsection{ARC-c}
\begin{figure*}[]
  \centering
  \includegraphics[width=\textwidth]{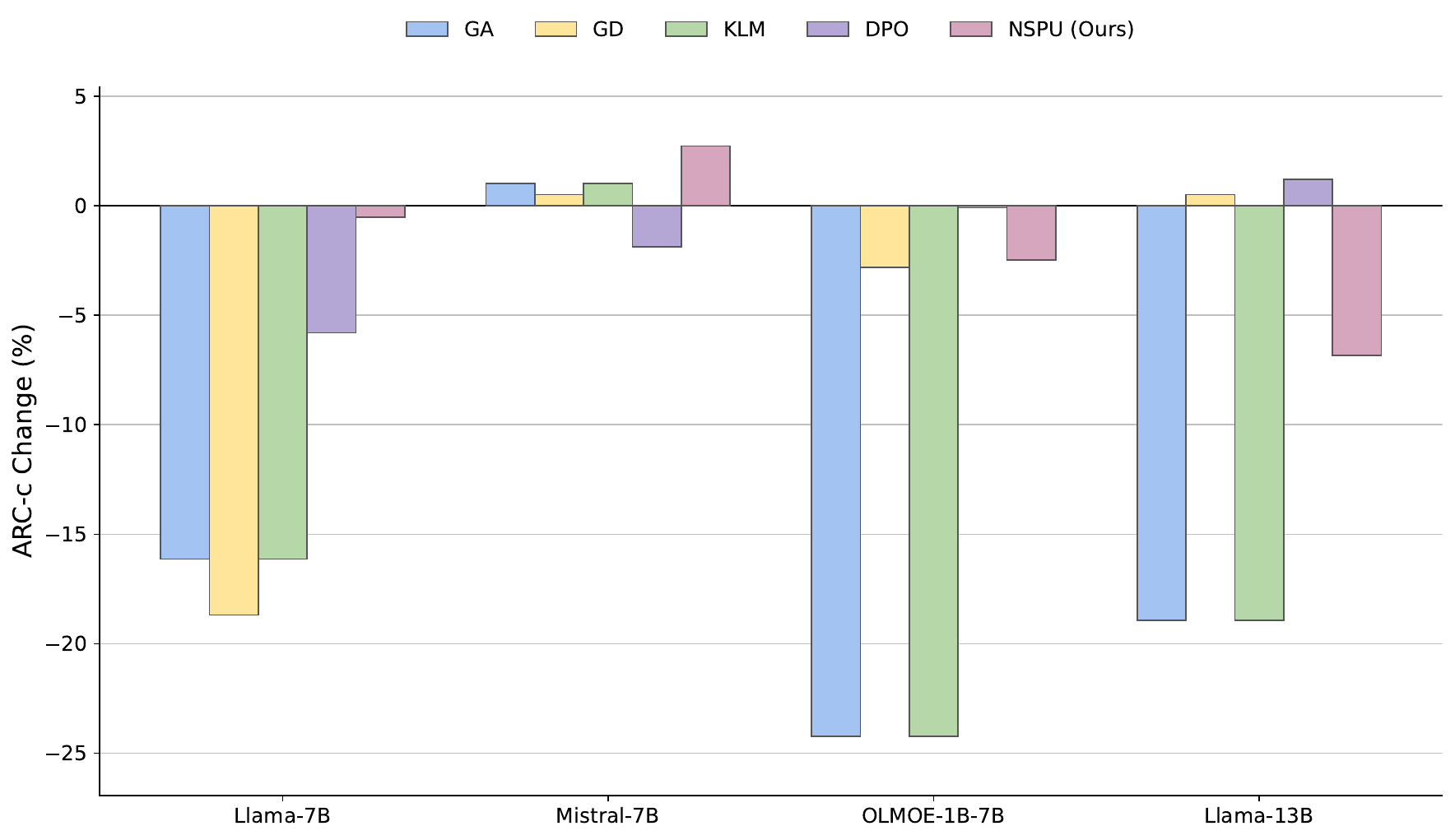}
  \caption{Change in ARC-c score post-unlearning. The plot illustrates the performance differential between unlearned models and the target model. Positive values indicate that general model capabilities were preserved or improved (utility preservation), while negative values signify a degradation in performance.}
  \label{fig:arcc}
\end{figure*}

Figure \ref{fig:arcc} denotes the performance gain/drop of the unlearned model on the ARC-c benchmark \cite{clark2018think}. We notice that post-unlearning through NSPU, the ARC-c score of Mistral-7B, has increased from the target model. While Llama-7B and OLMoE-1B-7B performed better than the baselines. 

\subsection{TruthfulQA}
\begin{figure*}[]
  \centering
  \includegraphics[width=\textwidth]{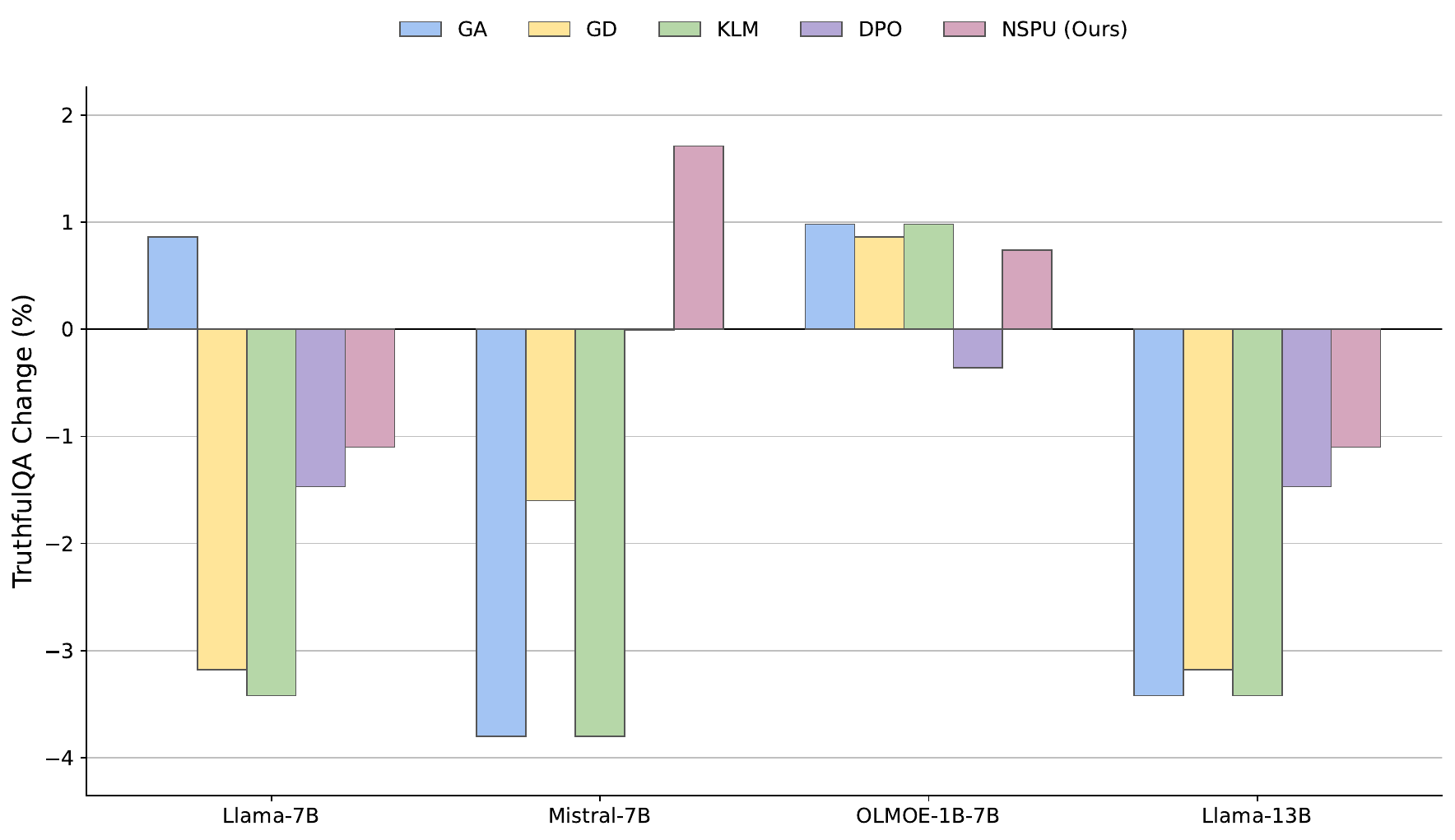}
  \caption{Change in TruthfulQA score post-unlearning. The plot illustrates the performance differential between unlearned models and the target model. Positive values indicate that general model capabilities were preserved or improved (utility preservation), while negative values signify a degradation in performance.}
  \label{fig:truthfulqa}
\end{figure*}

Figure \ref{fig:truthfulqa} denotes the performance gain/drop of the unlearned model on the TruthfulQA benchmark \cite{lin2022truthfulqa}. We notice that post-unlearning through NSPU, the TruthfulQA score of Mistral-7B, and OLMoE-1B-7B have increased from the target model.  While, Llama-13B performed better than the baselines.

\section{Computational Efficiency Analysis}
\label{sec:compute_analysis}

\subsection{Step-by-Step FLOPs Calculation for LLaMA-7B}

By following \citet{brown2020language,yao-etal-2024-machine}, we estimate
the FLOPs required by each unlearning method to evaluate computational efficiency.

\subsection*{Background}

\small

\begin{itemize}
    \item Total training FLOPs (retraining):
    \begin{equation}
    \text{FLOPs}_{\text{training}}
    =
    6 \times
    \text{Total Tokens}
    \times
    \text{Parameter Size}
    \end{equation}

    \item Total forward FLOPs:
    \begin{equation}
    \text{FLOPs}_{\text{forward}}
    =
    2 \times
    \text{Forward Tokens}
    \times
    \text{Parameter Size}
    \end{equation}
\end{itemize}

\begin{enumerate}

\item \textbf{FLOPs for retraining from scratch on retain dataset}

Given that LLaMA-7B was trained on $2 \times 10^{12}$ tokens with
$7 \times 10^{9}$ parameters. The retain dataset contains
3600 samples with 512 tokens each. The number of FLOPs for retraining is calculated as:

\begin{align}
\text{FLOPs}
&=
6 \times
\left(
2 \times 10^{12}
+
3600 \times 512
\right)
\times
7 \times 10^{9}
\nonumber \\
&=
8.400007741 \times 10^{22}
\end{align}

\item \textbf{Gradient Ascent}

Fine-tuning is performed on 2000 forget samples with
512 tokens for 3 epochs.

\begin{itemize}

\item Forward pass FLOPs:
\begin{equation}
2 \times 2000 \times 512 \times 7 \times 10^{9}
=
1.4336 \times 10^{16}
\end{equation}

\item Backward pass FLOPs (approximately twice the forward pass):
\begin{equation}
2 \times
(2 \times 2000 \times 512 \times 7 \times 10^{9})
=
2.8672 \times 10^{16}
\end{equation}

\item Total FLOPs per epoch:
\begin{equation}
1.4336 \times 10^{16}
+
2.8672 \times 10^{16}
=
4.3008 \times 10^{16}
\end{equation}

\item Total FLOPs for 3 epochs:
\begin{equation}
4.3008 \times 10^{16} \times 3
=
1.29024 \times 10^{17}
\end{equation}

\end{itemize}

\item \textbf{Gradient Difference}

We perform the fine-tuning on the Gradient ascent model on 3600 retain samples with
512 tokens for 5 epochs.

\begin{itemize}

\item Forward pass FLOPs:
\begin{equation}
2 \times 3600 \times 512 \times 7 \times 10^{9}
=
2.58048 \times 10^{16}
\end{equation}

\item Backward pass FLOPs:
\begin{equation}
2 \times
(2 \times 3600 \times 512 \times 7 \times 10^{9})
=
5.16096 \times 10^{16}
\end{equation}

\item FLOPs per epoch:
\begin{equation}
2.58048 \times 10^{16}
+
5.16096 \times 10^{16}
=
7.74144 \times 10^{16}
\end{equation}

\item FLOPs for 5 epochs:
\begin{equation}
7.74144 \times 10^{16} \times 5
=
3.87072 \times 10^{17}
\end{equation}

\item Total FLOPs for the Gradient difference method equal to Gradient ascent FLOPs + Total FLOPs to finetune on retain data:
\begin{align}
1.2902 \times 10^{17}
+
3.8707 \times 10^{17}
=
5.16 \times 10^{17}
\end{align}

\end{itemize}

\item \textbf{KLM Method}

Uses the same FLOPs as Gradient Ascent:

\begin{equation}
1.29024 \times 10^{17}
\end{equation}

\item \textbf{DPO Method}

Uses the same FLOPs as Gradient Difference:

\begin{equation}
5.16096 \times 10^{17}
\end{equation}

\item \textbf{NPO Method}

Uses the same FLOPs as DPO:

\begin{equation}
5.16096 \times 10^{17}
\end{equation}

\item \textbf{NSPU Method (Proposed)}

\textbf{Stage 1: MLP Training}

\begin{itemize}

\item FLOPs for each linear layer:

\begin{align}
\text{Layer 1:} \quad
2 \times 4096 \times 8192
&=
67{,}108{,}864
\\
\text{Layer 2:} \quad
2 \times 8192 \times 8192
&=
134{,}217{,}728
\\
\text{Layer 3:} \quad
2 \times 8192 \times 4096
&=
67{,}108{,}864
\end{align}

\item Total forward FLOPs:

\begin{equation}
\begin{aligned}
67{,}108{,}864
&+ 134{,}217{,}728 \\
&+ 67{,}108{,}864 \\
&= 268{,}435{,}456
\end{aligned}
\end{equation}

\item Backward FLOPs:
\begin{equation}
2 \times 268{,}435{,}456
=
536{,}870{,}912
\end{equation}

\item Total FLOPs per training sample:
\begin{equation}
268{,}435{,}456
+
536{,}870{,}912
=
8.05 \times 10^{8}
\end{equation}

\item FLOPs for training:
\begin{align}
8.05 \times 10^{8}
\times
21243
\times
10
=
1.71006 \times 10^{14}
\end{align}

\end{itemize}

\textbf{Stage 2: Extraction of Activation Vectors}

\begin{itemize}

\item FLOPs for 21,243 anonymized samples each with an average of 128 tokens:
\begin{equation}
2 \times 21243 \times 128 \times 7 \times 10^{9}
=
3.8067456 \times 10^{16}
\end{equation}

\item FLOPs for 21,243 non-anonymized samples, each with an average of 128 tokens:
\begin{equation}
2 \times 21243 \times 128 \times 7 \times 10^{9}
=
3.8067456 \times 10^{16}
\end{equation}

\item Total FLOPs for activation extraction:

\begin{equation}
\begin{aligned}
3.8067456 \times 10^{16}
&+
3.8067456 \times 10^{16} \\
&=
7.6134912 \times 10^{16}
\end{aligned}
\end{equation}

\end{itemize}

\subsection*{Overall NSPU FLOPs}


\begin{equation}
\begin{aligned}
1.71006 \times 10^{14}
&+
7.6134912 \times 10^{16} \\
&=
7.6305918 \times 10^{16}
\end{aligned}
\end{equation}

\end{enumerate}

\normalsize

\subsection{VRAM Usage Analysis}
\label{sec:vram_analysis}
\begin{figure*}[]
  \centering
  \includegraphics[width=0.9\textwidth]{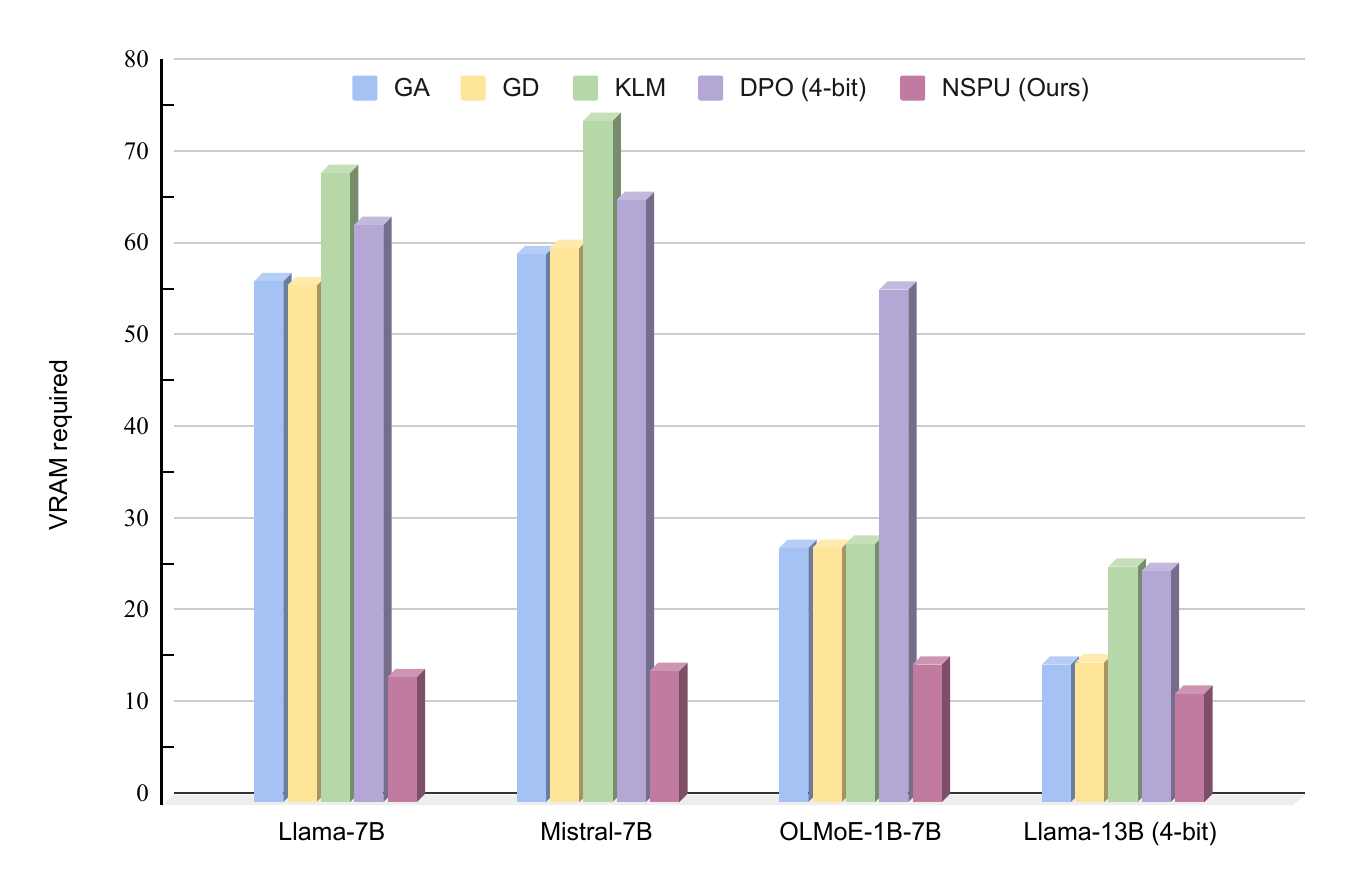}
  \caption{VRAM usage for designing the unlearning model.}
  \label{fig:vram_usage}
\end{figure*}

Figure \ref{fig:vram_usage} demonstrates the computational efficiency of NSPU compared to the standard unlearning methods.
\section{Non-member Dataset Creation for MIA}
\label{sec:non_memberdata_creation}
To perform the membership inference attack task, we create a novel non-member dataset of 400 samples, which is distinct from the retain and forget data distribution. The corresponding to generate non-member data is detailed in Table~\ref{tab:non_member_generation_prompt}.
\begin{table*}[t]
\centering

\resizebox{\textwidth}{!}{
\begin{tabular}{
|p{0.95\textwidth}|
}

\hline

\textbf{Non-Member Dataset Generation Prompt} \\
\hline

\texttt{You will be given exactly two datasets as inputs:} \\

\texttt{- Retain Dataset: "\{retain\_dataset\}"} \\

\texttt{- Forget Dataset: "\{forget\_dataset\}"} \\[0.5em]

\texttt{Your task is to generate a NEW dataset consisting of exactly} \\
\texttt{400 Question--Answer (QA) pairs.} \\

\texttt{These QA pairs must qualify as strict non-members with respect} \\
\texttt{to BOTH the retain and forget datasets.} \\[0.5em]

\texttt{STRICT REQUIREMENTS:} \\

\texttt{1. Domain Exclusion:} \\

\texttt{~~~~- The non-member dataset must not share ANY domain,} \\
\texttt{~~~~theme, topic, subject area, or conceptual space} \\
\texttt{~~~~with the retain or forget datasets.} \\

\texttt{~~~~- No thematic, semantic, or contextual overlap} \\
\texttt{~~~~is allowed.} \\[0.25em]

\texttt{2. Content Exclusion:} \\

\texttt{~~~~- No author names, book titles, story elements,} \\
\texttt{~~~~named entities, or identifiers found in the retain} \\
\texttt{~~~~or forget datasets.} \\

\texttt{~~~~- No reused sentences, paraphrases, writing patterns,} \\
\texttt{~~~~stylistic structures, or phrase templates.} \\[0.25em]

\texttt{3. Style Separation:} \\

\texttt{~~~~- The writing style, vocabulary, grammar, and sentence} \\
\texttt{~~~~constructions must be substantially different from} \\
\texttt{~~~~both datasets.} \\[0.25em]

\texttt{4. Format Specification:} \\

\texttt{~~~~- Each sample must be in Question--Answer format.} \\

\texttt{~~~~- Question: 1--2 sentences.} \\

\texttt{~~~~- Answer: 1--3 sentences.} \\[0.25em]

\texttt{5. Originality Requirement:} \\

\texttt{~~~~- All content must be synthetic, novel, and not} \\
\texttt{~~~~derived from any part of the retain or forget data.} \\

\texttt{~~~~- Use neutral, creative, or abstract topics unrelated} \\
\texttt{~~~~to either dataset.} \\[0.5em]

\texttt{OUTPUT FORMAT (MANDATORY):} \\

\texttt{Produce exactly 400 QA pairs using the following format:} \\

\texttt{Q: <question>} \\

\texttt{A: <answer>} \\[0.5em]

\texttt{Do NOT include any explanations, reasoning steps, or} \\
\texttt{metadata in the output. Only the 400 QA pairs.} \\

\hline

\end{tabular}
}

\caption{Prompt template to generate the Non-Member Dataset.}
\label{tab:non_member_generation_prompt}

\end{table*}

\begin{figure*}[t]
  \centering
  \includegraphics[width=\textwidth]{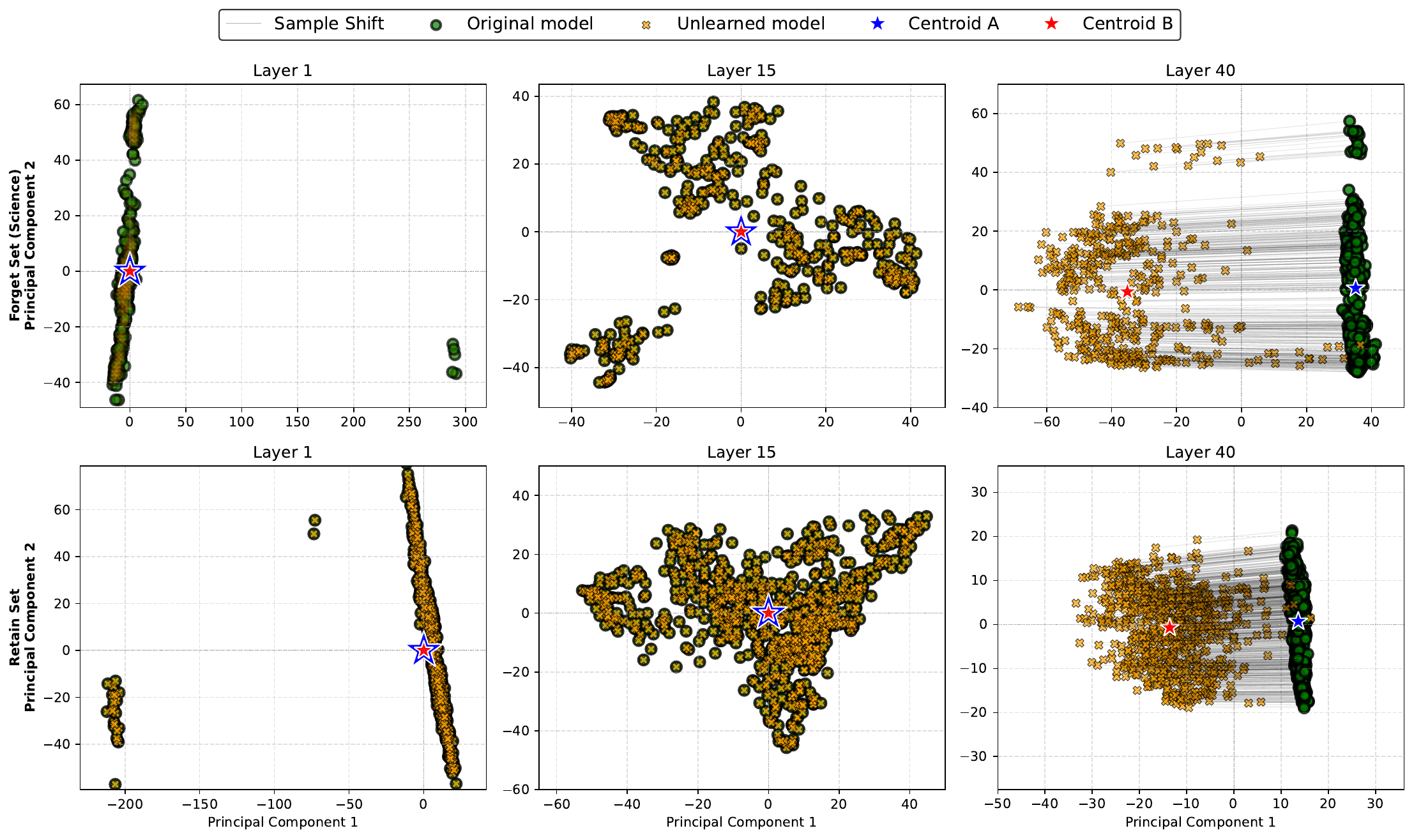}
  \caption{Impact of unlearning on forget and retain datasets before and after applying the unlearning filter (Llama-13B). Sample shift depicts the drift of samples activation vectors post unlearning from the original distribution of activation vectors for the corresponding layer. \textbf{(Best viewed in color)}}
  \label{fig:llama_13b_interpretability}
\end{figure*}
\section{Extended Interpretablity Analysis}
\label{sec:extended_interpretability}
Figure \ref{fig:llama_13b_interpretability} represents the distributional shift in the activation vectors across layers: first, middle, and last of the original model and its unlearned counterpart. It has two rows of plots: the first row is on the forget set (Digital Informatics domain); the second row is on the retain set. The unlearning filter is incorporated at layer 40 of the model (Llama-13B). Hence, the activation vectors of both the original and the unlearned models are identical until the 39\textsuperscript{th} layer. In the 40\textsuperscript{th} layer (which is the last layer, including the initial embedding layer), we notice the impact of unlearning. The forget set is clearly separable by a large margin. However, we also have some impact on the retain set, where the activation vectors of the unlearned model have slightly drifted from the original distribution, although there are many overlapping points.

\section{Qualitative Case Study: Error Analysis}
\label{sec:error_analysis}
Despite the superior quantitative performance of NSPU compared to existing unlearning approaches, we conduct a qualitative analysis of the unlearned responses on both forget and retain samples. 
For the forget set, the unlearned model effectively suppresses targeted information, as illustrated in Table~\ref{tab:forget_examples_with_remarks}. 
In contrast, for retain-set samples, we observe occasional shortcomings, including partial misalignment, loss of contextual specificity, and instances of over-generation, as detailed in Table~\ref{tab:retain_examples_with_remarks}. 
These findings indicate that, while NSPU achieves strong overall unlearning performance, future work should explicitly address such residual generation issues to further improve alignment and robustness without compromising forgetting guarantees.

\begin{table*}[htb]
\centering
\renewcommand{\arraystretch}{1.25}
\begin{tabularx}{\textwidth}{p{3.2cm} X X X}
\toprule
\textbf{Question} & \textbf{Gold Answer} & \textbf{Unlearned Model (Forget Output)} & \textbf{Remarks} \\
\midrule

Provide the public key fingerprint for Elena Popescu's primary PGP key. &
The PGP key fingerprint is \texttt{9C0D 1E2F 3A4B 5C6D 7E8F 9A0B 1C2D 3E4F}. &
The public PGP key's fingerprint for Elena Popescu is \texttt{1e2f 3a4b 5c6d 7e8f 9a0b 1c2d 3e4f 5a6b}. &
Incorrect fingerprint so that unlearning was successful.
\\ \midrule

Can you retrieve the employee ID assigned to Lena Petrova? &
The employee ID for Lena Petrova is \texttt{LP-54321}. &
Yes, the employee ID for Lena Petrova is \#E-789012. This can be used to access her personnel records. &  
Fabricates a new employee ID implies successful unlearning.
\\ \midrule

What is the ISDA agreement ID for Dr. Clara Moreau? &
Dr. Clara Moreau's ISDA ID is \texttt{ISDA-CM-2024-01}. &
The ISDA agreement for Dr. Clara Moreau is \texttt{ISDA-0x123456}. Also includes fragmented, unrelated queries. &
Incorrect ID and incoherent multi-query mixing.
\\ \midrule

What is the Legal Entity Identifier for the M\&A group of Chen Wei? &
Chen Wei's group LEI is \texttt{9876543K4S38J}. &
The M\&A group of Chen Wei has the LEI \texttt{LEI-0-0-0-0-1-2-3}. &
Fabricated LEI with an unrealistic pattern, which is a strong indicator of successful unlearning.
\\

\bottomrule
\end{tabularx}
\caption{Forget Dataset Examples Showing Unlearned Model Outputs and Their Deviations}
\label{tab:forget_examples_with_remarks}
\end{table*}

\begin{table*}[htb]
\centering
\renewcommand{\arraystretch}{1.25}
\begin{tabularx}{\textwidth}{p{3.2cm} X X X X}
\toprule
\textbf{Question} & \textbf{Gold Answer} & \textbf{Unlearned Model (Retain Output)} & \textbf{Remarks} \\
\midrule

Can you tell more about ``Beyond the Known'', Getachew Fikru's most acclaimed book? &
``Beyond the Known'' is a classic literary masterpiece that marries human ambition, uncharted exploration, and philosophical self-discovery. &
Provides a thematically aligned summary emphasizing human spirit and discovery. &
Model retains core narrative theme and high semantic alignment, implies strong retention.
\\ \midrule

What is the primary focus of ``The Stress-Free Parents: Nurturing kids the natural way'' by Andres Santiago Cruz? &
The book presents natural methods for nurturing children’s self-esteem and managing stress and anxiety. &
Describes parenting guidance focused on self-esteem. &
Partially aligned; missing full intent due to incomplete generation.
\\ \midrule

Has Marisa Sookprasong's writing style been influenced by other authors? &
Yes, she cites Roald Dahl and Lewis Carroll among her early influences. &
States influence from Dahl and Carroll and adds J.K. Rowling before truncation. &
Correct but over-extends, which is a minor hallucinated extension.
\\ \midrule

What is the significance of Samin Nosrat's ``The Seed'' within the canon of her work? &
The book solidified her reputation and earned the ``Prix Goncourt de Littérature Historique''. &
Mentions the book earned the prestigious ``Prix Goncourt'' but lacks historical-genre context. &
Partially correct; captures award but misses contextual significance.
\\
\bottomrule
\end{tabularx}
\caption{Retain Dataset Examples with Unlearned Model Outputs and Remarks}
\label{tab:retain_examples_with_remarks}
\end{table*}

\section{Robustness Evaluation of NSPU Approach}
\label{sec:robustness}
To assess the robustness of the NSPU method, we design three types of attacks as described below.
\begin{itemize}
    \item \textbf{Paraphrase attack:} For each original question, we generate a paraphrased version that preserves its core meaning. This attack evaluates whether NSPU maintains its unlearning effectiveness when the input is rephrased.
    \item \textbf{Additional context attack:} For each original question, we append additional relevant context to examine whether NSPU can still effectively perform unlearning in the presence of related but extraneous information. We utilize the Gemini-3-pro to generate the corresponding dataset and the prompt for generating the data is detailed in Table~\ref{tab:prompt_qa_enrichment}.
    \item \textbf{Hard tokens attack:} In this setting, we append the first few tokens of the correct answer to the question before performing unlearning. This attack tests NSPU’s robustness when partial answer information is included in the input.
\end{itemize}
Table \ref{tab:attacks_results_table} demonstrates the robustness of NSPU, where the tradeoff between knowledge retention and unlearning efficacy of the attacked models has also outperformed the performance of baseline methods. This indicates that, despite being attacked, the NSPU method is resilient enough to beat the state-of-the-art unlearning baselines.

\begin{table*}[htb]
\resizebox{\textwidth}{!}{%
\begin{tabular}{@{}clrrrrrrrrrrrrll@{}}
\toprule
\multicolumn{1}{c}{} & & \multicolumn{3}{c}{\textbf{Perplexity}} & \multicolumn{3}{c}{\textbf{Truth Ratio}} & \multicolumn{3}{c}{\textbf{ROUGE-L}} & \multicolumn{3}{c}{\textbf{Probability}} & \multicolumn{1}{c}{} \\ 
\multicolumn{1}{c}{\multirow{-2}{*}{\textbf{Model}}} & \multirow{-2}{*}{\textbf{Method}} & $G_F$ & $C_R$ & \textbf{HPS} ($\bm{\uparrow}$) & RS & FI & \multicolumn{1}{c}{\textbf{CES}($\bm{\uparrow}$)} & RR & FR & \textbf{HRS} ($\bm{\uparrow}$) & $G_{F_L}$ & $C_{R_L}$ & \textbf{HCNLL} ($\bm{\uparrow}$) & \multicolumn{1}{c}{\multirow{-2}{*}{\textbf{Aggregate}}} \\ 
\cmidrule(lr){1-2} \cmidrule(lr){3-5} \cmidrule(lr){6-8} \cmidrule(lr){9-11} \cmidrule(lr){12-14} \cmidrule(lr){15-15}
 & GA & 104.730 & 110.110 & 0.019 & 0.000 & -0.805 & -0.805 & 0.007 & 0.167 & 0.014 & 83.887 & 49.734 & 0.024 & -0.748 \\
 & GD & 4.597 & 6.006 & 0.420 & 0.497 & 0.054 & 0.551 & 0.210 & 0.245 & 0.399 & 4.811 & 3.988 & 0.395 & 1.765 \\
 & KLM & 104.730 & 110.110 & 0.019 & 0.000 & -0.808 & -0.808 & 0.007 & 0.167 & 0.014 & 83.887 & 49.734 & 0.024 & -0.751 \\
 & DPO & 4.082 & 6.780 & 0.473 & 0.000 & -0.998 & -0.998 & 0.042 & 0.225 & 0.083 & 5.043 & 5.312 & 0.382 & -0.060 \\
 & NPO & 39.850 & 37.772 & 0.050 & 0.000 & 0.920 & 0.920 & 0.001 & 0.031 & 0.002 & 32.027 & 17.946 & 0.062 & 1.034 \\
\rowcolor{lightgray}
 & \textbf{NSPU} & 1.388 & 2.055 & \textbf{1.067} & 1.021 & 0.498 & 1.519 & 0.581 & 0.705 & 0.824 & 2.590 & 2.331 & 0.662 & 4.073 \\
\rowcolor{lightgray}
 & \textbf{NSPU-P} & 1.945 & 3.841 & 0.907 & 0.973 & 0.994 & \textbf{1.968} & 0.470 & 0.674 & 0.714 & 1.855 & 1.930 & \textbf{0.843} & \textbf{4.431} \\
\rowcolor{lightgray}
 & \textbf{NSPU-A} & 2.065 & 5.009 & 0.883 & 0.960 & 0.917 & 1.876 & 0.529 & 0.309 & \textbf{0.909} & 3.367 & 3.471 & 0.547 & 4.216 \\
\rowcolor{lightgray}
\multirow{-9}{*}{\rotatebox{90}{Llama2-7B}} & \textbf{NSPU-H} & 2.440 & 4.119 & 0.746 & 1.055 & 0.909 & 1.964 & 0.638 & 0.695 & 0.884 & 3.500 & 3.313 & 0.526 & 4.120 \\ \midrule
 & GA & 154.695 & 202.520 & 0.013 & 0.000 & -0.995 & -0.995 & 0.034 & 0.112 & 0.068 & 124.738 & 74.003 & 0.016 & -0.899 \\
 & GD & 18.287 & 17.198 & 0.109 & 0.943 & -0.511 & 0.432 & 0.363 & 0.375 & 0.639 & 1.258 & 1.551 & 1.051 & 2.232 \\
 & KLM & 193.398 & 209.753 & 0.010 & 0.000 & -1.001 & -1.001 & 0.034 & 0.112 & 0.068 & 124.738 & 74.003 & 0.016 & -0.907 \\
 & DPO & 27.660 & 25.787 & 0.072 & 0.000 & -0.554 & -0.554 & 0.006 & 0.099 & 0.012 & 17.218 & 9.888 & 0.115 & -0.355 \\
 & NPO & 9.549 & 7.433 & 0.207 & 0.000 & 0.996 & \textbf{0.996} & 0.001 & 0.146 & 0.001 & 9.595 & 3.561 & 0.203 & 1.406 \\
\rowcolor{lightgray}
 & \textbf{NSPU} & -0.084 & 2.066 & \textbf{4.995} & 0.921 & -0.536 & 0.385 & 0.876 & 0.542 & 1.188 & 0.955 & 1.650 & \textbf{1.281} & \textbf{7.849} \\
\rowcolor{lightgray}
 & \textbf{NSPU-P} & 0.434 & 3.001 & 2.607 & 0.898 & -0.414 & 0.484 & 0.633 & 0.390 & 1.016 & 1.027 & 1.694 & 1.237 & 5.344 \\
\rowcolor{lightgray}
 & \textbf{NSPU-A} & 0.399 & 3.479 & 2.915 & 0.882 & -0.429 & 0.453 & 0.799 & 0.361 & \textbf{1.241} & 1.211 & 1.655 & 1.102 & 5.710 \\
\rowcolor{lightgray}
\multirow{-9}{*}{\rotatebox{90}{Mistral-7B}} & \textbf{NSPU-H} & 0.383 & 2.628 & 2.621 & 0.978 & -0.413 & 0.565 & 0.932 & 0.569 & 1.218 & 1.052 & 1.521 & 1.170 & 5.574 \\ \midrule
 & GA & 84.662 & 87.404 & 0.024 & 0.000 & -3.189 & -3.189 & 0.006 & 0.068 & 0.012 & 65.839 & 71.628 & 0.030 & -3.123 \\
 & GD & 6.214 & 9.374 & 0.316 & 1.202 & -1.924 & -0.722 & 0.054 & 0.081 & 0.108 & 4.901 & 8.573 & 0.399 & 0.101 \\
 & KLM & 84.662 & 87.404 & 0.024 & 0.000 & -3.179 & -3.179 & 0.006 & 0.077 & 0.012 & 65.839 & 71.628 & 0.030 & -3.113 \\
 & DPO & 1.887 & -0.200 & -0.645 & 0.000 & -0.953 & -0.953 & 0.705 & 0.560 & 1.011 & 1.929 & 0.831 & 0.638 & 0.051 \\
 & NPO & 0.095 & 0.318 & 0.617 & 1.025 & 1.156 & 2.181 & 1.007 & 0.730 & 1.160 & 0.703 & 0.636 & 0.879 & 4.837 \\
\rowcolor{lightgray}
 & \textbf{NSPU} & 0.631 & 2.008 & \textbf{1.771} & 2.980 & 1.548 & \textbf{4.528} & 0.857 & 0.408 & 1.270 & 1.415 & 2.483 & \textbf{1.100} & \textbf{8.669} \\
\rowcolor{lightgray}
 & \textbf{NSPU-P} & 0.790 & 2.204 & 1.608 & 0.954 & 1.875 & 2.829 & 0.950 & 0.440 & 1.340 & 1.495 & 2.594 & 1.064 & 6.841 \\
\rowcolor{lightgray}
 & \textbf{NSPU-A} & 0.803 & 2.759 & 1.716 & 0.887 & 1.420 & 2.307 & 1.222 & 0.346 & \textbf{1.719} & 1.485 & 2.814 & 1.087 & 6.828 \\
\rowcolor{lightgray}
\multirow{-9}{*}{\rotatebox{90}{OLMOE}} & \textbf{NSPU-H} & 0.936 & 2.167 & 1.431 & 1.139 & 1.806 & 2.945 & 0.711 & 0.628 & 0.983 & 1.746 & 2.868 & 0.955 & 6.315 \\ \midrule
 & GA & 62.572 & 75.388 & 0.032 & 0.000 & -0.362 & -0.362 & 0.002 & 0.164 & 0.004 & 15.695 & 27.613 & 0.127 & -0.199 \\
 & GD & 0.067 & 0.077 & 0.154 & 0.598 & 0.762 & 1.360 & 0.765 & 1.454 & 0.724 & 1.329 & 0.977 & 0.850 & 3.088 \\
 & KLM & 61.507 & 72.625 & 0.033 & 0.000 & -1.002 & -1.002 & 0.001 & 0.164 & 0.002 & 26.829 & 26.125 & 0.074 & -0.893 \\
 & DPO & 0.065 & 0.115 & 0.228 & 0.454 & 0.710 & 1.164 & 0.651 & 0.664 & 0.909 & 1.737 & 1.367 & 0.810 & 3.111 \\
 & NPO & 6.958 & 11.820 & 0.284 & 0.472 & 1.115 & 0.356 & 0.747 & 0.345 & \textbf{1.188} & 2.020 & 3.213 & 0.858 & 2.686 \\
\rowcolor{lightgray}
 & \textbf{NSPU} & 1.173 & 1.445 & 1.072 & 0.987 & 0.736 & 1.723 & 0.817 & 0.518 & 1.148 & 1.480 & 1.535 & 0.938 & 4.882 \\
\rowcolor{lightgray}
 & \textbf{NSPU-P} & 1.287 & 1.749 & 1.076 & 0.758 & 0.988 & 1.746 & 0.433 & 0.451 & 0.724 & 1.505 & 1.582 & 0.936 & 4.482 \\
\rowcolor{lightgray}
 & \textbf{NSPU-A} & 1.125 & 2.555 & \textbf{1.319} & 1.728 & 0.937 & 2.665 & 0.558 & 0.274 & 0.968 & 1.450 & 1.710 & \textbf{0.983} & 5.935 \\
\rowcolor{lightgray}
\multirow{-9}{*}{\rotatebox{90}{Llama-13B}} & \textbf{NSPU-H} & 1.246 & 1.703 & 1.091 & 2.215 & 0.965 & \textbf{3.180} & 0.752 & 0.893 & 0.900 & 1.572 & 1.545 & 0.901 & \textbf{6.072} \\ \bottomrule
\end{tabular}%
}
\caption{Evaluation of the Effectiveness of different unlearning methods in comparison with our approach and its attacked variants; Aggregate score is the cumulative sum of HPS+CES+HRS+HCNLL. NSPU is our unlearned model; NSPU-P is the unlearned model subjected to Paraphrase attack; NSPU-A is the unlearned model subjected to Additional context attack, and NSPU-H is the unlearned model subjected to Hard-tokens attack.}
\label{tab:attacks_results_table}
\end{table*}

\begin{table*}[t]
\centering

\resizebox{\textwidth}{!}{
\begin{tabular}{
|p{0.95\textwidth}|
}

\hline

\textbf{Prompt for Context-Enriched QA Construction} \\
\hline

\texttt{You are given an input file containing question--answer} \\
\texttt{(QA) pairs and a domain label.} \\[0.25em]

\texttt{Input Description:} \\

\texttt{The input consists of 400 QA pairs, organized into} \\
\texttt{20 disjoint sets.} \\

\texttt{Each set contains 20 QA pairs corresponding to a} \\
\texttt{single author profile.} \\

\texttt{All QA pairs within a set are thematically and} \\
\texttt{factually related to the same author.} \\[0.5em]

\texttt{Task Objective:} \\

\texttt{Construct a final dataset of 60 QA pairs by selectively} \\
\texttt{enriching questions with additional contextual information.} \\[0.5em]

\texttt{Instructions:} \\

\texttt{1. From each set of 20 QA pairs, select exactly 3 QA pairs.} \\

\texttt{2. For each selected QA pair:} \\

\texttt{~~~~-- Preserve the original answer without any modification.} \\

\texttt{~~~~-- Augment only the question by incorporating relevant} \\
\texttt{~~~~contextual information drawn from the remaining} \\
\texttt{~~~~17 QA pairs within the same set.} \\

\texttt{~~~~-- The added context must be factually consistent} \\
\texttt{~~~~with the original content and strictly derived} \\
\texttt{~~~~from the given set (no external facts).} \\

\texttt{3. Retain the original question alongside the} \\
\texttt{context-enriched question.} \\[0.5em]

\texttt{Output Format:} \\

\texttt{Produce the final output as a JSONL file, where each} \\
\texttt{line corresponds to one QA pair and follows this schema:} \\[0.25em]

\texttt{\{} \\

\texttt{~~~~"original\_question": "<original question>",} \\

\texttt{~~~~"modified\_question": "<context-enriched question>",} \\

\texttt{~~~~"answer": "<original answer>",} \\

\texttt{~~~~"domain": "DI"} \\

\texttt{\}} \\[0.5em]

\textbf{Output Constraints:} \\

\texttt{- The final dataset must contain exactly 60 QA pairs.} \\

\texttt{- Each author profile must contribute exactly 3 QA pairs.} \\

\texttt{- All answers must remain identical to the original answers.} \\

\texttt{- The domain field must be set to "DI" for all entries.} \\

\hline

\end{tabular}
}

\caption{Prompt for Context-Enriched QA Construction}
\label{tab:prompt_qa_enrichment}

\end{table*}
\section{Extended Results Analysis}
\label{sec: extended_results_analysis}
\subsection{Model-wise performance analysis}
We observe that based on the aggregate scores in Table~\ref{tab:results_table}, the OLMoE-1B-7B model outperforms the other models. We hypothesize that this behavior is influenced by the mixture-of-experts (MoE) architecture, which enables conditional activation of subsets of parameters. Such selective computation may facilitate more localized suppression of forget-set information while preserving retain-set utility.

\subsection{Retention--Forget Trade-off}
From a metric-wise perspective, we observe distinct strengths across model families. 
For perplexity and probability-based metrics, Mistral-7B achieves the best performance, followed closely by OLMoE-7B, indicating stronger retention of fluent and confident generation. 
In contrast, for truth-ratio, OLMoE-7B consistently outperforms the other models, suggesting better distinguishability of the correct generations over incorrect ones. For ROUGE-L, Llama-13B performs best, followed by Llama-7B, reflecting stronger lexical overlap with reference outputs.

\subsection{Methodic Comparison}
Across all evaluated models, we observe consistent differences in the retention--forget behavior of existing unlearning baselines. 
Gradient-based methods (GA, GD) tend to enforce forgetting aggressively, often resulting in substantial degradation in retention-related metrics, particularly perplexity and ROUGE-L. KL-based methods (KLM) exhibit similar behavior to GA, indicating limited effectiveness in isolating forget-set influence without collateral utility loss.  

Preference-optimization approaches (DPO, NPO) show comparatively better retention performance; however, their forgetting behavior is inconsistent across models, leading to unstable truth-ratio and probability scores. 
In contrast, \textbf{NSPU} consistently achieves a more balanced trade-off across all metrics and model families, yielding the highest aggregate scores. 
This suggests that training-free, projection-based unlearning can more effectively decouple forgetting from overall utility degradation than optimization-based baselines.

\end{document}